\documentclass[english,aps,prl,superscriptaddress,reprint,longbibliography]{revtex4-1}
\usepackage[T1]{fontenc}
\usepackage{color}
\usepackage{amsmath}
\usepackage{amssymb}
\usepackage{graphicx}
\usepackage{wasysym}

\makeatletter
\usepackage{amsfonts}
\usepackage{xcolor}
\usepackage{transparent}

\makeatother

\usepackage{babel}
\begin{document}

\title{Unconditional steady-state entanglement in \\macroscopic hybrid systems by coherent noise cancellation}

\author{Xinyao Huang}
\affiliation{State Key Laboratory of Mesoscopic Physics, School of Physics, Peking University, Collaborative Innovation Center of Quantum Matter, Beijing 100871, China}
\affiliation{Niels Bohr Institute, University of Copenhagen, DK-2100 Copenhagen,
Denmark}

\author{Emil Zeuthen}
\email[Corresponding author. E-mail: ]{zeuthen@nbi.ku.dk}
\affiliation{Niels Bohr Institute, University of Copenhagen, DK-2100 Copenhagen,
Denmark}

\author{Denis V.~Vasilyev}
\affiliation{Center for Quantum Physics, Faculty of Mathematics, Computer Science and Physics, University of Innsbruck, A-6020 Innsbruck, Austria}
\affiliation{Institute for Quantum Optics and Quantum Information of the Austrian Academy of Sciences, A-6020 Innsbruck, Austria}

\author{Qiongyi He}
\email[E-mail: ]{qiongyihe@pku.edu.cn}
\affiliation{State Key Laboratory of Mesoscopic Physics, School of Physics, Peking University, Collaborative Innovation Center of Quantum Matter, Beijing 100871, China}

\author{Klemens Hammerer}
\affiliation{Institute for Theoretical Physics and Institute for Gravitational Physics (Albert Einstein Institute), Leibniz Universit\"{a}t Hannover, Callinstra{\ss}e 38, 30167 Hannover, Germany}

\author{Eugene S.~Polzik}
\affiliation{Niels Bohr Institute, University of Copenhagen, DK-2100 Copenhagen,
Denmark}

\begin{abstract}

The generation of entanglement between disparate physical objects is a key ingredient in the field of quantum technologies, since they can have different functionalities in a quantum network.
Here we propose and analyze a generic approach to steady-state entanglement generation between two oscillators with different temperatures and decoherence properties coupled in cascade to a common unidirectional light field. The scheme is based on a combination of coherent noise cancellation and dynamical cooling techniques for two oscillators with effective masses of opposite signs, such as quasi-spin and motional degrees of freedom, respectively. The interference effect provided by the cascaded setup can be tuned to implement additional noise cancellation leading to improved entanglement even in the presence of a hot thermal environment. The unconditional entanglement generation is advantageous since it provides a ready-to-use quantum resource.  Remarkably, by comparing to the conditional entanglement achievable in the dynamically stable regime, we find our unconditional scheme to deliver a virtually identical performance when operated optimally.

\end{abstract}
\maketitle

Entanglement is a peculiar property of quantum physics and a key technological resource in quantum information processing~\cite{Horodecki2009} and 
quantum metrology~\cite{Giovannetti2004,QmetroRMP}, allowing improvements of atomic clocks~\cite{Leroux2010,Hosten2016} and optical magnetometers~\cite{Sewell2012,Wasilewski2010}.  Moreover, entanglement is often used to delineate the boundary between classical and quantum physics. 
Generating entanglement for ever-larger objects therefore  establishes the reach of quantum mechanics into the macroscopic realm. 
Entanglement between separate macroscopic systems has already been realized with pairs of atomic vapor ensembles~\cite{Julsgaard2001,Wasilewski2010,Krauter2011} and diamonds~\cite{Lee2011} at room temperature, and mechanical oscillators at cryogenic temperatures~\cite{Riedinger2018,Ockeloen-Korppi2018}. 
However, generation of entanglement in hybrid systems composed of disparate macroscopic objects is  an outstanding challenge---in particular due to the presence of the hot thermal environment. Such hybrid entanglement would combine attractive features of very different systems as required to realize
complex quantum information networks~\cite{Kurizki2015}. 

In this Letter, we devise an efficient scheme for unconditionally entangling two macroscopic systems with potentially very different decoherence properties. 
The scheme works for two generic bosonic oscillators coupled linearly to a unidirectional traveling light field, with the extra provision that their effective masses have opposite signs. 
A negative mass oscillator in the entanglement context was first used in Ref.~\cite{Julsgaard2001}, and further extensively developed for collective degrees of freedom in polarized spin ensembles prepared in an energetically inverted state~\cite{Tsang2012,Polzik2015} such as in atomic ensembles at room temperature in free space~\cite{Hammerer2010,Vasilakis2015}, cold atoms in Bose-Einstein condensates~\cite{Muessel2014}, optical cavities~\cite{McConnell2015,Kohler2018}, or trapped in 1-dimensional arrays~\cite{Vetsch2010,Beguin2014} as well as in solid-state ensembles of nitrogen-vacancy centers~\cite{Grezes2014} and quasi-spins of rare-earth-ion doped crystals~\cite{Jobez2015,Gundogan2015}.
The positive mass subsystem can, naturally, be implemented in a wider range of systems, in particular in motional degrees of freedom, e.g., the center-of-mass motion of ensembles of atoms~\cite{Joeckel2015,SPethmann2016,Kohler2018} or ions~\cite{Gilmore2017} and micromechanical oscillators~\cite{Teufel2011,Peterson2016,Nielsen2017,Riedinger2018,Ockeloen-Korppi2018}.
A motional degree of freedom can also implement an effective negative mass by employing two-tone driving schemes~\cite{Ockeloen-Korppi2016} (see also Ref.~\cite{Tan2013}).

An essential mechanism of our scheme is coherent quantum noise cancellation (CQNC) of the back action (BA) of light on the two oscillators. This hinges on the observation that for two oscillators with masses of opposite signs, $m_{+}=-m_{-}:=m>0$, we have $(d/dt)[\hat{X}_{+}+\hat{X}_{-}]=[\hat{P}_{+}-\hat{P}_{-}]/m$ for which $[\hat{X}_{+}+\hat{X}_{-},\hat{P}_{+}-\hat{P}_{-}]=0$, where $\hat{X}_{\pm}$ and $\hat{P}_{\pm}$ are canonical conjugate variables for the positive (negative) mass oscillator. Hence, this pair of variables 
is classical in the sense that the Heisenberg uncertainty relation imposes no constraint on the simultaneous knowledge of them~\cite{Julsgaard2001,Tsang2012,Polzik2015}. This is possible because (for ideally matched oscillators) the associated measurement BA goes into the canonically conjugate joint variables,
while interfering destructively in the BA-free variables. Measuring the latter beyond the Heisenberg limit of the individual systems entails entanglement between the two oscillators.
CQNC based on this principle has previously been analyzed theoretically in the context of sensing beyond the standard quantum limit (SQL)~\cite{Tsang2010,Tsang2012,Woolley2013,Wimmer2014,Bariani2015,Motazedifard2016} and realized experimentally using two mechanical oscillators~\cite{Ockeloen-Korppi2016} and in a spin-optomechanical hybrid system~\cite{Moller2017}.
It has also been analyzed as a means of entangling two atomic spin~\cite{Muschik2011,Vasilyev2013} or mechanical~\cite{Woolley2013,Woolley2014} systems as has been demonstrated in experiment~\cite{Julsgaard2001,Wasilewski2010,Krauter2011,Ockeloen-Korppi2018}. 

However, the theoretical studies have mostly focused on oscillators with identical or negligible intrinsic linewidths, a condition which is difficult to meet in practice for disparate hybrid systems. The present scheme circumvents this restriction by interfacing the two oscillators unidirectionally.
The resulting causal asymmetry permits efficient CQNC even for vastly different intrinsic linewidths, thereby facilitating entanglement generation.

\begin{figure}
\centering
\includegraphics[width=\columnwidth]{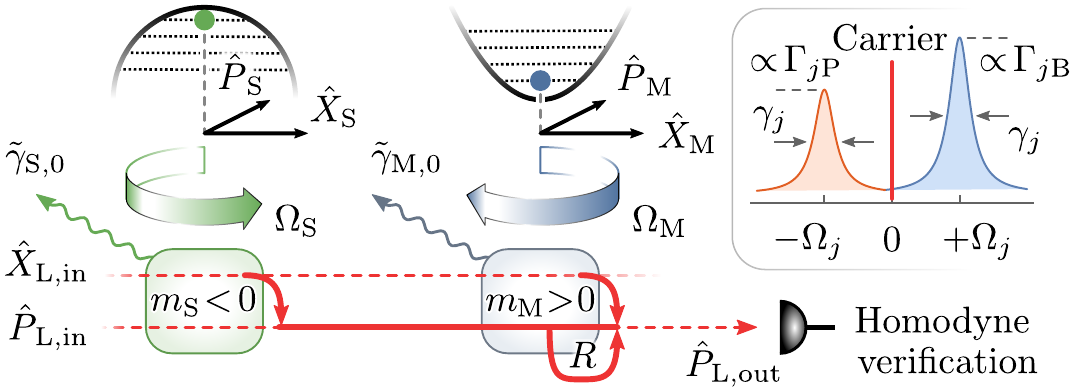}
\caption{Hybrid system consisting of two oscillators with negative and positive mass, respectively, typically implemented in collective spin (S) and motional (M) degrees of freedom. These are coupled in cascade to a common unidirectional light field via quadratic interactions  induced typically by a strong, classical carrier. For each oscillator, this results in Stokes and anti-Stokes sidebands proportional to rates $\Gamma_{j\text{P/B}}$ and of width $\gamma_{j}$, $j\in\{\text{S,M}\}$ (see inset); the effective resonance frequency $\Omega_j \equiv \text{sgn}(m_j)\omega_j$ accounts for, e.g., the fact that energy must be extracted from a negative-mass oscillator to excite it. The joint interaction with the light field is best described by homodyne quadratures $\hat{X}_{\text{L}},\hat{P}_{\text{L}}$ (symmetrized combinations of sidebands), 
whose initial state $\hat{X}_{\text{L,in}},\hat{P}_{\text{L,in}}$ is vacuum (see lower part of figure). The negative mass system is driven by $\hat{X}_{\text{L,in}}$ only ($\Gamma_{\text{SP}}= \Gamma_{\text{SB}}$) and its response is mapped onto $\hat{P}_{\text{L}}$. The positive mass system is likewise coupled to $\hat{X}_{\text{L,in}}$, but also to $\hat{P}_{\text{L}}$ ($\Gamma_{\text{MP}}< \Gamma_{\text{MB}}$) at an adjustable rate $R$, so that the response of the negative mass system  drives the positive mass system. Consequently the response of the positive mass system will interfere destructively with that of the negative mass system in the outgoing field quadrature $\hat{P}_{\text{L,out}}$ 
as can (optionally) be verified by homodyne detection. Additionally, the  oscillators are driven by distinct thermal reservoirs with decoherence rates $\tilde{\gamma}_{j\text{,0}}$ (wavy arrows).}
\label{fig:system}
\end{figure}

\textit{Model.---}
We consider a generic hybrid system composed of two subsystems with effective masses $\text{sgn}(m_{\text{S}})=-\text{sgn}(m_{\text{M}})<0$ coupled to a unidirectional optical field~[Fig.~\ref{fig:system}] (near-ideal unidirectionality has been achieved experimentally, e.g., see Refs.~\cite{Wasilewski2010,Moller2017}). Both subsystems are driven by individual thermal reservoirs. 
The positive/negative mass subsystem is referred to as a motional/collective spin (M/S) degree of freedom and represented by a localized bosonic mode with dimensionless canonical variables. These variables satisfy $[\hat{X}_j,\hat{P}_k]=i\delta_{j,k}$, ($j,k\in \{\text{M,S}\}$) resulting from a rescaling by the zero-point fluctuation amplitudes $x_{j,\text{zpf}}=\sqrt{\hbar/(|m_j| \omega_j)}$ and $p_{j,\text{zpf}}=\hbar/x_{j,\text{zpf}}$, where 
$\omega_j$ is the resonance frequency. 
The free evolution of the hybrid system is (setting $\hbar=1$)
\begin{equation}
\hat{H}_0 = \sum_{j\in\{\text{M,S}\}}  \text{sgn}(m_j) \frac{\omega_j}{2}(\hat{X}^2_j + \hat{P}^2_j),\label{eq:H_0}
\end{equation}
and hence a negative mass translates into a negative effective resonance frequency $\Omega_{j}\equiv\text{sgn}(m_j)\omega_j$ for the dimensionless variables, inverting the sense of rotation in the $\{\hat{X}_{j},\hat{P}_{j}\}$ phase space~[Fig.~\ref{fig:system}] and making the state with zero quanta its \emph{highest} energy state (not to be confused with a positive mass oscillator with an inverted potential, $\hat{H}_{0}\propto -\hat{X}^2_j + \hat{P}^2_j$). We specialize to the resonant 
scenario $\omega_{\text{M}} = \omega_{\text{S}} := \omega$.

We introduce annihilation operators for the localized modes, $\hat{X}_{j}=(\hat{a}_j+\hat{a}_{j}^{\dagger})/\sqrt{2}$ and $\hat{P}_{j}=(\hat{a}_j-\hat{a}_{j}^{\dagger})/(\sqrt{2}i)$, and the propagating field linking them, $\hat{b}(t)=(2\pi)^{-1/2}\int_{-\infty}^{\infty}\hat{b}(\Omega)e^{-i\Omega t}d\Omega$ (defined in a rotating frame with respect to the optical carrier).
The Hamiltonian for two-mode quadratic interaction between the localized oscillators and the light field is~\cite{Vasilyev2013,Muschik2011} 
\begin{equation}
\hat{H}_{\text{int}} = \sum_{j\in\{\text{M,S}\}}(\sqrt{\Gamma_{j\text{B}}}\hat{a}_{j}^{\dagger}\hat{b}
(t_j)+\sqrt{\Gamma_{j\text{P}}}\hat{a}_{j}^{\dagger}\hat{b}^{\dagger}(t_j)+\text{H.c.}),\label{eq:H-int2}
\end{equation}
where we assume $t_\text{S}<t_\text{M}$, i.e., the optical field interacts with S first.
Equation~\eqref{eq:H-int2} comprises two kinds of interaction: beam-splitter (B), $\propto(\hat{a}_{j}^{\dagger}\hat{b}+\text{H.c.})$, and parametric down-conversion (P), $\propto(\hat{a}_{j}^{\dagger}\hat{b}^{\dagger}+\text{H.c.})$, $j\in \{\text{M,S}\}$.
These processes produce 
sidebands at rates
$\Gamma_{j\text{B}}=\Gamma_{j}\sin^{2}\theta_j,\Gamma_{j\text{P}}=\Gamma_{j}\cos^{2}\theta_j$ [Fig.~\ref{fig:system}, inset], which we parametrize by   $\Gamma_{j}=\Gamma_{j\text{B}}+\Gamma_{j\text{P}}$ and $\theta_{j}\in{[0,\pi/2]}$, the coupling rates and angles.

An excitation in the upper sideband from the positive-(negative-)mass oscillator arises from $\sqrt{\Gamma_{\text{MB}}}\hat{a}_{\text{M}}\hat{b}^{\dagger}$ ($\sqrt{\Gamma_{\text{SP}}}\hat{a}_{\text{S}}^{\dagger}\hat{b}^{\dagger}$), simultaneously removing (adding) an oscillator quantum (analogously for the lower sideband).  
This indistinguishability of adding a quantum to one subsystem and removing one from the other as energy is either added or removed by the common light field permits the system to be driven into a two-mode-squeezed entangled state accompanied by CQNC of the BA contribution to the joint output field.
Perfect indistinguishability necessitates $\Gamma_{\text{MB}}=\Gamma_{\text{SP}}$ and $\Gamma_{\text{MP}}=\Gamma_{\text{SB}}$, i.e., $\theta_{\text{M}} = -\theta_{\text{S}} + \pi/2$ and $\Gamma_{\text{M}}=\Gamma_{\text{S}}$, but also the temporal responses of the subsystems must be suitably matched. However, whenever $\theta_{\text{M}} \neq  \theta_{\text{S}}$ there is an  overlap between the light quadrature reading out S  and the quadrature driving M , i.e., the spin response to light and thermal forces drives the motional mode.
This induces a tunable interference effect that can implement additional quantum and classical noise cancellation even for highly asymmetric subsystems, leading to unconditional entanglement generation competitive with conditional schemes---this is the main finding of this Letter.

The regime of interest is $\omega_{j} \gg \Gamma_{j} \gtrsim \tilde{\gamma}_{j,0}$, where  $\tilde{\gamma}_{j,0}$ is the thermal decoherence rate, providing the time-scale separation required for probing the system over several quantum-coherent oscillations.
This permits  treating Eq.~\eqref{eq:H-int2} in the rotating wave approximation (RWA), i.e., retaining only slowly varying terms, and implies that the interaction with light is confined to two disjoint sidebands $\hat{b}_{-}(t)+\hat{b}_{+}(t) := (2\pi)^{-1/2}(\int_{-\infty}^{0}+\int_{0}^{\infty})\hat{b}(\Omega)e^{-i\Omega t}d\Omega = \hat{b}(t)$, $[\hat{b}_{\pm}(t), \hat{b}_{\pm}^{\dagger}(t')]=\delta(t-t')$, centered at frequencies $\Omega=\mp \omega$ (relative to the carrier). We introduce the non-Hermitian homodyne two-mode quadratures $\hat{X}_{\text{L}}:=(\hat{b}_{+}+\hat{b}_{-}^{\dagger})/\sqrt{2}$ and $\hat{P}_{\text{L}}:=(\hat{b}_{+}-\hat{b}_{-}^{\dagger})/(\sqrt{2}i)$. Performing the RWA we find
\begin{equation}
\hat{H}_{\text{int}}\approx\hat{a}_{\text{M}}^{\dagger}\sqrt{\Gamma_{\text{M}}}\hat{Q}_{\theta'_{\text{M}}}(t_{\text{M}})+\hat{a}_{\text{S}}\sqrt{\Gamma_{\text{S}}}\hat{Q}_{-\theta'_{\text{S}}}(t_{\text{S}})+\text{H.c.},\label{eq:H-int3}
\end{equation}
where $\hat{Q}_{\theta'}:=\cos\theta' \hat{X}_{\text{L}}+i\sin\theta'\hat{P}_{\text{L}}$, $\theta_{j}'=\theta_{j}-\pi/4$.
Equation~\eqref{eq:H-int3} indicates that the cosine and sine components of the phase quadrature $\propto\hat{P}_{\text{L}}+\hat{P}_{\text{L}}^{\dagger}$ 
read out the (unnormalized) EPR-type variables $\sqrt{\Gamma_{\text{M}}}\cos\theta'_{\text{M}}\hat{X}_{\text{M}}+\sqrt{\Gamma_{\text{S}}}\cos\theta'_{\text{S}}\hat{X}_{\text{S}}$ and $\sqrt{\Gamma_{\text{M}}}\cos\theta'_{\text{M}}\hat{P}_{\text{M}}-\sqrt{\Gamma_{\text{S}}}\cos\theta'_{\text{S}}\hat{P}_{\text{S}}$, respectively.
These commute when $\sqrt{\Gamma_{\text{M}}}\cos\theta'_{\text{M}}=\sqrt{\Gamma_{\text{S}}}\cos\theta'_{\text{S}}$, in which case they can be BA-free.

Eliminating the light field and using a co-propagating time coordinate $t' = t-x/c$ (dropping the prime henceforth), the Heisenberg-Langevin equations can be expressed in terms of the forces $\hat{f}_{j}:=\sqrt{\gamma_{j,0}}\hat{a}_{j\text{,in}}+\hat{f}_{j\text{,ba}}$ as [henceforth $\hat{a}_{j}$ is in the rotating frame of $\hat{H}_0$~\eqref{eq:H_0}]~\cite{Gardiner1993,Gardiner2004,Carmichael1993}
\begin{align}
\frac{d}{d t}\hat{a}_{\text{S}}&=-\frac{\gamma_{\text{S}}}{2}\hat{a}_{\text{S}}+\hat{f}_{\text{S}},\nonumber\\
\frac{d}{d t}\hat{a}_{\text{M}}&=-\frac{\gamma_{\text{M}}}{2}\hat{a}_{\text{M}}+\hat{f}_{\text{M}} +\sqrt{1-\epsilon}R \hat{a}_{\text{S}}^{\dagger},\label{eq:aM-dot}
\end{align}
where
\begin{align}
\hat{f}_{\text{S,ba}} :={}& -i(\sqrt{\Gamma_{\text{SB}}}\hat{b}_{-,\text{in}}+\sqrt{\Gamma_{\text{SP}}}\hat{b}_{+,\text{in}}^{\dagger}),\nonumber\\
\hat{f}_{\text{M,ba}} :={}& -i\sqrt{1-\epsilon}(\sqrt{\Gamma_{\text{MB}}}\hat{b}_{+,\text{in}}+\sqrt{\Gamma_{\text{MP}}}\hat{b}_{-,\text{in}}^{\dagger})\nonumber\\
& -i\sqrt{\epsilon}(\sqrt{\Gamma_{\text{MB}}}\hat{b}_{+,\text{in}}'+\sqrt{\Gamma_{\text{MP}}}\hat{b}_{-,\text{in}}'^{\dagger}).\label{eq:fBA}
\end{align}
Here, an additional uncorrelated vacuum $\hat{b}_{\pm,\text{in}}'$ impinges on M due to transmission (power) loss $\epsilon > 0$ between the subsystems. The vacuum fields satisfy $\langle\hat{b}_{\pm,\text{in}}(t) \hat{b}_{\pm,\text{in}}^{\dagger}(t')\rangle=\langle\hat{b}_{\pm,\text{in}}'(t) \hat{b}_{\pm,\text{in}}'^{\dagger}(t')\rangle=\delta(t-t')$. 
$[\hat{a}_{j,\text{in}}\left(t\right),\hat{a}_{j,\text{in}}^{\dagger}\left(t'\right)]=\delta(t-t')$, $j\in\{\text{M,S}\}$, represent the thermal noise fluctuations with $\langle \hat{a}_{j,\text{in}}\left(t\right)\hat{a}_{j,\text{in}}^{\dagger}\left(t'\right)\rangle=(\bar{n}_j+1)\delta(t-t')$ in terms of the thermal occupancy $\bar{n}_j$. 
For example, for S, $\bar{n}_{\text{\text{S}}} > 0$ represents the additional noise present for an imperfectly polarized ensemble, while for M, $\bar{n}_{\text{M}}\approx k_{\text{B}}T_{\text{M}}/(\hbar\omega_{\text{M}})$ at ambient temperature $T_{\text{M}}$ ($k_{\text{B}}$ is the Boltzmann constant).
The effective linewidths (including dynamical broadening from the light field coupling) are denoted $\gamma_{j}=\gamma_{j,0}-\Gamma_{j}\cos(2\theta_j)$, where $\gamma_{j,0}$ is the linewidth in absence of dynamical broadening; dynamical stability requires $\gamma_j >0$.
Finally, due to the unidirectionality of the light field, information can only propagate from the first to the second subsystem in the cascade. 
The corresponding unidirectional coupling rate is $R=\sqrt{\Gamma_{\text{SB}}\Gamma_{\text{MP}}}-\sqrt{\Gamma_{\text{MB}}\Gamma_{\text{SP}}}=-\sqrt{\Gamma_{\text{M}}\Gamma_{\text{S}}}\sin(\theta_{\text{M}}-\theta_{\text{S}})$. For $R=0 \Leftrightarrow \theta_{\text{S}}= \theta_{\text{M}}$, Eqs.~\eqref{eq:aM-dot} decouple so that correlations build up solely due to those between $\hat{f}_{\text{S,ba}}$ and $\hat{f}_{\text{M,ba}}$, and the  ordering of oscillators becomes immaterial (assuming $\epsilon=0$). In contrast, $R\neq 0$ gives rise to a nontrivial asymmetry of the cascaded system~\eqref{eq:aM-dot}, which is exploited below for improved noise cancellation and entanglement generation.

\textit{Unconditional steady-state solution.---}The steady-state solution to Eqs.~\eqref{eq:aM-dot} is
\begin{align}
\hat{a}_{\text{S}}(t) ={}& \int_{-\infty}^{t} dt' e^{-(t-t') \gamma_{\text{S}}/2}\hat{f}_{\text{S}}(t')  ,\nonumber\\
\hat{a}_{\text{M}}(t) ={}& \int_{-\infty}^{t} dt' \{e^{-(t-t') \gamma_{\text{M}}/2}\hat{f}_{\text{M}}(t')\label{eq:response}\\
&+ \frac{2\sqrt{1-\epsilon}R}{\gamma_{\text{M}}-\gamma_{\text{S}}} [e^{-(t-t') \gamma_{\text{S}}/2}-e^{-(t-t') \gamma_{\text{M}}/2}]\hat{f}_{\text{S}}^{\dagger}(t')\}.\nonumber
\end{align}
For $R=0$, the steady states of the individual subsystems are determined solely by the (stochastic) driving forces in the past time interval of duration $\sim 1/\gamma_j$. Hence, whenever $\gamma_{\text{M}} \neq \gamma_{\text{S}}$ the different temporal responses to the BA $\hat{b}_{\pm,\text{in}}$ will result in imperfect CQNC.
However, if it is the second system (M) in the cascade which is relatively short-lived, $\gamma_{\text{M}} > \gamma_{\text{S}}$, then 
for $R\neq0$ the unidirectional coupling term $\propto R a_{\text{S}}^{\dagger}$ [Eq.~\eqref{eq:aM-dot}] effectively prolongs the memory time $1/\gamma_{\text{M}}$ by driving M with the spin response contained in the light field, resulting in improved CQNC for $R<0 \Leftrightarrow \theta_{\text{M}} > \theta_{\text{S}}$. Ideal cancellation can be achieved in the adiabatic limit $\gamma_{\text{M}} \gg \gamma_{\text{S}}$ and $2R/\gamma_{\text{M}}\rightarrow -1$ (for $\epsilon=0$)~[Eq.~\eqref{eq:response}], which is compatible with the demand for near-ground-state dynamical cooling of the motional mode $\gamma_{\text{M}} \gg \tilde{\gamma}_{\text{M},0}$, where  $\tilde{\gamma}_{j,0}:=\gamma_{j,0}(\bar{n}_j + 1/2)$. 
The additional interference arising for $R<0$ does not rely on the opposite signs of masses (in contrast to the scheme as a whole) 
and can simultaneously suppress both quantum noise and the spin thermal noise, thereby removing the need for dynamical spin cooling.

From Eqs.~\eqref{eq:response} the entries of the covariance matrix in steady state are
\begin{align}
\Delta^{2}\hat{X}_{\text{S}}&=\frac{1}{\gamma_{\text{S}}}(\frac{\Gamma_{\text{S}}}{2}+\tilde{\gamma}_{\text{S},0}),
\nonumber\\
\Delta^{2}\hat{X}_{\text{M}}&=\frac{1}{\gamma_{\text{M}}}(\frac{\Gamma_{\text{M}}}{2}+\tilde{\gamma}_{\text{M},0}+\sqrt{1-\epsilon}R\langle \hat{X}_{\text{S}},\hat{X}_{\text{M}}\rangle),\label{eq:ss_sol}\\
\langle \hat{X}_{\text{S}},\hat{X}_{\text{M}}\rangle &=-\frac{2\sqrt{1-\epsilon}}{\gamma_{\text{S}}+\gamma_{\text{M}}}(\sqrt{\Gamma_{\text{S}}\Gamma_{\text{M}}}\sin(\theta_{\text{M}}+\theta_{\text{S}})-2R\Delta^{2}\hat{X}_{\text{S}}),\nonumber
\end{align}
where $\langle \hat{X}_{\text{S}},\hat{X}_{\text{M}}\rangle:=\langle \hat{X}_{\text{S}}\hat{X}_{\text{M}}\rangle+\langle \hat{X}_{\text{M}}\hat{X}_{\text{S}}\rangle-2\langle \hat{X}_{\text{S}}\rangle \langle \hat{X}_{\text{M}}\rangle$.
As our entanglement figure of merit we consider the variance of generalized EPR variables of the form~\cite{Simon2000,Giovannetti2003}
\begin{equation}
\xi_{g}=\frac{\Delta^{2}(\hat{X}_{\text{S}}+g\hat{X}_{\text{M}})+\Delta^{2}(\hat{P}_{\text{S}}-g \hat{P}_{\text{M}})}{1+g^2}<1,\label{eq:ent_crit}
\end{equation}
which is the inseparability criterion for Gaussian states and any $g\in \mathbb{R}$. 
The steady-state value of $\xi_{g}$ can be evaluated using the solution~\eqref{eq:ss_sol}, noting that $\Delta^{2}(\hat{X}_{\text{S}}+g\hat{X}_{\text{M}})\approx\Delta^{2}(\hat{P}_{\text{S}}-g \hat{P}_{\text{M}})$ within the RWA. 
In principle, $\xi_g$ can be minimized over $g$, but verifying such entanglement experimentally requires individual, and hence destructive, readout of the two subsystems. Since our scheme automatically and non-destructively produces readout of the EPR variables with $g=\sqrt{\Gamma_{\text{M}}/[(1-\epsilon)\Gamma_{\text{S}}]}\cos\theta'_{\text{M}}/\cos\theta'_{\text{S}}$ [see discussion below Eq.~\eqref{eq:H-int3}], which can be BA-free when $g\rightarrow 1$ and $\epsilon \rightarrow 0$, we henceforth fix $g$ by the aforementioned expression. 

\textit{Spin-optomechanical implementation.---}Let us consider a spin-optomechanical implementation~\cite{Hammerer2010,Aspelmeyer2014} (see Ref.~\cite{SM} for a derivation of Eqs.~\eqref{eq:aM-dot},\,\eqref{eq:fBA} in this context).
 Optomechanical systems 
 are routinely operated in the quantum regime, allowing ground-state cooling by dynamical broadening ($\gamma_{\text{M}}>\gamma_{\text{M,0}} \Leftrightarrow \theta_{\text{M}}>\pi/4$) even for 
 $\bar{n}_{\text{M}}\gg 1$. For the mechanical system, $\gamma_{\text{M},0}$ is usually due to intrinsic dissipation alone, such as friction, whereas for the spin oscillator, $\gamma_{\text{S},0}$ (typically $\gg \gamma_{\text{M},0}$) is often dominated by optical power broadening 
induced by the coherent driving. 
For quantum cooperativities defined as $C_{j}:= \Gamma_{j}/\tilde{\gamma}_{j,0}$, the value of $C_{\text{S}}$ is independent of the drive power in this regime.

\begin{figure}
\includegraphics[width=0.8\columnwidth]{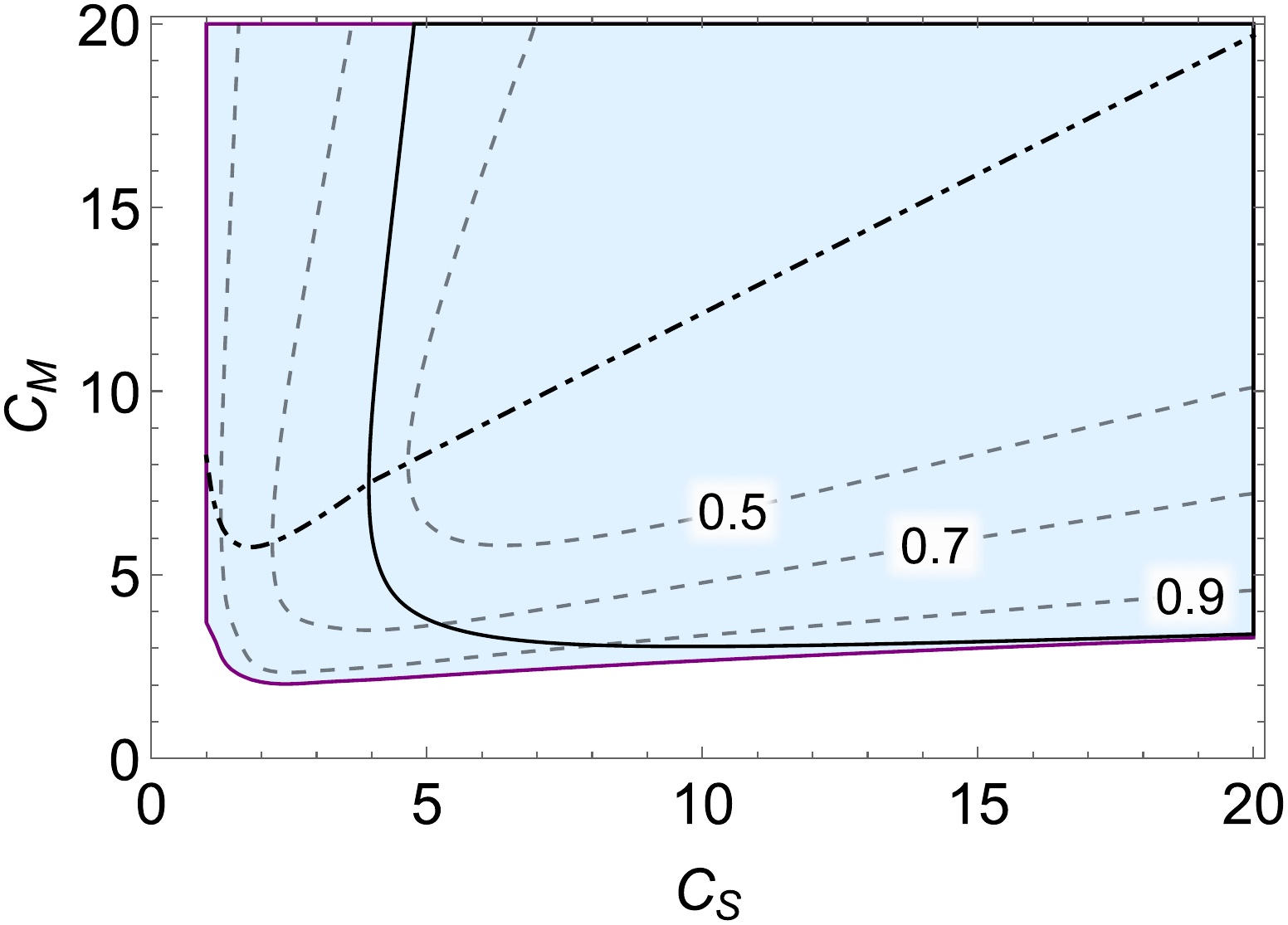}
\caption{Entanglement $\xi_{g}$ ($<1$ in the colored region) as a function of quantum cooperativities of the spin ($C_{\text{S}}$) and motional ($C_{\text{M}}$) subsystems for optimized coupling angles $\theta_{\text{S}}$ and $\theta_{\text{M}}$ while
fixing the parameters $\gamma_{\text{S,}0}=2\pi \times 5 \text{kHz}, \bar{n}_{\text{S}}=1, \gamma_{\text{M,}0}\bar{n}_{\text{M}}= 2\pi \times 10 \text{kHz}$, and assuming no transmission losses, $\epsilon =0$.
Optimal $C_{\text{M}}$ for given $C_{\text{S}}$ is indicated by the dashed-dotted curve.
Imposing the additional constraint $\theta_{\text{S}}=\theta_{\text{M}}\Rightarrow R=0$, entanglement $\xi_{g}<1$ is only possible in the subregion delineated by the solid contour.}
\label{fig:dissip-vs-OMIT}
\end{figure}

Conditional entanglement in a spin-optomechanical  system was previously analyzed for a pulsed quantum non-demolition (QND) measurement of the hybrid EPR variables~\cite{Hammerer2009} which projects the system into an entangled state fulfilling Eq.~\eqref{eq:ent_crit}; this approach has been demonstrated for two atomic spin ensembles~\cite{Julsgaard2001}.
In contrast to that protocol, steady-state unconditional entanglement is a ready-to-use resource \cite{Barreiro2011,Krauter2011} available on demand at any moment in time.

\begin{figure}[h!]
\hspace*{-0.2cm}
\includegraphics[width=.97\columnwidth]{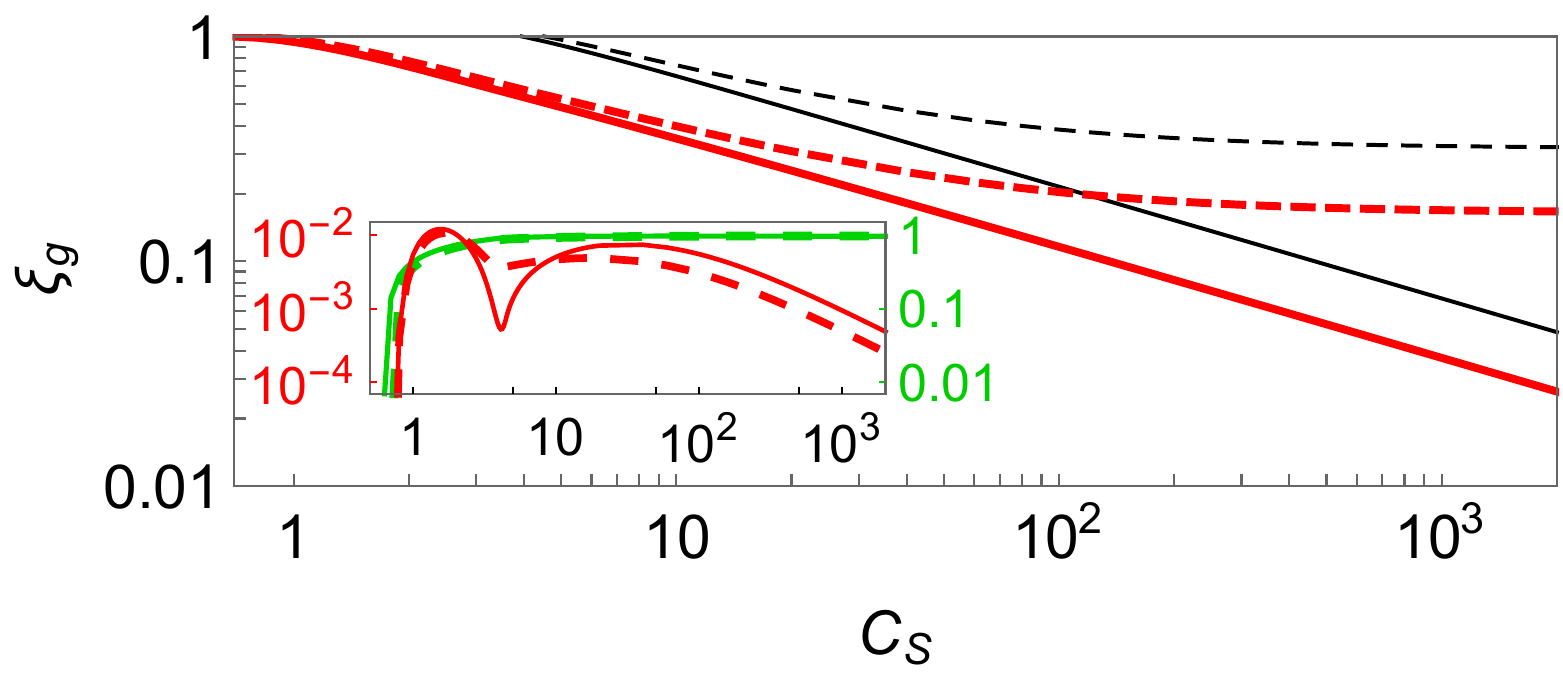}
\caption{Entanglement $\xi_{g}$ as a function of spin cooperativity $C_{\text{S}}$ for optimized coupling angles $\theta_{\text{S}},\theta_{\text{M}}$ and motional cooperativity $C_{\text{M}}$ when $R=0$ (thin black curves) and $R \neq 0$ (thick red curves), when transmission loss is absent, $\epsilon=0$ (solid), and present, $\epsilon = 0.1$ (dashed).
 (Inset) Plot of $-2\sqrt{1-\epsilon}R/\gamma_\text{M}$ (right scale, brighter green curves) as a function of $C_{\text{S}}$ used in evaluating the optimized curves of the main plot, and the
relative entanglement improvement (left scale, darker red curves) of the conditional scheme over the optimal unconditional scheme (referenced to the latter) for $\epsilon=0.1$; the conditional performance is evaluated using parameters optimized for the unconditional scheme (dashed) and optimal conditional parameters for QND readout $\theta_{\text{S}}=\theta_{\text{M}}=\pi/4$ (solid). The fixed parameters are $\gamma_{\text{S,}0}=2\pi \times 5 \text{kHz}, \bar{n}_{\text{S}}=1$, and $\gamma_{\text{M,}0}\bar{n}_{\text{M}}= 2\pi \times 10 \text{kHz}$.}
\label{fig:opt_ent}
\end{figure}

Figure~\ref{fig:dissip-vs-OMIT} presents the optimized unconditional steady-state entanglement~\eqref{eq:ent_crit} as a function of 
$C_{j}$, 
 illustrating the relaxation of parameter requirements compared to dissipative entanglement generation ($R=0$, both subsystems are driven optically only by the common vacuum field). Since the tunability of free-space spin systems is limited by the atomic density, we henceforth assume the bottleneck to be the spin system, characterized by a maximally attainable $C_{\text{S}}$, whereas 
$C_{\text{M}}$ is freely tunable and thus can be fixed at its optimal value [Fig.~\ref{fig:dissip-vs-OMIT}, dashed-dotted curve]. 
Under these conditions, optimization requires the two subsystems to be coupled asymmetrically to the field: The optimal $\theta_{\text{M}}$ favors beam-splitter interaction $\pi/2\geq\theta_{\text{M,opt}}> \pi/4$, while for S, the Stokes and anti-Stokes processes should be balanced, $\theta_{\text{S,opt}}\approx \pi/4\Leftrightarrow \Gamma_{\text{SB}}\approx \Gamma_{\text{SP}}$ (QND interaction) yielding $R<0$ [Fig.~\ref{fig:opt_ent}, inset]; this is the scenario illustrated in Fig.~\ref{fig:system}.
The resulting effective motional linewidth considerably exceeds that of S, $\gamma_{\text{M}} \gg \gamma_{\text{S}}$, in the regime of substantial entanglement, thereby reversing the hierarchy set by the intrinsic linewidths $\gamma_{\text{M,}0}\ll \gamma_{\text{S,}0}$ while providing strong dynamical cooling of the motional thermal noise $\tilde{\gamma}_{\text{M,}0}$, which is essential to unconditional operation. Since $\gamma_{\text{S}}\sim\gamma_{\text{S,0}}$ for $\theta_{\text{S}}\approx \pi/4$, the suppression of spin thermal noise is due mainly to coherent cancellation in contrast to previous work relying on dynamical spin cooling in the dissipative regime ($R\approx 0$)~\cite{Krauter2011,Muschik2011,Vasilyev2013}.

In the absence of transmission loss ($\epsilon=0$), the asymptotic scaling of the unconditional entanglement is
$\xi_{g}\approx\sqrt{[1+r+1/(2\bar{n}_{\text{S}}+1)]/(2C_{\text{S}})}$,
where $r=\tilde{\gamma}_{\text{M,0}}/\tilde{\gamma}_{\text{S,0}}$. An improvement by up to a factor of 
$2$ can be found when comparing to the dissipative case ($R=0$), $\xi_{g}\approx\sqrt{2(1+r)/C_{\text{S}}}$~[Fig.~\ref{fig:opt_ent}] (see Ref.~\cite{SM} for derivations of scalings). The presence of loss $\epsilon>0$ imposes a lower bound $\xi_{g} \geq \sqrt{\epsilon/(4-3\epsilon)}$, which is also an improvement of up to a factor of 2 compared to $R=0$.

\textit{Comparison with conditional scheme.---}Another benchmark is the conditional steady-state entanglement generated by performing a continuous homodyne measurement of the light field emanating from the hybrid system~\cite{Vasilyev2013}. 
The evolution of the system conditioned on the measurement record is described by a stochastic master equation~\cite{Wiseman2010} whose steady state can be found numerically and even analytically in our regime of interest, $\bar{n}_{\text{M}}\gg 1$ (see Ref.~\cite{SM} for mathematical details).
For the fixed parameters considered above~[Fig.~\ref{fig:opt_ent}], we find in the limit of substantial entanglement that, remarkably, the conditional steady-state entanglement matches that of our unconditional scheme within a few-percent margin, even when separately optimized under the same conditions in the dynamically stable regime (see Fig.~\ref{fig:opt_ent}, inset; supplementary details in Ref.~\cite{SM}). 
We thus conclude that our unconditional scheme leaves practically no information in the output light about the noise affecting the squeezed EPR variables.
From a practical standpoint this is beneficial as it allows optimal performance without the need to measure the output field nor perform the feedback required to make the conditional entanglement unconditional. Moreover, the dynamical cooling of the motional mode occurring in the unconditional scheme facilitates technical stability in the apparatus.

In conclusion, unconditional steady-state entanglement in a cascaded negative-positive mass hybrid system can be efficiently generated by engineering an asymmetric interaction between the subsystems via the light field connecting them.
Applications for such a resource of ready-to-use entanglement include quantum teleportation~\cite{He2015} and key distribution~\cite{Walk2016} in hybrid quantum networks.
The scheme can compete with conditional schemes, a fact which we speculate can be elucidated by formally framing our unconditional scheme in terms of a coherent-feedback master equation. 
The noise cancellation technique inherent to the scheme enables sub-SQL sensitivity when using the hybrid system as a continuous force sensor, as will be elaborated on elsewhere~\cite{EZInPrep}. Moreover, we have evidence that this sensing enhancement is closely linked to the generation of EPR-type entanglement studied here~\cite{SM}, warranting further study.

\begin{acknowledgments}
The authors acknowledge productive discussions with C.~M{\o}ller, R.~A.~Thomas, and A.~S{\o}rensen.
This work was supported by the ERC AdG grant Quantum-N, the ARO Grant No. W911NF and by John Templeton Foundation.
X.\,H. is supported partly by the China Scholarship Council (201606010044). E.\,Z. is supported by the Carlsberg Foundation. X.\,H. and Q.\,H. acknowledge the support of the National Key R\&D Program of China (Grant No. 2016YFA0301302) and National Natural Science Foundation of China (Grants No. 11622428 and
No. 61475006). D.\,V. acknowledges support of the
Austrian Science Fund SFB FoQuS (FWF Project
No. F4016-N23) and the European Research Council
(ERC) Synergy Grant UQUAM. K.\,H. acknowledges support
by Deutsche Forschungsgemeinschaft (DFG) through
CRC 1227 (DQ-mat), project A05. X.\,H. is grateful to
E.\,S.\,P. for hosting her during her stays at NBI.
\end{acknowledgments}
\bibliographystyle{apsrev4-1}
\bibliography{Entanglement}

\begin{thebibliography}{62}%
\makeatletter
\providecommand \@ifxundefined [1]{%
 \@ifx{#1\undefined}
}%
\providecommand \@ifnum [1]{%
 \ifnum #1\expandafter \@firstoftwo
 \else \expandafter \@secondoftwo
 \fi
}%
\providecommand \@ifx [1]{%
 \ifx #1\expandafter \@firstoftwo
 \else \expandafter \@secondoftwo
 \fi
}%
\providecommand \natexlab [1]{#1}%
\providecommand \enquote  [1]{``#1''}%
\providecommand \bibnamefont  [1]{#1}%
\providecommand \bibfnamefont [1]{#1}%
\providecommand \citenamefont [1]{#1}%
\providecommand \href@noop [0]{\@secondoftwo}%
\providecommand \href [0]{\begingroup \@sanitize@url \@href}%
\providecommand \@href[1]{\@@startlink{#1}\@@href}%
\providecommand \@@href[1]{\endgroup#1\@@endlink}%
\providecommand \@sanitize@url [0]{\catcode `\\12\catcode `\$12\catcode
  `\&12\catcode `\#12\catcode `\^12\catcode `\_12\catcode `\%12\relax}%
\providecommand \@@startlink[1]{}%
\providecommand \@@endlink[0]{}%
\providecommand \url  [0]{\begingroup\@sanitize@url \@url }%
\providecommand \@url [1]{\endgroup\@href {#1}{\urlprefix }}%
\providecommand \urlprefix  [0]{URL }%
\providecommand \Eprint [0]{\href }%
\providecommand \doibase [0]{http://dx.doi.org/}%
\providecommand \selectlanguage [0]{\@gobble}%
\providecommand \bibinfo  [0]{\@secondoftwo}%
\providecommand \bibfield  [0]{\@secondoftwo}%
\providecommand \translation [1]{[#1]}%
\providecommand \BibitemOpen [0]{}%
\providecommand \bibitemStop [0]{}%
\providecommand \bibitemNoStop [0]{.\EOS\space}%
\providecommand \EOS [0]{\spacefactor3000\relax}%
\providecommand \BibitemShut  [1]{\csname bibitem#1\endcsname}%
\let\auto@bib@innerbib\@empty
\bibitem [{\citenamefont {Horodecki}\ \emph {et~al.}(2009)\citenamefont
  {Horodecki}, \citenamefont {Horodecki}, \citenamefont {Horodecki},\ and\
  \citenamefont {Horodecki}}]{Horodecki2009}%
  \BibitemOpen
  \bibfield  {author} {\bibinfo {author} {\bibfnamefont {R.}~\bibnamefont
  {Horodecki}}, \bibinfo {author} {\bibfnamefont {P.}~\bibnamefont
  {Horodecki}}, \bibinfo {author} {\bibfnamefont {M.}~\bibnamefont
  {Horodecki}}, \ and\ \bibinfo {author} {\bibfnamefont {K.}~\bibnamefont
  {Horodecki}},\ }\href {\doibase 10.1103/RevModPhys.81.865} {\bibfield
  {journal} {\bibinfo  {journal} {Rev. Mod. Phys.}\ }\textbf {\bibinfo {volume}
  {81}},\ \bibinfo {pages} {865} (\bibinfo {year} {2009})}\BibitemShut
  {NoStop}%
\bibitem [{\citenamefont {Giovannetti}\ \emph {et~al.}(2004)\citenamefont
  {Giovannetti}, \citenamefont {Lloyd},\ and\ \citenamefont
  {Maccone}}]{Giovannetti2004}%
  \BibitemOpen
  \bibfield  {author} {\bibinfo {author} {\bibfnamefont {V.}~\bibnamefont
  {Giovannetti}}, \bibinfo {author} {\bibfnamefont {S.}~\bibnamefont {Lloyd}},
  \ and\ \bibinfo {author} {\bibfnamefont {L.}~\bibnamefont {Maccone}},\ }\href
  {\doibase 10.1126/science.1104149} {\bibfield  {journal} {\bibinfo  {journal}
  {Science}\ }\textbf {\bibinfo {volume} {306}},\ \bibinfo {pages} {1330}
  (\bibinfo {year} {2004})}\BibitemShut {NoStop}%
\bibitem [{\citenamefont {{Pezz{\`e}}}\ \emph {et~al.}(2016)\citenamefont
  {{Pezz{\`e}}}, \citenamefont {{Smerzi}}, \citenamefont {{Oberthaler}},
  \citenamefont {{Schmied}},\ and\ \citenamefont {{Treutlein}}}]{QmetroRMP}%
  \BibitemOpen
  \bibfield  {author} {\bibinfo {author} {\bibfnamefont {L.}~\bibnamefont
  {{Pezz{\`e}}}}, \bibinfo {author} {\bibfnamefont {A.}~\bibnamefont
  {{Smerzi}}}, \bibinfo {author} {\bibfnamefont {M.~K.}\ \bibnamefont
  {{Oberthaler}}}, \bibinfo {author} {\bibfnamefont {R.}~\bibnamefont
  {{Schmied}}}, \ and\ \bibinfo {author} {\bibfnamefont {P.}~\bibnamefont
  {{Treutlein}}},\ }\href@noop {} {\bibfield  {journal} {\bibinfo  {journal}
  {ArXiv e-prints}\ } (\bibinfo {year} {2016})},\ \Eprint
  {http://arxiv.org/abs/1609.01609} {arXiv:1609.01609 [quant-ph]} \BibitemShut
  {NoStop}%
\bibitem [{\citenamefont {Leroux}\ \emph {et~al.}(2010)\citenamefont {Leroux},
  \citenamefont {Schleier-Smith},\ and\ \citenamefont
  {Vuleti\ifmmode~\acute{c}\else \'{c}\fi{}}}]{Leroux2010}%
  \BibitemOpen
  \bibfield  {author} {\bibinfo {author} {\bibfnamefont {I.~D.}\ \bibnamefont
  {Leroux}}, \bibinfo {author} {\bibfnamefont {M.~H.}\ \bibnamefont
  {Schleier-Smith}}, \ and\ \bibinfo {author} {\bibfnamefont {V.}~\bibnamefont
  {Vuleti\ifmmode~\acute{c}\else \'{c}\fi{}}},\ }\href {\doibase
  10.1103/PhysRevLett.104.250801} {\bibfield  {journal} {\bibinfo  {journal}
  {Phys. Rev. Lett.}\ }\textbf {\bibinfo {volume} {104}},\ \bibinfo {pages}
  {250801} (\bibinfo {year} {2010})}\BibitemShut {NoStop}%
\bibitem [{\citenamefont {Hosten}\ \emph {et~al.}(2016)\citenamefont {Hosten},
  \citenamefont {Engelsen}, \citenamefont {Krishnakumar},\ and\ \citenamefont
  {Kasevich}}]{Hosten2016}%
  \BibitemOpen
  \bibfield  {author} {\bibinfo {author} {\bibfnamefont {O.}~\bibnamefont
  {Hosten}}, \bibinfo {author} {\bibfnamefont {N.~J.}\ \bibnamefont
  {Engelsen}}, \bibinfo {author} {\bibfnamefont {R.}~\bibnamefont
  {Krishnakumar}}, \ and\ \bibinfo {author} {\bibfnamefont {M.~A.}\
  \bibnamefont {Kasevich}},\ }\href {http://dx.doi.org/10.1038/nature16176}
  {\bibfield  {journal} {\bibinfo  {journal} {Nature}\ }\textbf {\bibinfo
  {volume} {529}},\ \bibinfo {pages} {505 EP } (\bibinfo {year}
  {2016})}\BibitemShut {NoStop}%
\bibitem [{\citenamefont {Sewell}\ \emph {et~al.}(2012)\citenamefont {Sewell},
  \citenamefont {Koschorreck}, \citenamefont {Napolitano}, \citenamefont
  {Dubost}, \citenamefont {Behbood},\ and\ \citenamefont
  {Mitchell}}]{Sewell2012}%
  \BibitemOpen
  \bibfield  {author} {\bibinfo {author} {\bibfnamefont {R.~J.}\ \bibnamefont
  {Sewell}}, \bibinfo {author} {\bibfnamefont {M.}~\bibnamefont {Koschorreck}},
  \bibinfo {author} {\bibfnamefont {M.}~\bibnamefont {Napolitano}}, \bibinfo
  {author} {\bibfnamefont {B.}~\bibnamefont {Dubost}}, \bibinfo {author}
  {\bibfnamefont {N.}~\bibnamefont {Behbood}}, \ and\ \bibinfo {author}
  {\bibfnamefont {M.~W.}\ \bibnamefont {Mitchell}},\ }\href {\doibase
  10.1103/PhysRevLett.109.253605} {\bibfield  {journal} {\bibinfo  {journal}
  {Phys. Rev. Lett.}\ }\textbf {\bibinfo {volume} {109}},\ \bibinfo {pages}
  {253605} (\bibinfo {year} {2012})}\BibitemShut {NoStop}%
\bibitem [{\citenamefont {Wasilewski}\ \emph {et~al.}(2010)\citenamefont
  {Wasilewski}, \citenamefont {Jensen}, \citenamefont {Krauter}, \citenamefont
  {Renema}, \citenamefont {Balabas},\ and\ \citenamefont
  {Polzik}}]{Wasilewski2010}%
  \BibitemOpen
  \bibfield  {author} {\bibinfo {author} {\bibfnamefont {W.}~\bibnamefont
  {Wasilewski}}, \bibinfo {author} {\bibfnamefont {K.}~\bibnamefont {Jensen}},
  \bibinfo {author} {\bibfnamefont {H.}~\bibnamefont {Krauter}}, \bibinfo
  {author} {\bibfnamefont {J.~J.}\ \bibnamefont {Renema}}, \bibinfo {author}
  {\bibfnamefont {M.~V.}\ \bibnamefont {Balabas}}, \ and\ \bibinfo {author}
  {\bibfnamefont {E.~S.}\ \bibnamefont {Polzik}},\ }\href {\doibase
  10.1103/PhysRevLett.104.133601} {\bibfield  {journal} {\bibinfo  {journal}
  {Phys. Rev. Lett.}\ }\textbf {\bibinfo {volume} {104}},\ \bibinfo {pages}
  {133601} (\bibinfo {year} {2010})}\BibitemShut {NoStop}%
\bibitem [{\citenamefont {Julsgaard}\ \emph {et~al.}(2001)\citenamefont
  {Julsgaard}, \citenamefont {Kozhekin},\ and\ \citenamefont
  {Polzik}}]{Julsgaard2001}%
  \BibitemOpen
  \bibfield  {author} {\bibinfo {author} {\bibfnamefont {B.}~\bibnamefont
  {Julsgaard}}, \bibinfo {author} {\bibfnamefont {A.}~\bibnamefont {Kozhekin}},
  \ and\ \bibinfo {author} {\bibfnamefont {E.~S.}\ \bibnamefont {Polzik}},\
  }\href {http://dx.doi.org/10.1038/35096524} {\bibfield  {journal} {\bibinfo
  {journal} {Nature}\ }\textbf {\bibinfo {volume} {413}},\ \bibinfo {pages}
  {400} (\bibinfo {year} {2001})}\BibitemShut {NoStop}%
\bibitem [{\citenamefont {Krauter}\ \emph {et~al.}(2011)\citenamefont
  {Krauter}, \citenamefont {Muschik}, \citenamefont {Jensen}, \citenamefont
  {Wasilewski}, \citenamefont {Petersen}, \citenamefont {Cirac},\ and\
  \citenamefont {Polzik}}]{Krauter2011}%
  \BibitemOpen
  \bibfield  {author} {\bibinfo {author} {\bibfnamefont {H.}~\bibnamefont
  {Krauter}}, \bibinfo {author} {\bibfnamefont {C.~A.}\ \bibnamefont
  {Muschik}}, \bibinfo {author} {\bibfnamefont {K.}~\bibnamefont {Jensen}},
  \bibinfo {author} {\bibfnamefont {W.}~\bibnamefont {Wasilewski}}, \bibinfo
  {author} {\bibfnamefont {J.~M.}\ \bibnamefont {Petersen}}, \bibinfo {author}
  {\bibfnamefont {J.~I.}\ \bibnamefont {Cirac}}, \ and\ \bibinfo {author}
  {\bibfnamefont {E.~S.}\ \bibnamefont {Polzik}},\ }\href {\doibase
  10.1103/PhysRevLett.107.080503} {\bibfield  {journal} {\bibinfo  {journal}
  {Phys. Rev. Lett.}\ }\textbf {\bibinfo {volume} {107}},\ \bibinfo {pages}
  {080503} (\bibinfo {year} {2011})}\BibitemShut {NoStop}%
\bibitem [{\citenamefont {Lee}\ \emph {et~al.}(2011)\citenamefont {Lee},
  \citenamefont {Sprague}, \citenamefont {Sussman}, \citenamefont {Nunn},
  \citenamefont {Langford}, \citenamefont {Jin}, \citenamefont {Champion},
  \citenamefont {Michelberger}, \citenamefont {Reim}, \citenamefont {England},
  \citenamefont {Jaksch},\ and\ \citenamefont {Walmsley}}]{Lee2011}%
  \BibitemOpen
  \bibfield  {author} {\bibinfo {author} {\bibfnamefont {K.~C.}\ \bibnamefont
  {Lee}}, \bibinfo {author} {\bibfnamefont {M.~R.}\ \bibnamefont {Sprague}},
  \bibinfo {author} {\bibfnamefont {B.~J.}\ \bibnamefont {Sussman}}, \bibinfo
  {author} {\bibfnamefont {J.}~\bibnamefont {Nunn}}, \bibinfo {author}
  {\bibfnamefont {N.~K.}\ \bibnamefont {Langford}}, \bibinfo {author}
  {\bibfnamefont {X.-M.}\ \bibnamefont {Jin}}, \bibinfo {author} {\bibfnamefont
  {T.}~\bibnamefont {Champion}}, \bibinfo {author} {\bibfnamefont
  {P.}~\bibnamefont {Michelberger}}, \bibinfo {author} {\bibfnamefont {K.~F.}\
  \bibnamefont {Reim}}, \bibinfo {author} {\bibfnamefont {D.}~\bibnamefont
  {England}}, \bibinfo {author} {\bibfnamefont {D.}~\bibnamefont {Jaksch}}, \
  and\ \bibinfo {author} {\bibfnamefont {I.~A.}\ \bibnamefont {Walmsley}},\
  }\href {\doibase 10.1126/science.1211914} {\bibfield  {journal} {\bibinfo
  {journal} {Science}\ }\textbf {\bibinfo {volume} {334}},\ \bibinfo {pages}
  {1253} (\bibinfo {year} {2011})}\BibitemShut {NoStop}%
\bibitem [{\citenamefont {Riedinger}\ \emph {et~al.}(2018)\citenamefont
  {Riedinger}, \citenamefont {Wallucks}, \citenamefont {Marinkovi{\'c}},
  \citenamefont {L{\"o}schnauer}, \citenamefont {Aspelmeyer}, \citenamefont
  {Hong},\ and\ \citenamefont {Gr{\"o}blacher}}]{Riedinger2018}%
  \BibitemOpen
  \bibfield  {author} {\bibinfo {author} {\bibfnamefont {R.}~\bibnamefont
  {Riedinger}}, \bibinfo {author} {\bibfnamefont {A.}~\bibnamefont {Wallucks}},
  \bibinfo {author} {\bibfnamefont {I.}~\bibnamefont {Marinkovi{\'c}}},
  \bibinfo {author} {\bibfnamefont {C.}~\bibnamefont {L{\"o}schnauer}},
  \bibinfo {author} {\bibfnamefont {M.}~\bibnamefont {Aspelmeyer}}, \bibinfo
  {author} {\bibfnamefont {S.}~\bibnamefont {Hong}}, \ and\ \bibinfo {author}
  {\bibfnamefont {S.}~\bibnamefont {Gr{\"o}blacher}},\ }\href
  {https://doi.org/10.1038/s41586-018-0036-z} {\bibfield  {journal} {\bibinfo
  {journal} {Nature}\ }\textbf {\bibinfo {volume} {556}},\ \bibinfo {pages}
  {473} (\bibinfo {year} {2018})}\BibitemShut {NoStop}%
\bibitem [{\citenamefont {Ockeloen-Korppi}\ \emph {et~al.}(2018)\citenamefont
  {Ockeloen-Korppi}, \citenamefont {Damsk{\"a}gg}, \citenamefont
  {Pirkkalainen}, \citenamefont {Asjad}, \citenamefont {Clerk}, \citenamefont
  {Massel}, \citenamefont {Woolley},\ and\ \citenamefont
  {Sillanp{\"a}{\"a}}}]{Ockeloen-Korppi2018}%
  \BibitemOpen
  \bibfield  {author} {\bibinfo {author} {\bibfnamefont {C.~F.}\ \bibnamefont
  {Ockeloen-Korppi}}, \bibinfo {author} {\bibfnamefont {E.}~\bibnamefont
  {Damsk{\"a}gg}}, \bibinfo {author} {\bibfnamefont {J.-M.}\ \bibnamefont
  {Pirkkalainen}}, \bibinfo {author} {\bibfnamefont {M.}~\bibnamefont {Asjad}},
  \bibinfo {author} {\bibfnamefont {A.~A.}\ \bibnamefont {Clerk}}, \bibinfo
  {author} {\bibfnamefont {F.}~\bibnamefont {Massel}}, \bibinfo {author}
  {\bibfnamefont {M.~J.}\ \bibnamefont {Woolley}}, \ and\ \bibinfo {author}
  {\bibfnamefont {M.~A.}\ \bibnamefont {Sillanp{\"a}{\"a}}},\ }\href
  {https://doi.org/10.1038/s41586-018-0038-x} {\bibfield  {journal} {\bibinfo
  {journal} {Nature}\ }\textbf {\bibinfo {volume} {556}},\ \bibinfo {pages}
  {478} (\bibinfo {year} {2018})}\BibitemShut {NoStop}%
\bibitem [{\citenamefont {Kurizki}\ \emph {et~al.}(2015)\citenamefont
  {Kurizki}, \citenamefont {Bertet}, \citenamefont {Kubo}, \citenamefont
  {M{\o}lmer}, \citenamefont {Petrosyan}, \citenamefont {Rabl},\ and\
  \citenamefont {Schmiedmayer}}]{Kurizki2015}%
  \BibitemOpen
  \bibfield  {author} {\bibinfo {author} {\bibfnamefont {G.}~\bibnamefont
  {Kurizki}}, \bibinfo {author} {\bibfnamefont {P.}~\bibnamefont {Bertet}},
  \bibinfo {author} {\bibfnamefont {Y.}~\bibnamefont {Kubo}}, \bibinfo {author}
  {\bibfnamefont {K.}~\bibnamefont {M{\o}lmer}}, \bibinfo {author}
  {\bibfnamefont {D.}~\bibnamefont {Petrosyan}}, \bibinfo {author}
  {\bibfnamefont {P.}~\bibnamefont {Rabl}}, \ and\ \bibinfo {author}
  {\bibfnamefont {J.}~\bibnamefont {Schmiedmayer}},\ }\href {\doibase
  10.1073/pnas.1419326112} {\bibfield  {journal} {\bibinfo  {journal}
  {Proceedings of the National Academy of Sciences}\ }\textbf {\bibinfo
  {volume} {112}},\ \bibinfo {pages} {3866} (\bibinfo {year}
  {2015})}\BibitemShut {NoStop}%
\bibitem [{\citenamefont {Tsang}\ and\ \citenamefont
  {Caves}(2012)}]{Tsang2012}%
  \BibitemOpen
  \bibfield  {author} {\bibinfo {author} {\bibfnamefont {M.}~\bibnamefont
  {Tsang}}\ and\ \bibinfo {author} {\bibfnamefont {C.~M.}\ \bibnamefont
  {Caves}},\ }\href {\doibase 10.1103/PhysRevX.2.031016} {\bibfield  {journal}
  {\bibinfo  {journal} {Phys. Rev. X}\ }\textbf {\bibinfo {volume} {2}},\
  \bibinfo {pages} {031016} (\bibinfo {year} {2012})}\BibitemShut {NoStop}%
\bibitem [{\citenamefont {Polzik}\ and\ \citenamefont
  {Hammerer}(2015)}]{Polzik2015}%
  \BibitemOpen
  \bibfield  {author} {\bibinfo {author} {\bibfnamefont {E.~S.}\ \bibnamefont
  {Polzik}}\ and\ \bibinfo {author} {\bibfnamefont {K.}~\bibnamefont
  {Hammerer}},\ }\href {\doibase 10.1002/andp.201400099} {\bibfield  {journal}
  {\bibinfo  {journal} {Annalen der Physik}\ }\textbf {\bibinfo {volume}
  {527}},\ \bibinfo {pages} {A15} (\bibinfo {year} {2015})}\BibitemShut
  {NoStop}%
\bibitem [{\citenamefont {Hammerer}\ \emph {et~al.}(2010)\citenamefont
  {Hammerer}, \citenamefont {S\o{}rensen},\ and\ \citenamefont
  {Polzik}}]{Hammerer2010}%
  \BibitemOpen
  \bibfield  {author} {\bibinfo {author} {\bibfnamefont {K.}~\bibnamefont
  {Hammerer}}, \bibinfo {author} {\bibfnamefont {A.~S.}\ \bibnamefont
  {S\o{}rensen}}, \ and\ \bibinfo {author} {\bibfnamefont {E.~S.}\ \bibnamefont
  {Polzik}},\ }\href {\doibase 10.1103/RevModPhys.82.1041} {\bibfield
  {journal} {\bibinfo  {journal} {Rev. Mod. Phys.}\ }\textbf {\bibinfo {volume}
  {82}},\ \bibinfo {pages} {1041} (\bibinfo {year} {2010})}\BibitemShut
  {NoStop}%
\bibitem [{\citenamefont {Vasilakis}\ \emph {et~al.}(2015)\citenamefont
  {Vasilakis}, \citenamefont {Shen}, \citenamefont {Jensen}, \citenamefont
  {Balabas}, \citenamefont {Salart}, \citenamefont {Chen},\ and\ \citenamefont
  {Polzik}}]{Vasilakis2015}%
  \BibitemOpen
  \bibfield  {author} {\bibinfo {author} {\bibfnamefont {G.}~\bibnamefont
  {Vasilakis}}, \bibinfo {author} {\bibfnamefont {H.}~\bibnamefont {Shen}},
  \bibinfo {author} {\bibfnamefont {K.}~\bibnamefont {Jensen}}, \bibinfo
  {author} {\bibfnamefont {M.}~\bibnamefont {Balabas}}, \bibinfo {author}
  {\bibfnamefont {D.}~\bibnamefont {Salart}}, \bibinfo {author} {\bibfnamefont
  {B.}~\bibnamefont {Chen}}, \ and\ \bibinfo {author} {\bibfnamefont {E.~S.}\
  \bibnamefont {Polzik}},\ }\href {http://dx.doi.org/10.1038/nphys3280}
  {\bibfield  {journal} {\bibinfo  {journal} {Nat Phys}\ }\textbf {\bibinfo
  {volume} {11}},\ \bibinfo {pages} {389} (\bibinfo {year} {2015})}\BibitemShut
  {NoStop}%
\bibitem [{\citenamefont {Muessel}\ \emph {et~al.}(2014)\citenamefont
  {Muessel}, \citenamefont {Strobel}, \citenamefont {Linnemann}, \citenamefont
  {Hume},\ and\ \citenamefont {Oberthaler}}]{Muessel2014}%
  \BibitemOpen
  \bibfield  {author} {\bibinfo {author} {\bibfnamefont {W.}~\bibnamefont
  {Muessel}}, \bibinfo {author} {\bibfnamefont {H.}~\bibnamefont {Strobel}},
  \bibinfo {author} {\bibfnamefont {D.}~\bibnamefont {Linnemann}}, \bibinfo
  {author} {\bibfnamefont {D.~B.}\ \bibnamefont {Hume}}, \ and\ \bibinfo
  {author} {\bibfnamefont {M.~K.}\ \bibnamefont {Oberthaler}},\ }\href
  {\doibase 10.1103/PhysRevLett.113.103004} {\bibfield  {journal} {\bibinfo
  {journal} {Phys. Rev. Lett.}\ }\textbf {\bibinfo {volume} {113}},\ \bibinfo
  {pages} {103004} (\bibinfo {year} {2014})}\BibitemShut {NoStop}%
\bibitem [{\citenamefont {McConnell}\ \emph {et~al.}(2015)\citenamefont
  {McConnell}, \citenamefont {Zhang}, \citenamefont {Hu}, \citenamefont {Cuk},\
  and\ \citenamefont {Vuletic}}]{McConnell2015}%
  \BibitemOpen
  \bibfield  {author} {\bibinfo {author} {\bibfnamefont {R.}~\bibnamefont
  {McConnell}}, \bibinfo {author} {\bibfnamefont {H.}~\bibnamefont {Zhang}},
  \bibinfo {author} {\bibfnamefont {J.}~\bibnamefont {Hu}}, \bibinfo {author}
  {\bibfnamefont {S.}~\bibnamefont {Cuk}}, \ and\ \bibinfo {author}
  {\bibfnamefont {V.}~\bibnamefont {Vuletic}},\ }\href
  {http://dx.doi.org/10.1038/nature14293} {\bibfield  {journal} {\bibinfo
  {journal} {Nature}\ }\textbf {\bibinfo {volume} {519}},\ \bibinfo {pages}
  {439} (\bibinfo {year} {2015})}\BibitemShut {NoStop}%
\bibitem [{\citenamefont {Kohler}\ \emph {et~al.}(2018)\citenamefont {Kohler},
  \citenamefont {Gerber}, \citenamefont {Dowd},\ and\ \citenamefont
  {Stamper-Kurn}}]{Kohler2018}%
  \BibitemOpen
  \bibfield  {author} {\bibinfo {author} {\bibfnamefont {J.}~\bibnamefont
  {Kohler}}, \bibinfo {author} {\bibfnamefont {J.~A.}\ \bibnamefont {Gerber}},
  \bibinfo {author} {\bibfnamefont {E.}~\bibnamefont {Dowd}}, \ and\ \bibinfo
  {author} {\bibfnamefont {D.~M.}\ \bibnamefont {Stamper-Kurn}},\ }\href
  {\doibase 10.1103/PhysRevLett.120.013601} {\bibfield  {journal} {\bibinfo
  {journal} {Phys. Rev. Lett.}\ }\textbf {\bibinfo {volume} {120}},\ \bibinfo
  {pages} {013601} (\bibinfo {year} {2018})}\BibitemShut {NoStop}%
\bibitem [{\citenamefont {Vetsch}\ \emph {et~al.}(2010)\citenamefont {Vetsch},
  \citenamefont {Reitz}, \citenamefont {Sagu\'e}, \citenamefont {Schmidt},
  \citenamefont {Dawkins},\ and\ \citenamefont {Rauschenbeutel}}]{Vetsch2010}%
  \BibitemOpen
  \bibfield  {author} {\bibinfo {author} {\bibfnamefont {E.}~\bibnamefont
  {Vetsch}}, \bibinfo {author} {\bibfnamefont {D.}~\bibnamefont {Reitz}},
  \bibinfo {author} {\bibfnamefont {G.}~\bibnamefont {Sagu\'e}}, \bibinfo
  {author} {\bibfnamefont {R.}~\bibnamefont {Schmidt}}, \bibinfo {author}
  {\bibfnamefont {S.~T.}\ \bibnamefont {Dawkins}}, \ and\ \bibinfo {author}
  {\bibfnamefont {A.}~\bibnamefont {Rauschenbeutel}},\ }\href {\doibase
  10.1103/PhysRevLett.104.203603} {\bibfield  {journal} {\bibinfo  {journal}
  {Phys. Rev. Lett.}\ }\textbf {\bibinfo {volume} {104}},\ \bibinfo {pages}
  {203603} (\bibinfo {year} {2010})}\BibitemShut {NoStop}%
\bibitem [{\citenamefont {B{\'e}guin}\ \emph {et~al.}(2014)\citenamefont
  {B{\'e}guin}, \citenamefont {Bookjans}, \citenamefont {Christensen},
  \citenamefont {S\o{}rensen}, \citenamefont {M{\"u}ller}, \citenamefont
  {Polzik},\ and\ \citenamefont {Appel}}]{Beguin2014}%
  \BibitemOpen
  \bibfield  {author} {\bibinfo {author} {\bibfnamefont {J.-B.}\ \bibnamefont
  {B{\'e}guin}}, \bibinfo {author} {\bibfnamefont {E.~M.}\ \bibnamefont
  {Bookjans}}, \bibinfo {author} {\bibfnamefont {S.~L.}\ \bibnamefont
  {Christensen}}, \bibinfo {author} {\bibfnamefont {H.~L.}\ \bibnamefont
  {S\o{}rensen}}, \bibinfo {author} {\bibfnamefont {J.~H.}\ \bibnamefont
  {M{\"u}ller}}, \bibinfo {author} {\bibfnamefont {E.~S.}\ \bibnamefont
  {Polzik}}, \ and\ \bibinfo {author} {\bibfnamefont {J.}~\bibnamefont
  {Appel}},\ }\href {\doibase 10.1103/PhysRevLett.113.263603} {\bibfield
  {journal} {\bibinfo  {journal} {Phys. Rev. Lett.}\ }\textbf {\bibinfo
  {volume} {113}},\ \bibinfo {pages} {263603} (\bibinfo {year}
  {2014})}\BibitemShut {NoStop}%
\bibitem [{\citenamefont {Grezes}\ \emph {et~al.}(2014)\citenamefont {Grezes},
  \citenamefont {Julsgaard}, \citenamefont {Kubo}, \citenamefont {Stern},
  \citenamefont {Umeda}, \citenamefont {Isoya}, \citenamefont {Sumiya},
  \citenamefont {Abe}, \citenamefont {Onoda}, \citenamefont {Ohshima},
  \citenamefont {Jacques}, \citenamefont {Esteve}, \citenamefont {Vion},
  \citenamefont {Esteve}, \citenamefont {M\o{}lmer},\ and\ \citenamefont
  {Bertet}}]{Grezes2014}%
  \BibitemOpen
  \bibfield  {author} {\bibinfo {author} {\bibfnamefont {C.}~\bibnamefont
  {Grezes}}, \bibinfo {author} {\bibfnamefont {B.}~\bibnamefont {Julsgaard}},
  \bibinfo {author} {\bibfnamefont {Y.}~\bibnamefont {Kubo}}, \bibinfo {author}
  {\bibfnamefont {M.}~\bibnamefont {Stern}}, \bibinfo {author} {\bibfnamefont
  {T.}~\bibnamefont {Umeda}}, \bibinfo {author} {\bibfnamefont
  {J.}~\bibnamefont {Isoya}}, \bibinfo {author} {\bibfnamefont
  {H.}~\bibnamefont {Sumiya}}, \bibinfo {author} {\bibfnamefont
  {H.}~\bibnamefont {Abe}}, \bibinfo {author} {\bibfnamefont {S.}~\bibnamefont
  {Onoda}}, \bibinfo {author} {\bibfnamefont {T.}~\bibnamefont {Ohshima}},
  \bibinfo {author} {\bibfnamefont {V.}~\bibnamefont {Jacques}}, \bibinfo
  {author} {\bibfnamefont {J.}~\bibnamefont {Esteve}}, \bibinfo {author}
  {\bibfnamefont {D.}~\bibnamefont {Vion}}, \bibinfo {author} {\bibfnamefont
  {D.}~\bibnamefont {Esteve}}, \bibinfo {author} {\bibfnamefont
  {K.}~\bibnamefont {M\o{}lmer}}, \ and\ \bibinfo {author} {\bibfnamefont
  {P.}~\bibnamefont {Bertet}},\ }\href {\doibase 10.1103/PhysRevX.4.021049}
  {\bibfield  {journal} {\bibinfo  {journal} {Phys. Rev. X}\ }\textbf {\bibinfo
  {volume} {4}},\ \bibinfo {pages} {021049} (\bibinfo {year}
  {2014})}\BibitemShut {NoStop}%
\bibitem [{\citenamefont {Jobez}\ \emph {et~al.}(2015)\citenamefont {Jobez},
  \citenamefont {Laplane}, \citenamefont {Timoney}, \citenamefont {Gisin},
  \citenamefont {Ferrier}, \citenamefont {Goldner},\ and\ \citenamefont
  {Afzelius}}]{Jobez2015}%
  \BibitemOpen
  \bibfield  {author} {\bibinfo {author} {\bibfnamefont {P.}~\bibnamefont
  {Jobez}}, \bibinfo {author} {\bibfnamefont {C.}~\bibnamefont {Laplane}},
  \bibinfo {author} {\bibfnamefont {N.}~\bibnamefont {Timoney}}, \bibinfo
  {author} {\bibfnamefont {N.}~\bibnamefont {Gisin}}, \bibinfo {author}
  {\bibfnamefont {A.}~\bibnamefont {Ferrier}}, \bibinfo {author} {\bibfnamefont
  {P.}~\bibnamefont {Goldner}}, \ and\ \bibinfo {author} {\bibfnamefont
  {M.}~\bibnamefont {Afzelius}},\ }\href {\doibase
  10.1103/PhysRevLett.114.230502} {\bibfield  {journal} {\bibinfo  {journal}
  {Phys. Rev. Lett.}\ }\textbf {\bibinfo {volume} {114}},\ \bibinfo {pages}
  {230502} (\bibinfo {year} {2015})}\BibitemShut {NoStop}%
\bibitem [{\citenamefont {G{\"undo\ifmmode \breve{g}\else \u{g}\fi{}}an}\ \emph
  {et~al.}(2015)\citenamefont {G{\"undo\ifmmode \breve{g}\else \u{g}\fi{}}an},
  \citenamefont {Ledingham}, \citenamefont {Kutluer}, \citenamefont {Mazzera},\
  and\ \citenamefont {de~Riedmatten}}]{Gundogan2015}%
  \BibitemOpen
  \bibfield  {author} {\bibinfo {author} {\bibfnamefont {M.}~\bibnamefont
  {G{\"undo\ifmmode \breve{g}\else \u{g}\fi{}}an}}, \bibinfo {author}
  {\bibfnamefont {P.~M.}\ \bibnamefont {Ledingham}}, \bibinfo {author}
  {\bibfnamefont {K.}~\bibnamefont {Kutluer}}, \bibinfo {author} {\bibfnamefont
  {M.}~\bibnamefont {Mazzera}}, \ and\ \bibinfo {author} {\bibfnamefont
  {H.}~\bibnamefont {de~Riedmatten}},\ }\href {\doibase
  10.1103/PhysRevLett.114.230501} {\bibfield  {journal} {\bibinfo  {journal}
  {Phys. Rev. Lett.}\ }\textbf {\bibinfo {volume} {114}},\ \bibinfo {pages}
  {230501} (\bibinfo {year} {2015})}\BibitemShut {NoStop}%
\bibitem [{\citenamefont {J{\"o}ckel}\ \emph {et~al.}(2015)\citenamefont
  {J{\"o}ckel}, \citenamefont {Faber}, \citenamefont {Kampschulte},
  \citenamefont {Korppi}, \citenamefont {Rakher},\ and\ \citenamefont
  {Treutlein}}]{Joeckel2015}%
  \BibitemOpen
  \bibfield  {author} {\bibinfo {author} {\bibfnamefont {A.}~\bibnamefont
  {J{\"o}ckel}}, \bibinfo {author} {\bibfnamefont {A.}~\bibnamefont {Faber}},
  \bibinfo {author} {\bibfnamefont {T.}~\bibnamefont {Kampschulte}}, \bibinfo
  {author} {\bibfnamefont {M.}~\bibnamefont {Korppi}}, \bibinfo {author}
  {\bibfnamefont {M.~T.}\ \bibnamefont {Rakher}}, \ and\ \bibinfo {author}
  {\bibfnamefont {P.}~\bibnamefont {Treutlein}},\ }\href
  {http://dx.doi.org/10.1038/nnano.2014.278} {\bibfield  {journal} {\bibinfo
  {journal} {Nat Nano}\ }\textbf {\bibinfo {volume} {10}},\ \bibinfo {pages}
  {55} (\bibinfo {year} {2015})}\BibitemShut {NoStop}%
\bibitem [{\citenamefont {Spethmann}\ \emph {et~al.}(2016)\citenamefont
  {Spethmann}, \citenamefont {Kohler}, \citenamefont {Schreppler},
  \citenamefont {Buchmann},\ and\ \citenamefont
  {Stamper-Kurn}}]{SPethmann2016}%
  \BibitemOpen
  \bibfield  {author} {\bibinfo {author} {\bibfnamefont {N.}~\bibnamefont
  {Spethmann}}, \bibinfo {author} {\bibfnamefont {J.}~\bibnamefont {Kohler}},
  \bibinfo {author} {\bibfnamefont {S.}~\bibnamefont {Schreppler}}, \bibinfo
  {author} {\bibfnamefont {L.}~\bibnamefont {Buchmann}}, \ and\ \bibinfo
  {author} {\bibfnamefont {D.~M.}\ \bibnamefont {Stamper-Kurn}},\ }\href
  {http://dx.doi.org/10.1038/nphys3515} {\bibfield  {journal} {\bibinfo
  {journal} {Nat Phys}\ }\textbf {\bibinfo {volume} {12}},\ \bibinfo {pages}
  {27} (\bibinfo {year} {2016})}\BibitemShut {NoStop}%
\bibitem [{\citenamefont {Gilmore}\ \emph {et~al.}(2017)\citenamefont
  {Gilmore}, \citenamefont {Bohnet}, \citenamefont {Sawyer}, \citenamefont
  {Britton},\ and\ \citenamefont {Bollinger}}]{Gilmore2017}%
  \BibitemOpen
  \bibfield  {author} {\bibinfo {author} {\bibfnamefont {K.~A.}\ \bibnamefont
  {Gilmore}}, \bibinfo {author} {\bibfnamefont {J.~G.}\ \bibnamefont {Bohnet}},
  \bibinfo {author} {\bibfnamefont {B.~C.}\ \bibnamefont {Sawyer}}, \bibinfo
  {author} {\bibfnamefont {J.~W.}\ \bibnamefont {Britton}}, \ and\ \bibinfo
  {author} {\bibfnamefont {J.~J.}\ \bibnamefont {Bollinger}},\ }\href {\doibase
  10.1103/PhysRevLett.118.263602} {\bibfield  {journal} {\bibinfo  {journal}
  {Phys. Rev. Lett.}\ }\textbf {\bibinfo {volume} {118}},\ \bibinfo {pages}
  {263602} (\bibinfo {year} {2017})}\BibitemShut {NoStop}%
\bibitem [{\citenamefont {Teufel}\ \emph {et~al.}(2011)\citenamefont {Teufel},
  \citenamefont {Donner}, \citenamefont {Li}, \citenamefont {Harlow},
  \citenamefont {Allman}, \citenamefont {Cicak}, \citenamefont {Sirois},
  \citenamefont {Whittaker}, \citenamefont {Lehnert},\ and\ \citenamefont
  {Simmonds}}]{Teufel2011}%
  \BibitemOpen
  \bibfield  {author} {\bibinfo {author} {\bibfnamefont {J.~D.}\ \bibnamefont
  {Teufel}}, \bibinfo {author} {\bibfnamefont {T.}~\bibnamefont {Donner}},
  \bibinfo {author} {\bibfnamefont {D.}~\bibnamefont {Li}}, \bibinfo {author}
  {\bibfnamefont {J.~W.}\ \bibnamefont {Harlow}}, \bibinfo {author}
  {\bibfnamefont {M.~S.}\ \bibnamefont {Allman}}, \bibinfo {author}
  {\bibfnamefont {K.}~\bibnamefont {Cicak}}, \bibinfo {author} {\bibfnamefont
  {A.~J.}\ \bibnamefont {Sirois}}, \bibinfo {author} {\bibfnamefont {J.~D.}\
  \bibnamefont {Whittaker}}, \bibinfo {author} {\bibfnamefont {K.~W.}\
  \bibnamefont {Lehnert}}, \ and\ \bibinfo {author} {\bibfnamefont {R.~W.}\
  \bibnamefont {Simmonds}},\ }\href {http://dx.doi.org/10.1038/nature10261}
  {\bibfield  {journal} {\bibinfo  {journal} {Nature}\ }\textbf {\bibinfo
  {volume} {475}},\ \bibinfo {pages} {359} (\bibinfo {year}
  {2011})}\BibitemShut {NoStop}%
\bibitem [{\citenamefont {Peterson}\ \emph {et~al.}(2016)\citenamefont
  {Peterson}, \citenamefont {Purdy}, \citenamefont {Kampel}, \citenamefont
  {Andrews}, \citenamefont {Yu}, \citenamefont {Lehnert},\ and\ \citenamefont
  {Regal}}]{Peterson2016}%
  \BibitemOpen
  \bibfield  {author} {\bibinfo {author} {\bibfnamefont {R.~W.}\ \bibnamefont
  {Peterson}}, \bibinfo {author} {\bibfnamefont {T.~P.}\ \bibnamefont {Purdy}},
  \bibinfo {author} {\bibfnamefont {N.~S.}\ \bibnamefont {Kampel}}, \bibinfo
  {author} {\bibfnamefont {R.~W.}\ \bibnamefont {Andrews}}, \bibinfo {author}
  {\bibfnamefont {P.-L.}\ \bibnamefont {Yu}}, \bibinfo {author} {\bibfnamefont
  {K.~W.}\ \bibnamefont {Lehnert}}, \ and\ \bibinfo {author} {\bibfnamefont
  {C.~A.}\ \bibnamefont {Regal}},\ }\href {\doibase
  10.1103/PhysRevLett.116.063601} {\bibfield  {journal} {\bibinfo  {journal}
  {Phys. Rev. Lett.}\ }\textbf {\bibinfo {volume} {116}},\ \bibinfo {pages}
  {063601} (\bibinfo {year} {2016})}\BibitemShut {NoStop}%
\bibitem [{\citenamefont {Nielsen}\ \emph {et~al.}(2017)\citenamefont
  {Nielsen}, \citenamefont {Tsaturyan}, \citenamefont {M{\o}ller},
  \citenamefont {Polzik},\ and\ \citenamefont {Schliesser}}]{Nielsen2017}%
  \BibitemOpen
  \bibfield  {author} {\bibinfo {author} {\bibfnamefont {W.~H.~P.}\
  \bibnamefont {Nielsen}}, \bibinfo {author} {\bibfnamefont {Y.}~\bibnamefont
  {Tsaturyan}}, \bibinfo {author} {\bibfnamefont {C.~B.}\ \bibnamefont
  {M{\o}ller}}, \bibinfo {author} {\bibfnamefont {E.~S.}\ \bibnamefont
  {Polzik}}, \ and\ \bibinfo {author} {\bibfnamefont {A.}~\bibnamefont
  {Schliesser}},\ }\href {\doibase 10.1073/pnas.1608412114} {\bibfield
  {journal} {\bibinfo  {journal} {Proceedings of the National Academy of
  Sciences}\ }\textbf {\bibinfo {volume} {114}},\ \bibinfo {pages} {62}
  (\bibinfo {year} {2017})}\BibitemShut {NoStop}%
\bibitem [{\citenamefont {Ockeloen-Korppi}\ \emph {et~al.}(2016)\citenamefont
  {Ockeloen-Korppi}, \citenamefont {Damsk{\"a}gg}, \citenamefont
  {Pirkkalainen}, \citenamefont {Clerk}, \citenamefont {Woolley},\ and\
  \citenamefont {Sillanp{\"a}{\"a}}}]{Ockeloen-Korppi2016}%
  \BibitemOpen
  \bibfield  {author} {\bibinfo {author} {\bibfnamefont {C.~F.}\ \bibnamefont
  {Ockeloen-Korppi}}, \bibinfo {author} {\bibfnamefont {E.}~\bibnamefont
  {Damsk{\"a}gg}}, \bibinfo {author} {\bibfnamefont {J.-M.}\ \bibnamefont
  {Pirkkalainen}}, \bibinfo {author} {\bibfnamefont {A.~A.}\ \bibnamefont
  {Clerk}}, \bibinfo {author} {\bibfnamefont {M.~J.}\ \bibnamefont {Woolley}},
  \ and\ \bibinfo {author} {\bibfnamefont {M.~A.}\ \bibnamefont
  {Sillanp{\"a}{\"a}}},\ }\href {\doibase 10.1103/PhysRevLett.117.140401}
  {\bibfield  {journal} {\bibinfo  {journal} {Phys. Rev. Lett.}\ }\textbf
  {\bibinfo {volume} {117}},\ \bibinfo {pages} {140401} (\bibinfo {year}
  {2016})}\BibitemShut {NoStop}%
\bibitem [{\citenamefont {Tan}\ \emph {et~al.}(2013)\citenamefont {Tan},
  \citenamefont {Buchmann}, \citenamefont {Seok},\ and\ \citenamefont
  {Li}}]{Tan2013}%
  \BibitemOpen
  \bibfield  {author} {\bibinfo {author} {\bibfnamefont {H.}~\bibnamefont
  {Tan}}, \bibinfo {author} {\bibfnamefont {L.~F.}\ \bibnamefont {Buchmann}},
  \bibinfo {author} {\bibfnamefont {H.}~\bibnamefont {Seok}}, \ and\ \bibinfo
  {author} {\bibfnamefont {G.}~\bibnamefont {Li}},\ }\href {\doibase
  10.1103/PhysRevA.87.022318} {\bibfield  {journal} {\bibinfo  {journal} {Phys.
  Rev. A}\ }\textbf {\bibinfo {volume} {87}},\ \bibinfo {pages} {022318}
  (\bibinfo {year} {2013})}\BibitemShut {NoStop}%
\bibitem [{\citenamefont {Tsang}\ and\ \citenamefont
  {Caves}(2010)}]{Tsang2010}%
  \BibitemOpen
  \bibfield  {author} {\bibinfo {author} {\bibfnamefont {M.}~\bibnamefont
  {Tsang}}\ and\ \bibinfo {author} {\bibfnamefont {C.~M.}\ \bibnamefont
  {Caves}},\ }\href {\doibase 10.1103/PhysRevLett.105.123601} {\bibfield
  {journal} {\bibinfo  {journal} {Phys. Rev. Lett.}\ }\textbf {\bibinfo
  {volume} {105}},\ \bibinfo {pages} {123601} (\bibinfo {year}
  {2010})}\BibitemShut {NoStop}%
\bibitem [{\citenamefont {Woolley}\ and\ \citenamefont
  {Clerk}(2013)}]{Woolley2013}%
  \BibitemOpen
  \bibfield  {author} {\bibinfo {author} {\bibfnamefont {M.~J.}\ \bibnamefont
  {Woolley}}\ and\ \bibinfo {author} {\bibfnamefont {A.~A.}\ \bibnamefont
  {Clerk}},\ }\href {\doibase 10.1103/PhysRevA.87.063846} {\bibfield  {journal}
  {\bibinfo  {journal} {Phys. Rev. A}\ }\textbf {\bibinfo {volume} {87}},\
  \bibinfo {pages} {063846} (\bibinfo {year} {2013})}\BibitemShut {NoStop}%
\bibitem [{\citenamefont {Wimmer}\ \emph {et~al.}(2014)\citenamefont {Wimmer},
  \citenamefont {Steinmeyer}, \citenamefont {Hammerer},\ and\ \citenamefont
  {Heurs}}]{Wimmer2014}%
  \BibitemOpen
  \bibfield  {author} {\bibinfo {author} {\bibfnamefont {M.~H.}\ \bibnamefont
  {Wimmer}}, \bibinfo {author} {\bibfnamefont {D.}~\bibnamefont {Steinmeyer}},
  \bibinfo {author} {\bibfnamefont {K.}~\bibnamefont {Hammerer}}, \ and\
  \bibinfo {author} {\bibfnamefont {M.}~\bibnamefont {Heurs}},\ }\href
  {\doibase 10.1103/PhysRevA.89.053836} {\bibfield  {journal} {\bibinfo
  {journal} {Phys. Rev. A}\ }\textbf {\bibinfo {volume} {89}},\ \bibinfo
  {pages} {053836} (\bibinfo {year} {2014})}\BibitemShut {NoStop}%
\bibitem [{\citenamefont {Bariani}\ \emph {et~al.}(2015)\citenamefont
  {Bariani}, \citenamefont {Seok}, \citenamefont {Singh}, \citenamefont
  {Vengalattore},\ and\ \citenamefont {Meystre}}]{Bariani2015}%
  \BibitemOpen
  \bibfield  {author} {\bibinfo {author} {\bibfnamefont {F.}~\bibnamefont
  {Bariani}}, \bibinfo {author} {\bibfnamefont {H.}~\bibnamefont {Seok}},
  \bibinfo {author} {\bibfnamefont {S.}~\bibnamefont {Singh}}, \bibinfo
  {author} {\bibfnamefont {M.}~\bibnamefont {Vengalattore}}, \ and\ \bibinfo
  {author} {\bibfnamefont {P.}~\bibnamefont {Meystre}},\ }\href {\doibase
  10.1103/PhysRevA.92.043817} {\bibfield  {journal} {\bibinfo  {journal} {Phys.
  Rev. A}\ }\textbf {\bibinfo {volume} {92}},\ \bibinfo {pages} {043817}
  (\bibinfo {year} {2015})}\BibitemShut {NoStop}%
\bibitem [{\citenamefont {Motazedifard}\ \emph {et~al.}(2016)\citenamefont
  {Motazedifard}, \citenamefont {Bemani}, \citenamefont {Naderi}, \citenamefont
  {Roknizadeh},\ and\ \citenamefont {Vitali}}]{Motazedifard2016}%
  \BibitemOpen
  \bibfield  {author} {\bibinfo {author} {\bibfnamefont {A.}~\bibnamefont
  {Motazedifard}}, \bibinfo {author} {\bibfnamefont {F.}~\bibnamefont
  {Bemani}}, \bibinfo {author} {\bibfnamefont {M.~H.}\ \bibnamefont {Naderi}},
  \bibinfo {author} {\bibfnamefont {R.}~\bibnamefont {Roknizadeh}}, \ and\
  \bibinfo {author} {\bibfnamefont {D.}~\bibnamefont {Vitali}},\ }\href
  {http://stacks.iop.org/1367-2630/18/i=7/a=073040} {\bibfield  {journal}
  {\bibinfo  {journal} {New Journal of Physics}\ }\textbf {\bibinfo {volume}
  {18}},\ \bibinfo {pages} {073040} (\bibinfo {year} {2016})}\BibitemShut
  {NoStop}%
\bibitem [{\citenamefont {M{\o}ller}\ \emph {et~al.}(2017)\citenamefont
  {M{\o}ller}, \citenamefont {Thomas}, \citenamefont {Vasilakis}, \citenamefont
  {Zeuthen}, \citenamefont {Tsaturyan}, \citenamefont {Balabas}, \citenamefont
  {Jensen}, \citenamefont {Schliesser}, \citenamefont {Hammerer},\ and\
  \citenamefont {Polzik}}]{Moller2017}%
  \BibitemOpen
  \bibfield  {author} {\bibinfo {author} {\bibfnamefont {C.~B.}\ \bibnamefont
  {M{\o}ller}}, \bibinfo {author} {\bibfnamefont {R.~A.}\ \bibnamefont
  {Thomas}}, \bibinfo {author} {\bibfnamefont {G.}~\bibnamefont {Vasilakis}},
  \bibinfo {author} {\bibfnamefont {E.}~\bibnamefont {Zeuthen}}, \bibinfo
  {author} {\bibfnamefont {Y.}~\bibnamefont {Tsaturyan}}, \bibinfo {author}
  {\bibfnamefont {M.}~\bibnamefont {Balabas}}, \bibinfo {author} {\bibfnamefont
  {K.}~\bibnamefont {Jensen}}, \bibinfo {author} {\bibfnamefont
  {A.}~\bibnamefont {Schliesser}}, \bibinfo {author} {\bibfnamefont
  {K.}~\bibnamefont {Hammerer}}, \ and\ \bibinfo {author} {\bibfnamefont
  {E.~S.}\ \bibnamefont {Polzik}},\ }\href
  {http://dx.doi.org/10.1038/nature22980} {\bibfield  {journal} {\bibinfo
  {journal} {Nature}\ }\textbf {\bibinfo {volume} {547}},\ \bibinfo {pages}
  {191} (\bibinfo {year} {2017})}\BibitemShut {NoStop}%
\bibitem [{\citenamefont {Muschik}\ \emph {et~al.}(2011)\citenamefont
  {Muschik}, \citenamefont {Polzik},\ and\ \citenamefont
  {Cirac}}]{Muschik2011}%
  \BibitemOpen
  \bibfield  {author} {\bibinfo {author} {\bibfnamefont {C.~A.}\ \bibnamefont
  {Muschik}}, \bibinfo {author} {\bibfnamefont {E.~S.}\ \bibnamefont {Polzik}},
  \ and\ \bibinfo {author} {\bibfnamefont {J.~I.}\ \bibnamefont {Cirac}},\
  }\href {\doibase 10.1103/PhysRevA.83.052312} {\bibfield  {journal} {\bibinfo
  {journal} {Phys. Rev. A}\ }\textbf {\bibinfo {volume} {83}},\ \bibinfo
  {pages} {052312} (\bibinfo {year} {2011})}\BibitemShut {NoStop}%
\bibitem [{\citenamefont {Vasilyev}\ \emph {et~al.}(2013)\citenamefont
  {Vasilyev}, \citenamefont {Muschik},\ and\ \citenamefont
  {Hammerer}}]{Vasilyev2013}%
  \BibitemOpen
  \bibfield  {author} {\bibinfo {author} {\bibfnamefont {D.~V.}\ \bibnamefont
  {Vasilyev}}, \bibinfo {author} {\bibfnamefont {C.~A.}\ \bibnamefont
  {Muschik}}, \ and\ \bibinfo {author} {\bibfnamefont {K.}~\bibnamefont
  {Hammerer}},\ }\href {\doibase 10.1103/PhysRevA.87.053820} {\bibfield
  {journal} {\bibinfo  {journal} {Phys. Rev. A}\ }\textbf {\bibinfo {volume}
  {87}},\ \bibinfo {pages} {053820} (\bibinfo {year} {2013})}\BibitemShut
  {NoStop}%
\bibitem [{\citenamefont {Woolley}\ and\ \citenamefont
  {Clerk}(2014)}]{Woolley2014}%
  \BibitemOpen
  \bibfield  {author} {\bibinfo {author} {\bibfnamefont {M.~J.}\ \bibnamefont
  {Woolley}}\ and\ \bibinfo {author} {\bibfnamefont {A.~A.}\ \bibnamefont
  {Clerk}},\ }\href {\doibase 10.1103/PhysRevA.89.063805} {\bibfield  {journal}
  {\bibinfo  {journal} {Phys. Rev. A}\ }\textbf {\bibinfo {volume} {89}},\
  \bibinfo {pages} {063805} (\bibinfo {year} {2014})}\BibitemShut {NoStop}%
\bibitem [{\citenamefont {Gardiner}(1993)}]{Gardiner1993}%
  \BibitemOpen
  \bibfield  {author} {\bibinfo {author} {\bibfnamefont {C.~W.}\ \bibnamefont
  {Gardiner}},\ }\href {\doibase 10.1103/PhysRevLett.70.2269} {\bibfield
  {journal} {\bibinfo  {journal} {Phys. Rev. Lett.}\ }\textbf {\bibinfo
  {volume} {70}},\ \bibinfo {pages} {2269} (\bibinfo {year}
  {1993})}\BibitemShut {NoStop}%
\bibitem [{\citenamefont {Gardiner}\ and\ \citenamefont
  {Zoller}(2004)}]{Gardiner2004}%
  \BibitemOpen
  \bibfield  {author} {\bibinfo {author} {\bibfnamefont {C.}~\bibnamefont
  {Gardiner}}\ and\ \bibinfo {author} {\bibfnamefont {P.}~\bibnamefont
  {Zoller}},\ }\href
  {https://www.ebook.de/de/product/13905493/crispin_gardiner_peter_zoller_quantum_noise.html}
  {\emph {\bibinfo {title} {Quantum Noise}}},\ 0172-7389\ (\bibinfo
  {publisher} {Springer-Verlag Berlin Heidelberg},\ \bibinfo {year}
  {2004})\BibitemShut {NoStop}%
\bibitem [{\citenamefont {Carmichael}(1993)}]{Carmichael1993}%
  \BibitemOpen
  \bibfield  {author} {\bibinfo {author} {\bibfnamefont {H.~J.}\ \bibnamefont
  {Carmichael}},\ }\href {\doibase 10.1103/PhysRevLett.70.2273} {\bibfield
  {journal} {\bibinfo  {journal} {Phys. Rev. Lett.}\ }\textbf {\bibinfo
  {volume} {70}},\ \bibinfo {pages} {2273} (\bibinfo {year}
  {1993})}\BibitemShut {NoStop}%
\bibitem [{\citenamefont {Simon}(2000)}]{Simon2000}%
  \BibitemOpen
  \bibfield  {author} {\bibinfo {author} {\bibfnamefont {R.}~\bibnamefont
  {Simon}},\ }\href {\doibase 10.1103/PhysRevLett.84.2726} {\bibfield
  {journal} {\bibinfo  {journal} {Phys. Rev. Lett.}\ }\textbf {\bibinfo
  {volume} {84}},\ \bibinfo {pages} {2726} (\bibinfo {year}
  {2000})}\BibitemShut {NoStop}%
\bibitem [{\citenamefont {Giovannetti}\ \emph {et~al.}(2003)\citenamefont
  {Giovannetti}, \citenamefont {Mancini}, \citenamefont {Vitali},\ and\
  \citenamefont {Tombesi}}]{Giovannetti2003}%
  \BibitemOpen
  \bibfield  {author} {\bibinfo {author} {\bibfnamefont {V.}~\bibnamefont
  {Giovannetti}}, \bibinfo {author} {\bibfnamefont {S.}~\bibnamefont
  {Mancini}}, \bibinfo {author} {\bibfnamefont {D.}~\bibnamefont {Vitali}}, \
  and\ \bibinfo {author} {\bibfnamefont {P.}~\bibnamefont {Tombesi}},\ }\href
  {\doibase 10.1103/PhysRevA.67.022320} {\bibfield  {journal} {\bibinfo
  {journal} {Phys. Rev. A}\ }\textbf {\bibinfo {volume} {67}},\ \bibinfo
  {pages} {022320} (\bibinfo {year} {2003})}\BibitemShut {NoStop}%
\bibitem [{\citenamefont {Aspelmeyer}\ \emph {et~al.}(2014)\citenamefont
  {Aspelmeyer}, \citenamefont {Kippenberg},\ and\ \citenamefont
  {Marquardt}}]{Aspelmeyer2014}%
  \BibitemOpen
  \bibfield  {author} {\bibinfo {author} {\bibfnamefont {M.}~\bibnamefont
  {Aspelmeyer}}, \bibinfo {author} {\bibfnamefont {T.~J.}\ \bibnamefont
  {Kippenberg}}, \ and\ \bibinfo {author} {\bibfnamefont {F.}~\bibnamefont
  {Marquardt}},\ }\href {\doibase 10.1103/RevModPhys.86.1391} {\bibfield
  {journal} {\bibinfo  {journal} {Rev. Mod. Phys.}\ }\textbf {\bibinfo {volume}
  {86}},\ \bibinfo {pages} {1391} (\bibinfo {year} {2014})}\BibitemShut
  {NoStop}%
\bibitem [{SM()}]{SM}%
  \BibitemOpen
  \href@noop {} {}\bibinfo {note} {See Supplementary Material, which includes
  Refs.~\cite{Holstein1940,Wasilewski2009,Stannigel2012,Belavkin1992,Cernotik2015,Clerk2010,TurinJune1960},
  for details on the spin-optomechanical implementation, conditional stochastic
  master equation, steady-state entanglement optimization, and a comparison
  between unconditional entanglement and force sensitivity.}\BibitemShut
  {Stop}%
\bibitem [{\citenamefont {Hammerer}\ \emph {et~al.}(2009)\citenamefont
  {Hammerer}, \citenamefont {Aspelmeyer}, \citenamefont {Polzik},\ and\
  \citenamefont {Zoller}}]{Hammerer2009}%
  \BibitemOpen
  \bibfield  {author} {\bibinfo {author} {\bibfnamefont {K.}~\bibnamefont
  {Hammerer}}, \bibinfo {author} {\bibfnamefont {M.}~\bibnamefont
  {Aspelmeyer}}, \bibinfo {author} {\bibfnamefont {E.~S.}\ \bibnamefont
  {Polzik}}, \ and\ \bibinfo {author} {\bibfnamefont {P.}~\bibnamefont
  {Zoller}},\ }\href {\doibase 10.1103/PhysRevLett.102.020501} {\bibfield
  {journal} {\bibinfo  {journal} {Phys. Rev. Lett.}\ }\textbf {\bibinfo
  {volume} {102}},\ \bibinfo {pages} {020501} (\bibinfo {year}
  {2009})}\BibitemShut {NoStop}%
\bibitem [{\citenamefont {Barreiro}\ \emph {et~al.}(2011)\citenamefont
  {Barreiro}, \citenamefont {M{\"u}ller}, \citenamefont {Schindler},
  \citenamefont {Nigg}, \citenamefont {Monz}, \citenamefont {Chwalla},
  \citenamefont {Hennrich}, \citenamefont {Roos}, \citenamefont {Zoller},\ and\
  \citenamefont {Blatt}}]{Barreiro2011}%
  \BibitemOpen
  \bibfield  {author} {\bibinfo {author} {\bibfnamefont {J.~T.}\ \bibnamefont
  {Barreiro}}, \bibinfo {author} {\bibfnamefont {M.}~\bibnamefont
  {M{\"u}ller}}, \bibinfo {author} {\bibfnamefont {P.}~\bibnamefont
  {Schindler}}, \bibinfo {author} {\bibfnamefont {D.}~\bibnamefont {Nigg}},
  \bibinfo {author} {\bibfnamefont {T.}~\bibnamefont {Monz}}, \bibinfo {author}
  {\bibfnamefont {M.}~\bibnamefont {Chwalla}}, \bibinfo {author} {\bibfnamefont
  {M.}~\bibnamefont {Hennrich}}, \bibinfo {author} {\bibfnamefont {C.~F.}\
  \bibnamefont {Roos}}, \bibinfo {author} {\bibfnamefont {P.}~\bibnamefont
  {Zoller}}, \ and\ \bibinfo {author} {\bibfnamefont {R.}~\bibnamefont
  {Blatt}},\ }\href {http://dx.doi.org/10.1038/nature09801} {\bibfield
  {journal} {\bibinfo  {journal} {Nature}\ }\textbf {\bibinfo {volume} {470}},\
  \bibinfo {pages} {486 EP } (\bibinfo {year} {2011})}\BibitemShut {NoStop}%
\bibitem [{\citenamefont {Wiseman}\ and\ \citenamefont
  {Milburn}(2010)}]{Wiseman2010}%
  \BibitemOpen
  \bibfield  {author} {\bibinfo {author} {\bibfnamefont {H.~M.}\ \bibnamefont
  {Wiseman}}\ and\ \bibinfo {author} {\bibfnamefont {G.~J.}\ \bibnamefont
  {Milburn}},\ }\href
  {https://www.ebook.de/de/product/8631634/howard_m_griffith_university_queensland_wiseman_gerard_j_professor_university_of_queensland_milburn_quantum_measurement_and_control.html}
  {\emph {\bibinfo {title} {Quantum Measurement and Control}}}\ (\bibinfo
  {publisher} {Cambridge University Press},\ \bibinfo {year}
  {2010})\BibitemShut {NoStop}%
\bibitem [{\citenamefont {He}\ \emph {et~al.}(2015)\citenamefont {He},
  \citenamefont {Rosales-Z\'arate}, \citenamefont {Adesso},\ and\ \citenamefont
  {Reid}}]{He2015}%
  \BibitemOpen
  \bibfield  {author} {\bibinfo {author} {\bibfnamefont {Q.}~\bibnamefont
  {He}}, \bibinfo {author} {\bibfnamefont {L.}~\bibnamefont
  {Rosales-Z\'arate}}, \bibinfo {author} {\bibfnamefont {G.}~\bibnamefont
  {Adesso}}, \ and\ \bibinfo {author} {\bibfnamefont {M.~D.}\ \bibnamefont
  {Reid}},\ }\href {\doibase 10.1103/PhysRevLett.115.180502} {\bibfield
  {journal} {\bibinfo  {journal} {Phys. Rev. Lett.}\ }\textbf {\bibinfo
  {volume} {115}},\ \bibinfo {pages} {180502} (\bibinfo {year}
  {2015})}\BibitemShut {NoStop}%
\bibitem [{\citenamefont {Walk}\ \emph {et~al.}(2016)\citenamefont {Walk},
  \citenamefont {Hosseini}, \citenamefont {Geng}, \citenamefont {Thearle},
  \citenamefont {Haw}, \citenamefont {Armstrong}, \citenamefont {Assad},
  \citenamefont {Janousek}, \citenamefont {Ralph}, \citenamefont {Symul},
  \citenamefont {Wiseman},\ and\ \citenamefont {Lam}}]{Walk2016}%
  \BibitemOpen
  \bibfield  {author} {\bibinfo {author} {\bibfnamefont {N.}~\bibnamefont
  {Walk}}, \bibinfo {author} {\bibfnamefont {S.}~\bibnamefont {Hosseini}},
  \bibinfo {author} {\bibfnamefont {J.}~\bibnamefont {Geng}}, \bibinfo {author}
  {\bibfnamefont {O.}~\bibnamefont {Thearle}}, \bibinfo {author} {\bibfnamefont
  {J.~Y.}\ \bibnamefont {Haw}}, \bibinfo {author} {\bibfnamefont
  {S.}~\bibnamefont {Armstrong}}, \bibinfo {author} {\bibfnamefont {S.~M.}\
  \bibnamefont {Assad}}, \bibinfo {author} {\bibfnamefont {J.}~\bibnamefont
  {Janousek}}, \bibinfo {author} {\bibfnamefont {T.~C.}\ \bibnamefont {Ralph}},
  \bibinfo {author} {\bibfnamefont {T.}~\bibnamefont {Symul}}, \bibinfo
  {author} {\bibfnamefont {H.~M.}\ \bibnamefont {Wiseman}}, \ and\ \bibinfo
  {author} {\bibfnamefont {P.~K.}\ \bibnamefont {Lam}},\ }\href {\doibase
  10.1364/OPTICA.3.000634} {\bibfield  {journal} {\bibinfo  {journal} {Optica}\
  }\textbf {\bibinfo {volume} {3}},\ \bibinfo {pages} {634} (\bibinfo {year}
  {2016})}\BibitemShut {NoStop}%
\bibitem [{\citenamefont {Zeuthen~et al{.}}()}]{EZInPrep}%
  \BibitemOpen
  \bibfield  {author} {\bibinfo {author} {\bibfnamefont {E.}~\bibnamefont
  {Zeuthen~et al{.}}},\ }\href@noop {} {}\bibinfo {note} {{in
  preparation}}\BibitemShut {NoStop}%
\bibitem [{\citenamefont {Holstein}\ and\ \citenamefont
  {Primakoff}(1940)}]{Holstein1940}%
  \BibitemOpen
  \bibfield  {author} {\bibinfo {author} {\bibfnamefont {T.}~\bibnamefont
  {Holstein}}\ and\ \bibinfo {author} {\bibfnamefont {H.}~\bibnamefont
  {Primakoff}},\ }\href {\doibase 10.1103/PhysRev.58.1098} {\bibfield
  {journal} {\bibinfo  {journal} {Phys. Rev.}\ }\textbf {\bibinfo {volume}
  {58}},\ \bibinfo {pages} {1098} (\bibinfo {year} {1940})}\BibitemShut
  {NoStop}%
\bibitem [{\citenamefont {Wasilewski}\ \emph {et~al.}(2009)\citenamefont
  {Wasilewski}, \citenamefont {Fernholz}, \citenamefont {Jensen}, \citenamefont
  {Madsen}, \citenamefont {Krauter}, \citenamefont {Muschik},\ and\
  \citenamefont {Polzik}}]{Wasilewski2009}%
  \BibitemOpen
  \bibfield  {author} {\bibinfo {author} {\bibfnamefont {W.}~\bibnamefont
  {Wasilewski}}, \bibinfo {author} {\bibfnamefont {T.}~\bibnamefont
  {Fernholz}}, \bibinfo {author} {\bibfnamefont {K.}~\bibnamefont {Jensen}},
  \bibinfo {author} {\bibfnamefont {L.~S.}\ \bibnamefont {Madsen}}, \bibinfo
  {author} {\bibfnamefont {H.}~\bibnamefont {Krauter}}, \bibinfo {author}
  {\bibfnamefont {C.}~\bibnamefont {Muschik}}, \ and\ \bibinfo {author}
  {\bibfnamefont {E.~S.}\ \bibnamefont {Polzik}},\ }\href {\doibase
  10.1364/OE.17.014444} {\bibfield  {journal} {\bibinfo  {journal} {Opt.
  Express}\ }\textbf {\bibinfo {volume} {17}},\ \bibinfo {pages} {14444}
  (\bibinfo {year} {2009})}\BibitemShut {NoStop}%
\bibitem [{\citenamefont {Stannigel}\ \emph {et~al.}(2012)\citenamefont
  {Stannigel}, \citenamefont {Rabl},\ and\ \citenamefont
  {Zoller}}]{Stannigel2012}%
  \BibitemOpen
  \bibfield  {author} {\bibinfo {author} {\bibfnamefont {K.}~\bibnamefont
  {Stannigel}}, \bibinfo {author} {\bibfnamefont {P.}~\bibnamefont {Rabl}}, \
  and\ \bibinfo {author} {\bibfnamefont {P.}~\bibnamefont {Zoller}},\ }\href
  {http://stacks.iop.org/1367-2630/14/i=6/a=063014} {\bibfield  {journal}
  {\bibinfo  {journal} {New Journal of Physics}\ }\textbf {\bibinfo {volume}
  {14}},\ \bibinfo {pages} {063014} (\bibinfo {year} {2012})}\BibitemShut
  {NoStop}%
\bibitem [{\citenamefont {Belavkin}(1992)}]{Belavkin1992}%
  \BibitemOpen
  \bibfield  {author} {\bibinfo {author} {\bibfnamefont {V.~P.}\ \bibnamefont
  {Belavkin}},\ }\href {\doibase 10.1007/BF02097018} {\bibfield  {journal}
  {\bibinfo  {journal} {Communications in Mathematical Physics}\ }\textbf
  {\bibinfo {volume} {146}},\ \bibinfo {pages} {611} (\bibinfo {year}
  {1992})}\BibitemShut {NoStop}%
\bibitem [{\citenamefont {{\ifmmode \check{C}\else \v{C}\fi{}ernot\'{\i}k}}\
  \emph {et~al.}(2015)\citenamefont {{\ifmmode \check{C}\else
  \v{C}\fi{}ernot\'{\i}k}}, \citenamefont {Vasilyev},\ and\ \citenamefont
  {Hammerer}}]{Cernotik2015}%
  \BibitemOpen
  \bibfield  {author} {\bibinfo {author} {\bibfnamefont {O.}~\bibnamefont
  {{\ifmmode \check{C}\else \v{C}\fi{}ernot\'{\i}k}}}, \bibinfo {author}
  {\bibfnamefont {D.~V.}\ \bibnamefont {Vasilyev}}, \ and\ \bibinfo {author}
  {\bibfnamefont {K.}~\bibnamefont {Hammerer}},\ }\href {\doibase
  10.1103/PhysRevA.92.012124} {\bibfield  {journal} {\bibinfo  {journal} {Phys.
  Rev. A}\ }\textbf {\bibinfo {volume} {92}},\ \bibinfo {pages} {012124}
  (\bibinfo {year} {2015})}\BibitemShut {NoStop}%
\bibitem [{\citenamefont {Clerk}\ \emph {et~al.}(2010)\citenamefont {Clerk},
  \citenamefont {Devoret}, \citenamefont {Girvin}, \citenamefont {Marquardt},\
  and\ \citenamefont {Schoelkopf}}]{Clerk2010}%
  \BibitemOpen
  \bibfield  {author} {\bibinfo {author} {\bibfnamefont {A.~A.}\ \bibnamefont
  {Clerk}}, \bibinfo {author} {\bibfnamefont {M.~H.}\ \bibnamefont {Devoret}},
  \bibinfo {author} {\bibfnamefont {S.~M.}\ \bibnamefont {Girvin}}, \bibinfo
  {author} {\bibfnamefont {F.}~\bibnamefont {Marquardt}}, \ and\ \bibinfo
  {author} {\bibfnamefont {R.~J.}\ \bibnamefont {Schoelkopf}},\ }\href
  {\doibase 10.1103/RevModPhys.82.1155} {\bibfield  {journal} {\bibinfo
  {journal} {Rev. Mod. Phys.}\ }\textbf {\bibinfo {volume} {82}},\ \bibinfo
  {pages} {1155} (\bibinfo {year} {2010})}\BibitemShut {NoStop}%
\bibitem [{\citenamefont {Turin}(1960)}]{TurinJune1960}%
  \BibitemOpen
  \bibfield  {author} {\bibinfo {author} {\bibfnamefont {G.}~\bibnamefont
  {Turin}},\ }\bibfield  {booktitle} {\emph {\bibinfo {booktitle} {IRE
  Transactions on Information Theory}},\ }\href@noop {} {\bibfield  {journal}
  {\bibinfo  {journal} {IRE Transactions on Information Theory}\ }\textbf
  {\bibinfo {volume} {6}},\ \bibinfo {pages} {311} (\bibinfo {year} {June
  1960})}\BibitemShut {NoStop}%
\end{thebibliography}%


\begin{thebibliography}{20}%
\makeatletter
\providecommand \@ifxundefined [1]{%
 \@ifx{#1\undefined}
}%
\providecommand \@ifnum [1]{%
 \ifnum #1\expandafter \@firstoftwo
 \else \expandafter \@secondoftwo
 \fi
}%
\providecommand \@ifx [1]{%
 \ifx #1\expandafter \@firstoftwo
 \else \expandafter \@secondoftwo
 \fi
}%
\providecommand \natexlab [1]{#1}%
\providecommand \enquote  [1]{``#1''}%
\providecommand \bibnamefont  [1]{#1}%
\providecommand \bibfnamefont [1]{#1}%
\providecommand \citenamefont [1]{#1}%
\providecommand \href@noop [0]{\@secondoftwo}%
\providecommand \href [0]{\begingroup \@sanitize@url \@href}%
\providecommand \@href[1]{\@@startlink{#1}\@@href}%
\providecommand \@@href[1]{\endgroup#1\@@endlink}%
\providecommand \@sanitize@url [0]{\catcode `\\12\catcode `\$12\catcode
  `\&12\catcode `\#12\catcode `\^12\catcode `\_12\catcode `\%12\relax}%
\providecommand \@@startlink[1]{}%
\providecommand \@@endlink[0]{}%
\providecommand \url  [0]{\begingroup\@sanitize@url \@url }%
\providecommand \@url [1]{\endgroup\@href {#1}{\urlprefix }}%
\providecommand \urlprefix  [0]{URL }%
\providecommand \Eprint [0]{\href }%
\providecommand \doibase [0]{http://dx.doi.org/}%
\providecommand \selectlanguage [0]{\@gobble}%
\providecommand \bibinfo  [0]{\@secondoftwo}%
\providecommand \bibfield  [0]{\@secondoftwo}%
\providecommand \translation [1]{[#1]}%
\providecommand \BibitemOpen [0]{}%
\providecommand \bibitemStop [0]{}%
\providecommand \bibitemNoStop [0]{.\EOS\space}%
\providecommand \EOS [0]{\spacefactor3000\relax}%
\providecommand \BibitemShut  [1]{\csname bibitem#1\endcsname}%
\let\auto@bib@innerbib\@empty
\bibitem [{\citenamefont {Holstein}\ and\ \citenamefont
  {Primakoff}(1940)}]{Holstein1940}%
  \BibitemOpen
  \bibfield  {author} {\bibinfo {author} {\bibfnamefont {T.}~\bibnamefont
  {Holstein}}\ and\ \bibinfo {author} {\bibfnamefont {H.}~\bibnamefont
  {Primakoff}},\ }\href {\doibase 10.1103/PhysRev.58.1098} {\bibfield
  {journal} {\bibinfo  {journal} {Phys. Rev.}\ }\textbf {\bibinfo {volume}
  {58}},\ \bibinfo {pages} {1098} (\bibinfo {year} {1940})}\BibitemShut
  {NoStop}%
\bibitem [{\citenamefont {Hammerer}\ \emph {et~al.}(2010)\citenamefont
  {Hammerer}, \citenamefont {S\o{}rensen},\ and\ \citenamefont
  {Polzik}}]{Hammerer2010}%
  \BibitemOpen
  \bibfield  {author} {\bibinfo {author} {\bibfnamefont {K.}~\bibnamefont
  {Hammerer}}, \bibinfo {author} {\bibfnamefont {A.~S.}\ \bibnamefont
  {S\o{}rensen}}, \ and\ \bibinfo {author} {\bibfnamefont {E.~S.}\ \bibnamefont
  {Polzik}},\ }\href {\doibase 10.1103/RevModPhys.82.1041} {\bibfield
  {journal} {\bibinfo  {journal} {Rev. Mod. Phys.}\ }\textbf {\bibinfo {volume}
  {82}},\ \bibinfo {pages} {1041} (\bibinfo {year} {2010})}\BibitemShut
  {NoStop}%
\bibitem [{\citenamefont {Wasilewski}\ \emph {et~al.}(2009)\citenamefont
  {Wasilewski}, \citenamefont {Fernholz}, \citenamefont {Jensen}, \citenamefont
  {Madsen}, \citenamefont {Krauter}, \citenamefont {Muschik},\ and\
  \citenamefont {Polzik}}]{Wasilewski2009}%
  \BibitemOpen
  \bibfield  {author} {\bibinfo {author} {\bibfnamefont {W.}~\bibnamefont
  {Wasilewski}}, \bibinfo {author} {\bibfnamefont {T.}~\bibnamefont
  {Fernholz}}, \bibinfo {author} {\bibfnamefont {K.}~\bibnamefont {Jensen}},
  \bibinfo {author} {\bibfnamefont {L.~S.}\ \bibnamefont {Madsen}}, \bibinfo
  {author} {\bibfnamefont {H.}~\bibnamefont {Krauter}}, \bibinfo {author}
  {\bibfnamefont {C.}~\bibnamefont {Muschik}}, \ and\ \bibinfo {author}
  {\bibfnamefont {E.~S.}\ \bibnamefont {Polzik}},\ }\href {\doibase
  10.1364/OE.17.014444} {\bibfield  {journal} {\bibinfo  {journal} {Opt.
  Express}\ }\textbf {\bibinfo {volume} {17}},\ \bibinfo {pages} {14444}
  (\bibinfo {year} {2009})}\BibitemShut {NoStop}%
\bibitem [{\citenamefont {Muschik}\ \emph {et~al.}(2011)\citenamefont
  {Muschik}, \citenamefont {Polzik},\ and\ \citenamefont
  {Cirac}}]{Muschik2011}%
  \BibitemOpen
  \bibfield  {author} {\bibinfo {author} {\bibfnamefont {C.~A.}\ \bibnamefont
  {Muschik}}, \bibinfo {author} {\bibfnamefont {E.~S.}\ \bibnamefont {Polzik}},
  \ and\ \bibinfo {author} {\bibfnamefont {J.~I.}\ \bibnamefont {Cirac}},\
  }\href {\doibase 10.1103/PhysRevA.83.052312} {\bibfield  {journal} {\bibinfo
  {journal} {Phys. Rev. A}\ }\textbf {\bibinfo {volume} {83}},\ \bibinfo
  {pages} {052312} (\bibinfo {year} {2011})}\BibitemShut {NoStop}%
\bibitem [{\citenamefont {Gardiner}\ and\ \citenamefont
  {Zoller}(2004)}]{Gardiner2004}%
  \BibitemOpen
  \bibfield  {author} {\bibinfo {author} {\bibfnamefont {C.}~\bibnamefont
  {Gardiner}}\ and\ \bibinfo {author} {\bibfnamefont {P.}~\bibnamefont
  {Zoller}},\ }\href
  {https://www.ebook.de/de/product/13905493/crispin_gardiner_peter_zoller_quantum_noise.html}
  {\emph {\bibinfo {title} {Quantum Noise}}},\ 0172-7389\ (\bibinfo
  {publisher} {Springer-Verlag Berlin Heidelberg},\ \bibinfo {year}
  {2004})\BibitemShut {NoStop}%
\bibitem [{\citenamefont {Aspelmeyer}\ \emph {et~al.}(2014)\citenamefont
  {Aspelmeyer}, \citenamefont {Kippenberg},\ and\ \citenamefont
  {Marquardt}}]{Aspelmeyer2014}%
  \BibitemOpen
  \bibfield  {author} {\bibinfo {author} {\bibfnamefont {M.}~\bibnamefont
  {Aspelmeyer}}, \bibinfo {author} {\bibfnamefont {T.~J.}\ \bibnamefont
  {Kippenberg}}, \ and\ \bibinfo {author} {\bibfnamefont {F.}~\bibnamefont
  {Marquardt}},\ }\href {\doibase 10.1103/RevModPhys.86.1391} {\bibfield
  {journal} {\bibinfo  {journal} {Rev. Mod. Phys.}\ }\textbf {\bibinfo {volume}
  {86}},\ \bibinfo {pages} {1391} (\bibinfo {year} {2014})}\BibitemShut
  {NoStop}%
\bibitem [{\citenamefont {M{\o}ller}\ \emph {et~al.}(2017)\citenamefont
  {M{\o}ller}, \citenamefont {Thomas}, \citenamefont {Vasilakis}, \citenamefont
  {Zeuthen}, \citenamefont {Tsaturyan}, \citenamefont {Balabas}, \citenamefont
  {Jensen}, \citenamefont {Schliesser}, \citenamefont {Hammerer},\ and\
  \citenamefont {Polzik}}]{Moller2017}%
  \BibitemOpen
  \bibfield  {author} {\bibinfo {author} {\bibfnamefont {C.~B.}\ \bibnamefont
  {M{\o}ller}}, \bibinfo {author} {\bibfnamefont {R.~A.}\ \bibnamefont
  {Thomas}}, \bibinfo {author} {\bibfnamefont {G.}~\bibnamefont {Vasilakis}},
  \bibinfo {author} {\bibfnamefont {E.}~\bibnamefont {Zeuthen}}, \bibinfo
  {author} {\bibfnamefont {Y.}~\bibnamefont {Tsaturyan}}, \bibinfo {author}
  {\bibfnamefont {M.}~\bibnamefont {Balabas}}, \bibinfo {author} {\bibfnamefont
  {K.}~\bibnamefont {Jensen}}, \bibinfo {author} {\bibfnamefont
  {A.}~\bibnamefont {Schliesser}}, \bibinfo {author} {\bibfnamefont
  {K.}~\bibnamefont {Hammerer}}, \ and\ \bibinfo {author} {\bibfnamefont
  {E.~S.}\ \bibnamefont {Polzik}},\ }\href
  {http://dx.doi.org/10.1038/nature22980} {\bibfield  {journal} {\bibinfo
  {journal} {Nature}\ }\textbf {\bibinfo {volume} {547}},\ \bibinfo {pages}
  {191} (\bibinfo {year} {2017})}\BibitemShut {NoStop}%
\bibitem [{\citenamefont {Wiseman}\ and\ \citenamefont
  {Milburn}(2010)}]{Wiseman2010}%
  \BibitemOpen
  \bibfield  {author} {\bibinfo {author} {\bibfnamefont {H.~M.}\ \bibnamefont
  {Wiseman}}\ and\ \bibinfo {author} {\bibfnamefont {G.~J.}\ \bibnamefont
  {Milburn}},\ }\href
  {https://www.ebook.de/de/product/8631634/howard_m_griffith_university_queensland_wiseman_gerard_j_professor_university_of_queensland_milburn_quantum_measurement_and_control.html}
  {\emph {\bibinfo {title} {Quantum Measurement and Control}}}\ (\bibinfo
  {publisher} {Cambridge University Press},\ \bibinfo {year}
  {2010})\BibitemShut {NoStop}%
\bibitem [{\citenamefont {Vasilyev}\ \emph {et~al.}(2013)\citenamefont
  {Vasilyev}, \citenamefont {Muschik},\ and\ \citenamefont
  {Hammerer}}]{Vasilyev2013}%
  \BibitemOpen
  \bibfield  {author} {\bibinfo {author} {\bibfnamefont {D.~V.}\ \bibnamefont
  {Vasilyev}}, \bibinfo {author} {\bibfnamefont {C.~A.}\ \bibnamefont
  {Muschik}}, \ and\ \bibinfo {author} {\bibfnamefont {K.}~\bibnamefont
  {Hammerer}},\ }\href {\doibase 10.1103/PhysRevA.87.053820} {\bibfield
  {journal} {\bibinfo  {journal} {Phys. Rev. A}\ }\textbf {\bibinfo {volume}
  {87}},\ \bibinfo {pages} {053820} (\bibinfo {year} {2013})}\BibitemShut
  {NoStop}%
\bibitem [{\citenamefont {Stannigel}\ \emph {et~al.}(2012)\citenamefont
  {Stannigel}, \citenamefont {Rabl},\ and\ \citenamefont
  {Zoller}}]{Stannigel2012}%
  \BibitemOpen
  \bibfield  {author} {\bibinfo {author} {\bibfnamefont {K.}~\bibnamefont
  {Stannigel}}, \bibinfo {author} {\bibfnamefont {P.}~\bibnamefont {Rabl}}, \
  and\ \bibinfo {author} {\bibfnamefont {P.}~\bibnamefont {Zoller}},\ }\href
  {http://stacks.iop.org/1367-2630/14/i=6/a=063014} {\bibfield  {journal}
  {\bibinfo  {journal} {New Journal of Physics}\ }\textbf {\bibinfo {volume}
  {14}},\ \bibinfo {pages} {063014} (\bibinfo {year} {2012})}\BibitemShut
  {NoStop}%
\bibitem [{\citenamefont {Belavkin}(1992)}]{Belavkin1992}%
  \BibitemOpen
  \bibfield  {author} {\bibinfo {author} {\bibfnamefont {V.~P.}\ \bibnamefont
  {Belavkin}},\ }\href {\doibase 10.1007/BF02097018} {\bibfield  {journal}
  {\bibinfo  {journal} {Communications in Mathematical Physics}\ }\textbf
  {\bibinfo {volume} {146}},\ \bibinfo {pages} {611} (\bibinfo {year}
  {1992})}\BibitemShut {NoStop}%
\bibitem [{\citenamefont {{\ifmmode \check{C}\else \v{C}\fi{}ernot\'{\i}k}}\
  \emph {et~al.}(2015)\citenamefont {{\ifmmode \check{C}\else
  \v{C}\fi{}ernot\'{\i}k}}, \citenamefont {Vasilyev},\ and\ \citenamefont
  {Hammerer}}]{Cernotik2015}%
  \BibitemOpen
  \bibfield  {author} {\bibinfo {author} {\bibfnamefont {O.}~\bibnamefont
  {{\ifmmode \check{C}\else \v{C}\fi{}ernot\'{\i}k}}}, \bibinfo {author}
  {\bibfnamefont {D.~V.}\ \bibnamefont {Vasilyev}}, \ and\ \bibinfo {author}
  {\bibfnamefont {K.}~\bibnamefont {Hammerer}},\ }\href {\doibase
  10.1103/PhysRevA.92.012124} {\bibfield  {journal} {\bibinfo  {journal} {Phys.
  Rev. A}\ }\textbf {\bibinfo {volume} {92}},\ \bibinfo {pages} {012124}
  (\bibinfo {year} {2015})}\BibitemShut {NoStop}%
\bibitem [{\citenamefont {Tsang}\ and\ \citenamefont
  {Caves}(2010)}]{Tsang2010}%
  \BibitemOpen
  \bibfield  {author} {\bibinfo {author} {\bibfnamefont {M.}~\bibnamefont
  {Tsang}}\ and\ \bibinfo {author} {\bibfnamefont {C.~M.}\ \bibnamefont
  {Caves}},\ }\href {\doibase 10.1103/PhysRevLett.105.123601} {\bibfield
  {journal} {\bibinfo  {journal} {Phys. Rev. Lett.}\ }\textbf {\bibinfo
  {volume} {105}},\ \bibinfo {pages} {123601} (\bibinfo {year}
  {2010})}\BibitemShut {NoStop}%
\bibitem [{\citenamefont {Wimmer}\ \emph {et~al.}(2014)\citenamefont {Wimmer},
  \citenamefont {Steinmeyer}, \citenamefont {Hammerer},\ and\ \citenamefont
  {Heurs}}]{Wimmer2014}%
  \BibitemOpen
  \bibfield  {author} {\bibinfo {author} {\bibfnamefont {M.~H.}\ \bibnamefont
  {Wimmer}}, \bibinfo {author} {\bibfnamefont {D.}~\bibnamefont {Steinmeyer}},
  \bibinfo {author} {\bibfnamefont {K.}~\bibnamefont {Hammerer}}, \ and\
  \bibinfo {author} {\bibfnamefont {M.}~\bibnamefont {Heurs}},\ }\href
  {\doibase 10.1103/PhysRevA.89.053836} {\bibfield  {journal} {\bibinfo
  {journal} {Phys. Rev. A}\ }\textbf {\bibinfo {volume} {89}},\ \bibinfo
  {pages} {053836} (\bibinfo {year} {2014})}\BibitemShut {NoStop}%
\bibitem [{\citenamefont {Bariani}\ \emph {et~al.}(2015)\citenamefont
  {Bariani}, \citenamefont {Seok}, \citenamefont {Singh}, \citenamefont
  {Vengalattore},\ and\ \citenamefont {Meystre}}]{Bariani2015}%
  \BibitemOpen
  \bibfield  {author} {\bibinfo {author} {\bibfnamefont {F.}~\bibnamefont
  {Bariani}}, \bibinfo {author} {\bibfnamefont {H.}~\bibnamefont {Seok}},
  \bibinfo {author} {\bibfnamefont {S.}~\bibnamefont {Singh}}, \bibinfo
  {author} {\bibfnamefont {M.}~\bibnamefont {Vengalattore}}, \ and\ \bibinfo
  {author} {\bibfnamefont {P.}~\bibnamefont {Meystre}},\ }\href {\doibase
  10.1103/PhysRevA.92.043817} {\bibfield  {journal} {\bibinfo  {journal} {Phys.
  Rev. A}\ }\textbf {\bibinfo {volume} {92}},\ \bibinfo {pages} {043817}
  (\bibinfo {year} {2015})}\BibitemShut {NoStop}%
\bibitem [{\citenamefont {Motazedifard}\ \emph {et~al.}(2016)\citenamefont
  {Motazedifard}, \citenamefont {Bemani}, \citenamefont {Naderi}, \citenamefont
  {Roknizadeh},\ and\ \citenamefont {Vitali}}]{Motazedifard2016}%
  \BibitemOpen
  \bibfield  {author} {\bibinfo {author} {\bibfnamefont {A.}~\bibnamefont
  {Motazedifard}}, \bibinfo {author} {\bibfnamefont {F.}~\bibnamefont
  {Bemani}}, \bibinfo {author} {\bibfnamefont {M.~H.}\ \bibnamefont {Naderi}},
  \bibinfo {author} {\bibfnamefont {R.}~\bibnamefont {Roknizadeh}}, \ and\
  \bibinfo {author} {\bibfnamefont {D.}~\bibnamefont {Vitali}},\ }\href
  {http://stacks.iop.org/1367-2630/18/i=7/a=073040} {\bibfield  {journal}
  {\bibinfo  {journal} {New Journal of Physics}\ }\textbf {\bibinfo {volume}
  {18}},\ \bibinfo {pages} {073040} (\bibinfo {year} {2016})}\BibitemShut
  {NoStop}%
\bibitem [{\citenamefont {Wasilewski}\ \emph {et~al.}(2010)\citenamefont
  {Wasilewski}, \citenamefont {Jensen}, \citenamefont {Krauter}, \citenamefont
  {Renema}, \citenamefont {Balabas},\ and\ \citenamefont
  {Polzik}}]{Wasilewski2010}%
  \BibitemOpen
  \bibfield  {author} {\bibinfo {author} {\bibfnamefont {W.}~\bibnamefont
  {Wasilewski}}, \bibinfo {author} {\bibfnamefont {K.}~\bibnamefont {Jensen}},
  \bibinfo {author} {\bibfnamefont {H.}~\bibnamefont {Krauter}}, \bibinfo
  {author} {\bibfnamefont {J.~J.}\ \bibnamefont {Renema}}, \bibinfo {author}
  {\bibfnamefont {M.~V.}\ \bibnamefont {Balabas}}, \ and\ \bibinfo {author}
  {\bibfnamefont {E.~S.}\ \bibnamefont {Polzik}},\ }\href {\doibase
  10.1103/PhysRevLett.104.133601} {\bibfield  {journal} {\bibinfo  {journal}
  {Phys. Rev. Lett.}\ }\textbf {\bibinfo {volume} {104}},\ \bibinfo {pages}
  {133601} (\bibinfo {year} {2010})}\BibitemShut {NoStop}%
\bibitem [{\citenamefont {Sewell}\ \emph {et~al.}(2012)\citenamefont {Sewell},
  \citenamefont {Koschorreck}, \citenamefont {Napolitano}, \citenamefont
  {Dubost}, \citenamefont {Behbood},\ and\ \citenamefont
  {Mitchell}}]{Sewell2012}%
  \BibitemOpen
  \bibfield  {author} {\bibinfo {author} {\bibfnamefont {R.~J.}\ \bibnamefont
  {Sewell}}, \bibinfo {author} {\bibfnamefont {M.}~\bibnamefont {Koschorreck}},
  \bibinfo {author} {\bibfnamefont {M.}~\bibnamefont {Napolitano}}, \bibinfo
  {author} {\bibfnamefont {B.}~\bibnamefont {Dubost}}, \bibinfo {author}
  {\bibfnamefont {N.}~\bibnamefont {Behbood}}, \ and\ \bibinfo {author}
  {\bibfnamefont {M.~W.}\ \bibnamefont {Mitchell}},\ }\href {\doibase
  10.1103/PhysRevLett.109.253605} {\bibfield  {journal} {\bibinfo  {journal}
  {Phys. Rev. Lett.}\ }\textbf {\bibinfo {volume} {109}},\ \bibinfo {pages}
  {253605} (\bibinfo {year} {2012})}\BibitemShut {NoStop}%
\bibitem [{\citenamefont {Clerk}\ \emph {et~al.}(2010)\citenamefont {Clerk},
  \citenamefont {Devoret}, \citenamefont {Girvin}, \citenamefont {Marquardt},\
  and\ \citenamefont {Schoelkopf}}]{Clerk2010}%
  \BibitemOpen
  \bibfield  {author} {\bibinfo {author} {\bibfnamefont {A.~A.}\ \bibnamefont
  {Clerk}}, \bibinfo {author} {\bibfnamefont {M.~H.}\ \bibnamefont {Devoret}},
  \bibinfo {author} {\bibfnamefont {S.~M.}\ \bibnamefont {Girvin}}, \bibinfo
  {author} {\bibfnamefont {F.}~\bibnamefont {Marquardt}}, \ and\ \bibinfo
  {author} {\bibfnamefont {R.~J.}\ \bibnamefont {Schoelkopf}},\ }\href
  {\doibase 10.1103/RevModPhys.82.1155} {\bibfield  {journal} {\bibinfo
  {journal} {Rev. Mod. Phys.}\ }\textbf {\bibinfo {volume} {82}},\ \bibinfo
  {pages} {1155} (\bibinfo {year} {2010})}\BibitemShut {NoStop}%
\bibitem [{\citenamefont {Turin}(1960)}]{TurinJune1960}%
  \BibitemOpen
  \bibfield  {author} {\bibinfo {author} {\bibfnamefont {G.}~\bibnamefont
  {Turin}},\ }\bibfield  {booktitle} {\emph {\bibinfo {booktitle} {IRE
  Transactions on Information Theory}},\ }\href@noop {} {\bibfield  {journal}
  {\bibinfo  {journal} {IRE Transactions on Information Theory}\ }\textbf
  {\bibinfo {volume} {6}},\ \bibinfo {pages} {311} (\bibinfo {year} {June
  1960})}\BibitemShut {NoStop}%
\end{thebibliography}%

\end{document}


\title{Supplemental Material for\\
``Unconditional steady-state entanglement in \\macroscopic hybrid systems by coherent noise cancellation''}

\author{Xinyao Huang}
\affiliation{State Key Laboratory of Mesoscopic Physics, School of Physics, Peking University, Collaborative Innovation Center of Quantum Matter, Beijing 100871, China}
\affiliation{Niels Bohr Institute, University of Copenhagen, DK-2100 Copenhagen,
Denmark}

\author{Emil Zeuthen}
\email[Corresponding author. E-mail: ]{zeuthen@nbi.ku.dk}
\affiliation{Niels Bohr Institute, University of Copenhagen, DK-2100 Copenhagen,
Denmark}

\author{Denis V.~Vasilyev}
\affiliation{Center for Quantum Physics, Faculty of Mathematics, Computer Science and Physics, University of Innsbruck, A-6020 Innsbruck, Austria}
\affiliation{Institute for Quantum Optics and Quantum Information of the Austrian Academy of Sciences, A-6020 Innsbruck, Austria}

\author{Qiongyi He}
\email[E-mail: ]{qiongyihe@pku.edu.cn}
\affiliation{State Key Laboratory of Mesoscopic Physics, School of Physics, Peking University, Collaborative Innovation Center of Quantum Matter, Beijing 100871, China}

\author{Klemens Hammerer}
\affiliation{Institute for Theoretical Physics and Institute for Gravitational Physics (Albert Einstein Institute), Leibniz Universit\"{a}t Hannover, Callinstra{\ss}e 38, 30167 Hannover, Germany}

\author{Eugene S.~Polzik}
\affiliation{Niels Bohr Institute, University of Copenhagen, DK-2100 Copenhagen,
Denmark}

\maketitle

\section{Heisenberg-Langevin equations for spin-optomechanical hybrid system}

In this section, we discuss how to realize and tune the two-mode quadratic interaction [Eq.~(2) in the main text] in the specific hybrid system composed of a Cesium ensemble and an optomechanical system. The equations of motion of the cascaded hybrid setup are derived and mapped to the generic Eqs.~(4,5) presented in the main text.

\subsection{Spin subsystem}\label{sec:spin-sys}
\begin{figure}[!h]
\def\svgwidth{0.3\columnwidth}
 \centering{
	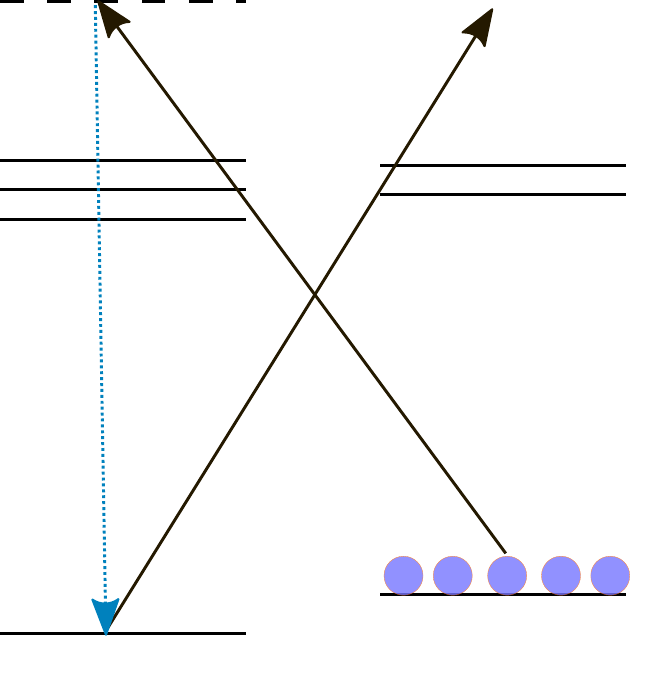
}
\caption{Level diagram of $^{133}$Cs ensemble polarized in the energetically maximal state (illustrated by purple circles), where the polarization points along the applied magnetic field, and the ground state levels are split by the Larmor frequency $\omega_{\text{S}}$ determined by the strength of the magnetic field. The atomic ensemble is driven by an off-resonant, linearly polarized laser beam (solid lines), which is far-detuned from the atomic resonance (detuning $\Delta_{\text{S}}$). The scattering sidebands (dotted arrows) induced by the drive light are described by field operators $\hat{b}_{\pm}$.}
\label{fig:atoms}
\end{figure}

We consider a free-space ensemble of $^{133}$Cs atoms driven by strong off-resonant laser light. As illustrated in Figure~\ref{fig:atoms}, for an appropriate collective spin degree of freedom of the ensemble, the required oscillator with negative mass can be achieved by initially pumping the atoms into the state $\left|F=4,m=4\right\rangle$, where an applied uniform magnetic field defines the quantization axis. The magnitude of the magnetic field can be used to tune the Larmor frequency $\omega_{\text{S}}$ of the collective spin precession~[Fig.~\ref{fig:atoms}]. Since energy must be extracted from this system to ``excite'' it from its initial state, it acts approximately as a bosonic oscillator with negative mass as can be seen from the Holstein-Primakoff transformation~\cite{Holstein1940, Hammerer2010}, provided that the system remains close to the fully polarized state. 
The negative mass property can conveniently be captured by letting the effective resonance frequency of the dimensionless bosonic spin operator be negative; 
this frequency is defined as $\Omega_{\text{S}} \equiv \text{sgn}(m_{\text{S}})\omega_{\text{S}}$ in the discussion below Eq.~\eqref{eq:H_0} in the main text.
We introduce the position-dependent atomic excitation mode in the frame rotating at the effective resonance frequency $\Omega_{\text{S}}$ as $\hat{a}_{\text{S}}(z,t)=\frac{1}{n}\sqrt{\frac{N_{\text{A}}}{L}}e^{i\Omega_{\text{S}}t}\sum_{i=1}^{n}\left|F=4,m=4\right\rangle _{i}\left\langle F=4,m=3\right|$ with $[\hat{a}_{\text{S}}(z,t),\hat{a}^{\dagger}_{\text{S}}(z',t)]=\delta(z-z')$, describing an annihilation of the atomic excitation in an atomic slice around a given $z$, where $n$ is the number of atoms per unit length, $N_{\text{A}}$ is the total number of atoms, and $L$ is the length of the ensemble. The effective interaction Hamiltonian describing the coupling of the collective spin to the light field in the rotating wave approximation can be derived as ($\hbar=1$)~\cite{Hammerer2010}
\begin{equation}
\hat{H}_{\text{S,int}}=-\frac{1}{\sqrt{L}}\int_{0}^{L}dz[g_{\text{SB}}\hat{b}_{-}(z,t)\hat{a}_{\text{S}}^{\dagger}(z,t)+g_{\text{SP}}\hat{b}_{+}(z,t)\hat{a}_{\text{S}}(z,t)+\text{H.c.}],
\label{eq:atoms}
\end{equation}
where $\hat{b}_{+}=(2\pi)^{-1/2}
e^{-i\Omega_{\text{S}} t}\int_{0}^{\infty}\hat{b}(\Omega)e^{-i\Omega t}d\Omega
$, $\hat{b}_{-}=(2\pi)^{-1/2}e^{i\Omega_{\text{S}} t}\int_{-\infty}^{0}\hat{b}(\Omega)e^{-i\Omega t}d\Omega
$ are the slowly varying annihilation operators of the sideband fields with $[\hat{b}_{\pm}(z,t), \hat{b}_{\pm}^{\dagger}(z,t')] = \delta(t-t')$. The coupling constants $g_{\text{SB/P}}$ are determined by evaluating the appropriate Clebsch-Gordan coefficients, which can be tuned by $\Delta_{\text{S}}$~\cite{Wasilewski2009,Muschik2011}. Without loss of generality, in the following we assume $g_{\text{SB/P}}$ to be real numbers.

As the light travels through the ensemble fast compared to the characteristic evolution time of the atomic state ($\tau \gg L/c$), 
we can introduce the spatially averaged atomic annihilation operator $\hat{a}_{\text{S}}(t)=\frac{1}{\sqrt{L}}\int_{0}^{L}dz\hat{a}_{\text{S}}(z,t)$ with $[\hat{a}_{\text{S}},\hat{a}^{\dagger}_{\text{S}}]=1$ 
and derive the equation of motion~\cite{Gardiner2004}
\begin{equation}
\frac{d}{d t}\hat{a}_{\text{S}}(t)=-\frac{\gamma_{\text{S}}}{2}\hat{a}_{\text{S}}+\sqrt{\gamma_{\text{S,0}}}\hat{a}_{\text{S,in}}+i(g_{\text{SB}}\hat{b}_{-,\text{in}}+g_{\text{SP}}\hat{b}_{+,\text{in}}^{\dagger}),\label{eq:aS-EOM}
\end{equation}
where $\hat{b}_{\pm,\text{in}}(t)=\hat{b}_{\pm}(0,t)$ with vacuum expectation value $\langle \hat{b}_{\pm,\text{in}}(t) \hat{b}_{\pm,\text{in}}^{\dagger}(t')\rangle=\delta(t-t')$. Eq.~\eqref{eq:aS-EOM} is seen to have the same form as in Eqs.~(\ref{eq:aM-dot},\ref{eq:fBA}) of the main text. The damping rate of the spin oscillator is $\gamma_{\text{S}}=\gamma_{\text{S,0}}+g_{\text{SB}}^{2}-g_{\text{SP}}^{2}$, where $\gamma_{\text{S,0}}=\gamma_{\text{S,i}}+\gamma_{\text{S,p}}$ is the damping rate in absence of dynamical broadening, composed of the intrinsic damping rate $\gamma_{\text{S,i}}$ due to decoherence processes like atomic collisions and imperfect optical pumping, and the power broadening $\gamma_{\text{S,p}}$ due to the drive-induced spontaneous emission of the atoms. These decoherence processes are accounted for by the thermal expectation value $\langle\hat{a}_{\text{S,in}}(t)\hat{a}^{\dagger}_{\text{S,in}}(t')\rangle=(\bar{n}_{\text{S}}+1)\delta(t-t')$ with $\bar{n}_{\text{S}}$ being the corresponding effective thermal occupation number. The total decoherence rate due to spontaneous emission and intrinsic damping can then be defined as $\tilde{\gamma}_{\text{S,0}}:=\gamma_{\text{S,0}}(\bar{n}_{\text{S}}+\frac{1}{2})=\gamma_{\text{S,i}}(\bar{n}_{\text{S,i}}+\frac{1}{2})+\frac{1}{2}\gamma_{\text{S,p}}$, where in the last expression we have decomposed it according to the contributions to $\gamma_{\text{S,0}}$. 

The corresponding input-output relations of the optical modes are given as
\begin{eqnarray}
&&\hat{b}_{+\text{S,out}}(t)=\hat{b}_{+,\text{in}}(t)+ig_{\text{SP}}\hat{a}_{\text{S}}^{\dagger}(t),
\nonumber \\
&&\hat{b}_{-\text{S,out}}(t)=\hat{b}_{-,\text{in}}(t)+ig_{\text{SB}}\hat{a}_{\text{S}}(t),\label{eq:S-IO}
\end{eqnarray}
where the output from the spin subsystem is $\hat{b}_{\pm,\text{S,out}}(t):=\hat{b}_{\pm}(L,t)$, which will serve as the input field to the optomechanical system below.

To make connection to the quantities used in the main text, we introduce the coupling rates of the atomic mode into the two sidebands $\Gamma_{\text{SB/P}}$ to reexpress the coupling constants as $g_{\text{SB/P}}=\sqrt{\Gamma_{\text{SB/P}}}$. The effective atomic quantum cooperativity is defined as $C_{\text{S}}:=\Gamma_{\text{S}}/\tilde{\gamma}_{\text{S,0}}$, which is the ratio of coupling rate $\Gamma_{\text{S}}:=\Gamma_{\text{SB}}+\Gamma_{\text{SP}}$ to the decoherence rate $\tilde{\gamma}_{\text{S,0}}$ for the spin system.

\subsection{Optomechanical subsystem}

\begin{figure}[!h]
\def\svgwidth{0.5\columnwidth}
\centering{
	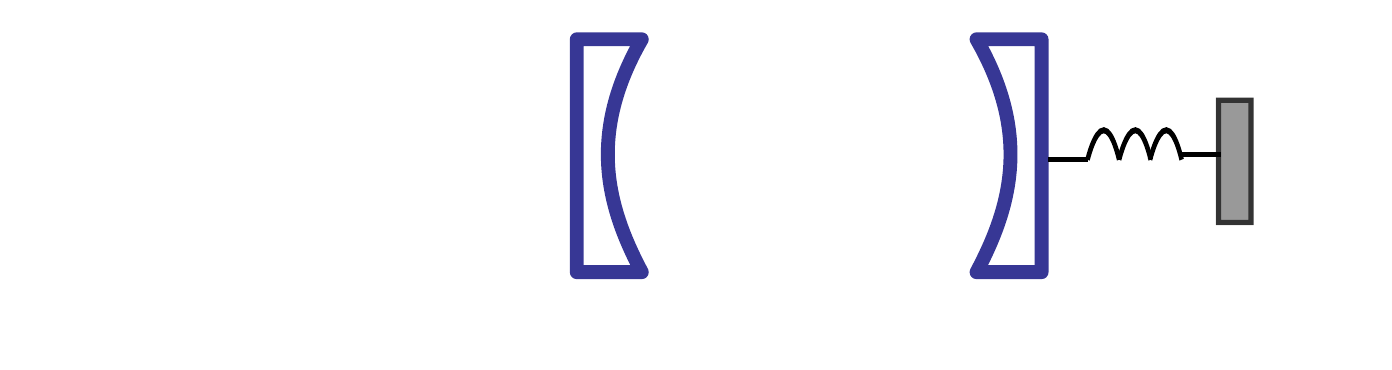
	\includegraphics[width=0.4\columnwidth]{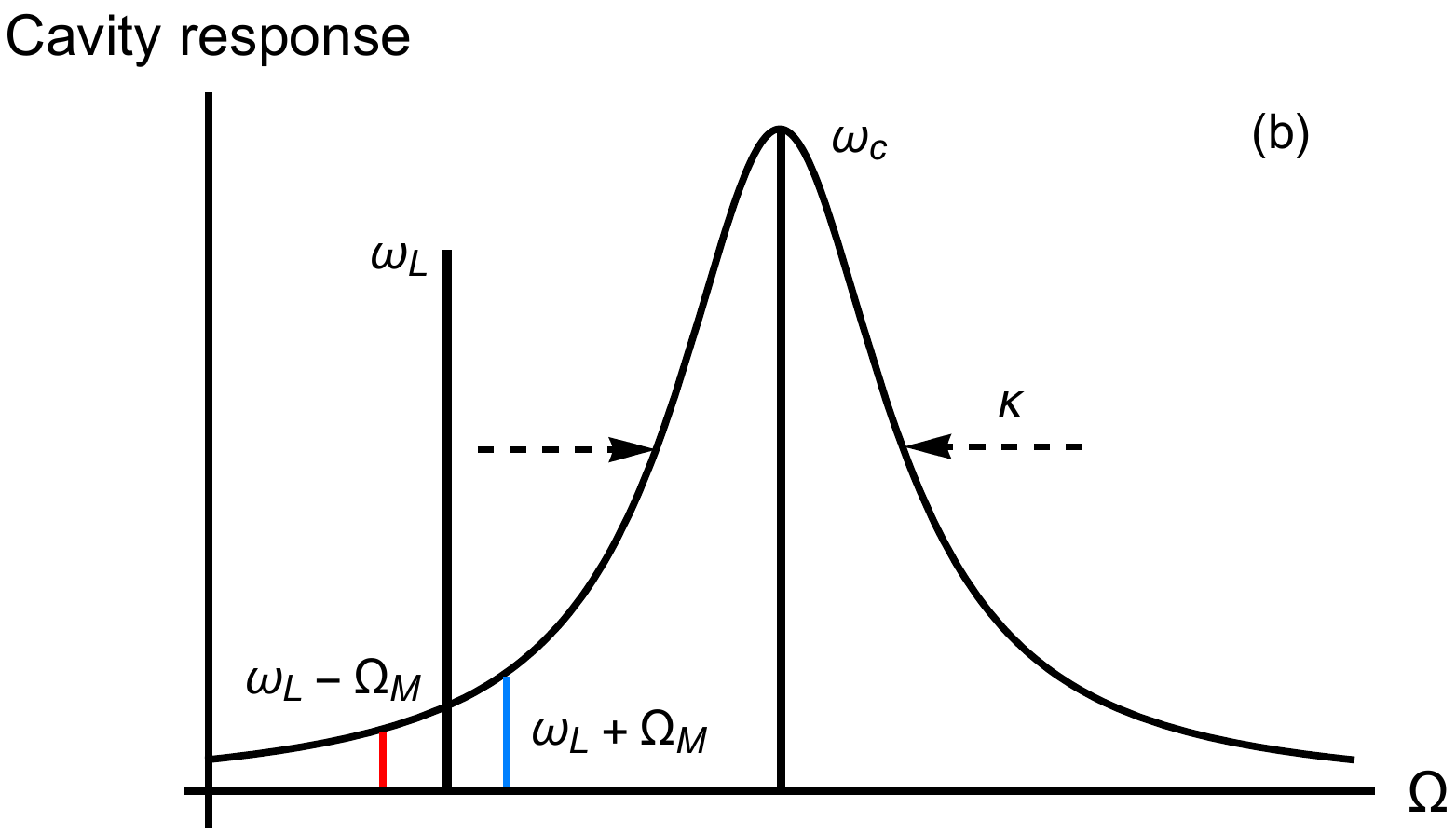}
}
\caption{(a) Standard optomechanical setup consisting of a single-sided Fabry-P\'{e}rot cavity (steady-state resonance frequency $\omega_{\text{c}}$) with one moving mirror (position coordinate $\hat{X}_{\text{M}}\propto (\hat{a}_{\text{Mo}}^{\dagger}+\hat{a}_{\text{Mo}})$). A classical drive laser of frequency $\omega_{\text{L}}$ (thick arrow) enhances the linear optomechanical interaction, resulting in the Stokes  ($\omega_{\text{L}}-\Omega_{\text{M}}$) and anti-Stokes ($\omega_{\text{L}}+\Omega_{\text{M}}$) scattering into sidebands depicted in (b) (red and blue lines, respectively). The relative sideband strengths are determined by the Lorentzian function with cavity decay rate $\kappa$. The upper/lower sidebands are mapped onto the traveling fields $\hat{b}_{\pm \text{M,in/out}}$ (thin arrows in (a)) that drive (in) and readout (out) the mechanical mode.}
\label{fig:optomechanics}
\end{figure}

Considering a driven optical cavity coupled by radiation pressure to a mechanical oscillator (see Fig.~\ref{fig:optomechanics}), the Hamiltonian in a frame rotating at the laser frequency $\omega_{\text{L}}$ is given by~\cite{Aspelmeyer2014}
\begin{eqnarray}
\hat{H}_{\text{OM}}=\Delta\hat{b}_{\text{c}}^{\dagger}\hat{b}_{\text{c}}+\Omega_{\text{M}}\hat{a}_{\text{Mo}}^{\dagger}\hat{a}_{\text{Mo}}-g_{0}\hat{b}_{\text{c}}^{\dagger}\hat{b}_{\text{c}}(\hat{a}_{\text{Mo}}^{\dagger}+\hat{a}_{\text{Mo}})+ \hat{H}_{\text{diss}},
\label{eq:om hamiltonian}
\end{eqnarray}
where $\Delta:=\omega_{\text{c}}-\omega_{\text{L}}$ is the drive detuning, $\omega_{\text{c}}/\Omega_{\text{M}}$ is the cavity/mechanical resonance frequency, $g_{0}$ is a suitable single-photon coupling constant, and $\hat{H}_\text{diss}$ accounts for the mechanical coupling to its thermal bath and the optical cavity coupling to external fields.
The bosonic annihilation operators for the optical cavity and mechanical modes in the lab frame are $\hat{b}_{\text{o}}$ and $\hat{a}_{\text{Mo}}$, respectively, whereas $\hat{b}_{\text{c}}=e^{i\omega_{\text{L}}t}\hat{b}_{\text{o}}$ is the slowly varying cavity operator relative to $\omega_{\text{L}}$. For a large, classical drive amplitude $\beta_{\text{in}}$, we may 
linearize Eq.~\eqref{eq:om hamiltonian} according to the standard procedure~\cite{Aspelmeyer2014}, permitting us to consider the fluctuations 
$\delta \hat{b}_{\text{c}}$ and $\delta \hat{a}_{\text{Mo}}$ around the resulting steady state  (the `$\delta$' will be dropped below for simplicity of notation, and static shifts of $\omega_{\text{c}}$, $\Delta$, and $\Omega_{\text{M}}$ are absorbed by suitable redefinition of these quantities). The effective equation of motion for the optical cavity mode after linearization has the solution in the frequency domain 
\begin{eqnarray}
\hat{b}_{\text{c}}(\Omega)=\frac{L(\Omega)}{\kappa/2}[ig_{\text{om}}(\hat{a}_{\text{Mo}}(\Omega)+\hat{a}_{\text{Mo}}^{\dagger}(-\Omega))+\sqrt{\kappa}\hat{b}_{\text{M,in}}(\Omega)],\label{eq:bc}
\end{eqnarray}
where $L(\Omega)=\frac{\kappa/2}{\kappa/2+i\left(\Delta-\Omega\right)}$ is the complex cavity Lorentzian, $\hat{b}_{\text{M,in}}(\Omega)$ is the input field, $\kappa$ is the decay rate (FWHM) of the optical cavity, and $g_{\text{om}}=\sqrt{\kappa}g_{0}\beta_{\text{in}}/\sqrt{(\kappa/2)^{2}+\Delta^{2}}$ is the drive-enhanced coupling constant (in the following, we consider $g_{\text{om}}$ to be a real number without loss of generality). 

Assuming $\gamma_{\text{M}}\ll \kappa$, we proceed by adiabatically eliminating the intracavity field~\eqref{eq:bc}. This amounts to approximating 
\begin{eqnarray}
L(\Omega)&\approx& L(\tilde{\Omega}_{\text{M}}) \;\text{for }\Omega>0,
\nonumber\\
L(\Omega)&\approx& L(-\tilde{\Omega}_{\text{M}}) \;\text{for }\Omega<0,
\end{eqnarray}
where $\tilde{\Omega}_{\text{M}}=2g_{\text{om}}^{2}\text{Im}[L(\tilde{\Omega}_{\text{M}})-L^{\ast}(-\tilde{\Omega}_{\text{M}})]/\kappa+\Omega_{\text{M}}$ is the dynamically shifted mechanical resonance frequency. By defining slowly varying operators for the intracavity and external optical fields
\begin{eqnarray}
\hat{b}_{\pm,\text{cav}}(t)&=&\frac{1}{\sqrt{2\pi}}\begin{cases}
e^{i\tilde{\Omega}_{\text{M}} t}\int_{0}^{\infty} \hat{b}_{\text{c}}(\Omega)e^{-i\Omega t}d\Omega\\
e^{-i\tilde{\Omega}_{\text{M}} t}\int_{-\infty}^{0} \hat{b}_{\text{c}}(\Omega)e^{-i\Omega t}d\Omega
\end{cases},
\nonumber\\
\hat{b}_{\pm \text{M,out/in}}(t)&=&\frac{1}{\sqrt{2\pi}}\begin{cases}
e^{i\tilde{\Omega}_{\text{M}} t}\int_{0}^{\infty}\hat{b}_{\text{M,out/in}}(\Omega)e^{-i\Omega t}d\Omega\\
e^{-i\tilde{\Omega}_{\text{M}} t}\int_{-\infty}^{0}\hat{b}_{\text{M,out/in}}(\Omega)e^{-i\Omega t}d\Omega
\end{cases},
\end{eqnarray}
and the mechanical mode $\hat{a}_{\text{M}}=e^{i\tilde{\Omega}_{\text{M}}t}\hat{a}_{\text{Mo}}$ with $[\hat{a}_{\text{M}},\hat{a}^{\dagger}_{\text{M}}]=1$, the mechanical equation of motion is~\cite{Gardiner2004,Aspelmeyer2014}
\begin{equation}
\frac{d}{d t}\hat{a}_{\text{M}}(t)=-\frac{\gamma_{\text{M}}}{2}\hat{a}_{\text{M}}+\sqrt{\gamma_{\text{M,0}}}\hat{a}_{\text{M,in}}+i\frac{2g_{\text{om}}}{\sqrt{\kappa}}[L(\tilde{\Omega}_{\text{M}})\hat{b}_{+\text{M,in}}+L^{\ast}(-\tilde{\Omega}_{\text{M}})\hat{b}_{-\text{M,in}}^{\dagger}],
\end{equation}
where $\gamma_{\text{M}}=\gamma_{\text{M,0}}+4g_{\text{om}}^{2}\text{Re}[L(\tilde{\Omega}_{\text{M}})-L^{\ast}(-\tilde{\Omega}_{\text{M}})]/\kappa$, composed of the intrinsic damping rate $\gamma_{\text{M,0}}$ and the optical broadening. We will use the thermal expectation value $\langle\hat{a}_{\text{M,in}}(t)\hat{a}^{\dagger}_{\text{M,in}}(t')\rangle=(\bar{n}_{\text{M}}+1)\delta(t-t')$ with $\bar{n}_{\text{M}}$ being the thermal occupation number of the environment, which corresponds to the thermal decoherence rate of the mechanical oscillator $\tilde{\gamma}_{\text{M,0}}:=\gamma_{\text{M,0}}(\bar{n}_{\text{M}}+1/2)\approx\gamma_{\text{M,0}}\bar{n}_{\text{M}}$ in the limit of large thermal occupation number $\bar{n}_{\text{M}} \gg1$. In addition, we have assumed a high-$Q$ mechanical oscillator, in the sense that $Q=\tilde{\Omega}_{\text{M}}/\gamma_{\text{M}}\gg1$, so that the response of $\hat{a}_{\text{M}}(\Omega)$ is confined to Fourier frequencies $\Omega\approx\tilde{\Omega}_{\text{M}}$. 

To make connection to the main text, we introduce sideband rates $\Gamma_{\text{MB/P}}$ which are related to the coupling constants as $g_{\text{MB/P}}=2g_{\text{om}}L(\pm \tilde{\Omega}_{\text{M}})/\sqrt{\kappa}=\sqrt{\Gamma_{\text{MB/P}}}e^{i\theta_{\pm}}$ and the phase of the upper/lower sideband $\theta_{\pm}=-\arctan(2(\Delta \mp \tilde{\Omega}_{\text{M}})/\kappa)$. We define the mechanical quantum cooperativity as $C_{\text{M}}:=\Gamma_{\text{M}}/\tilde{\gamma}_{\text{M,0}}$, where $\Gamma_{\text{M}}:=\Gamma_{\text{MB}}+\Gamma_{\text{MP}}$ is the coupling rate. 
As defined here, the quantum cooperativity $C_{\text{M}}$ contains Lorentzian penalty factors via $\Gamma_{\text{M}}$ accounting for the off-resonant readout from the optical cavity. This reflects the difference in readout rates resulting when changing the drive detuning while keeping the drive-induced intracavity population fixed. 

The effective input-output relations of the optomechanical system with the cavity mode adiabatically eliminated are given as
\begin{eqnarray}
&&\hat{b}_{+\text{M,out}}(t)=-\hat{b}_{+\text{M,in}}(t) e^{2i\theta_{+}}-ig_{\text{MB}}\hat{a}_{\text{M}}(t),
\nonumber \\
&&\hat{b}_{-\text{M,out}}(t)=-\hat{b}_{-\text{M,in}}(t) e^{2i\theta_{-}} -ig_{\text{MP}}\hat{a}_{\text{M}}^{\dagger}(t).\label{eq:M-IO}
\end{eqnarray}

\subsection{Cascaded hybrid setup}

Having established the theoretical description of the individual subsystems, we arrive at the equations governing the cascaded hybrid system by setting $\hat{b}_{\pm \text{M,in}}=e^{i\phi}\sqrt{1-\epsilon}\hat{b}_{\pm \text{S,out}}+\sqrt{\epsilon}\hat{b}_{\pm,\text{in}}'$. Here $\phi$ is a quadrature rotation phase factor between the two subsystems (as can be chosen experimentally by adjusting the relative phase between the spin and optomechanical drive fields~\cite{Moller2017}), and $\hat{b}_{\pm,\text{in}}'$ is the additional vacuum noise field due to the transmission loss $\epsilon$ between the two subsystems.
Specializing to the case of $\left|\Omega_{\text{S}}\right|=\tilde{\Omega}_{\text{M}}$, the equations of motion for the cascaded setup are given by 
\begin{eqnarray}
\frac{d}{d t}\hat{a}_{\text{S}}(t)&=&-\frac{\gamma_{\text{S}}}{2}\hat{a}_{\text{S}}+\sqrt{\gamma_{\text{S,0}}}\hat{a}_{\text{S,in}}+i(\sqrt{\Gamma_{\text{SB}}}\hat{b}_{-,\text{in}}+\sqrt{\Gamma_{\text{SP}}}\hat{b}_{+,\text{in}}^{\dagger}),
\nonumber\\
\frac{d}{d t}\hat{a}_{\text{M}}(t)&=&
-\frac{\gamma_{\text{M}}}{2}\hat{a}_{\text{M}}+\sqrt{\gamma_{\text{M,0}}}\hat{a}_{\text{M,in}}+\sqrt{1-\epsilon}R_{0}\hat{a}_{\text{S}}^{\dagger}+i\sqrt{1-\epsilon}(\sqrt{\Gamma_{\text{MB}}}e^{i(\theta_{+}+\phi)}\hat{b}_{+,\text{in}}+\sqrt{\Gamma_{\text{MP}}}e^{-i(\theta_{-}+\phi)}\hat{b}_{-,\text{in}}^{\dagger})\nonumber\\
&&+i\sqrt{\epsilon}(\sqrt{\Gamma_{\text{MB}}}e^{i\theta_{+}}\hat{b}_{+,\text{in}}'+\sqrt{\Gamma_{\text{MP}}}e^{-i\theta_{-}}\hat{b}_{-,\text{in}}'^{\dagger}),
\label{eq:Le}
\end{eqnarray}
where $R_{0}=\sqrt{\Gamma_{\text{SB}}\Gamma_{\text{MP}}}e^{-i(\theta_{-}+\phi)}-\sqrt{\Gamma_{\text{MB}}\Gamma_{\text{SP}}}e^{i(\theta_{+}+\phi)}$ is determined by the difference of the light field coupling with spin and mechanical oscillators. 

By numerical optimization of $\xi_{g}$ [Eq.~(8) in the main text], we find the optimal choice to be $\phi=-(\theta_{+}+\theta_{-})/2$ so that $R_{0}=\exp(i(\theta_{+}-\theta_{-})/2)(\sqrt{\Gamma_{\text{SB}}\Gamma_{\text{MP}}}-\sqrt{\Gamma_{\text{MB}}\Gamma_{\text{SP}}}):=\exp(i(\theta_{+}-\theta_{-})/2)R$, where $R$ is real with units of frequency. By defining $\hat{b}_{\pm,\text{in}}''\equiv \hat{b}_{\pm,\text{in}}'e^{i(\theta_{+}+\theta_{-})/2}$, the equations of motion can then be rewritten as
\begin{eqnarray}
\frac{d}{d t}\hat{a}_{\text{S}}(t) & = & -\frac{\gamma_{\text{S}}}{2}\hat{a}_{\text{S}}+\sqrt{\gamma_{\text{S,0}}}\hat{a}_{\text{S,in}} +i(\sqrt{\Gamma_{\text{SB}}}\hat{b}_{-,\text{in}}+\sqrt{\Gamma_{\text{SP}}}\hat{b}_{+,\text{in}}^{\dagger}),
\nonumber \\
\frac{d}{d t}\hat{a}_{\text{M}}'(t) & = &-\frac{\gamma_{\text{M}}}{2}\hat{a}_{\text{M}}'+\sqrt{\gamma_{\text{M,0}}}\hat{a}_{\text{M,in}}'+\sqrt{1-\epsilon}R\hat{a}_{\text{S}}^{\dagger}+ i\sqrt{1-\epsilon}(\sqrt{\Gamma_{\text{MB}}}\hat{b}_{+,\text{in}}+\sqrt{\Gamma_{\text{MP}}}\hat{b}_{-,\text{in}}^{\dagger})\nonumber\\
&&+i\sqrt{\epsilon}(\sqrt{\Gamma_{\text{MB}}}\hat{b}_{+,\text{in}}''+\sqrt{\Gamma_{\text{MP}}}\hat{b}_{-,\text{in}}''^{\dagger}),\label{eq:Le-nophase}
\end{eqnarray}
where an immaterial phase factor has been absorbed by introducing $\hat{a}_{\text{M/M,in}}'(t) := e^{-i(\theta_{+}-\theta_{-})/2}\hat{a}_{\text{M/M,in}}(t)$. Equations~\eqref{eq:Le-nophase} correspond to Eq.~(4,5) in the main text.  

In the present work, we focus on optimizing the entanglement $\xi_{g}$ of the particular EPR-type variables that are automatically mapped into the joint output of the hybrid system by the entangling dynamics (thereby fixing the parameter $g$ to be considered). To identify these EPR-type variables, we now 
derive the input/output relation in the frequency domain for the two-mode (homodyne) light quadrature operators
\begin{eqnarray}
\hat{X}_{\text{L,in/out}}(\Omega)&=&\frac{1}{\sqrt{2}}[\hat{b}_{\text{in/out}}(\Omega)+\hat{b}_{\text{in/out}}^{\dagger}(-\Omega)]=\frac{1}{\sqrt{2}}[\hat{X}_{\text{L,in/out}}^{\cos}(\Omega)+i\hat{X}_{\text{L,in/out}}^{\sin}(\Omega)],
\nonumber\\
\hat{P}_{\text{L,in/out}}(\Omega)&=&\frac{1}{\sqrt{2}i}[\hat{b}_{\text{in/out}}(\Omega)-\hat{b}_{\text{in/out}}^{\dagger}(-\Omega)]=\frac{1}{\sqrt{2}}[\hat{P}_{\text{L,in/out}}^{\cos}(\Omega)+i\hat{P}_{\text{L,in/out}}^{\sin}(\Omega)],
\label{eq:two-mode frequency}
\end{eqnarray}
where $[\hat{b}_{\text{in,out}}(\Omega),\hat{b}^{\dagger}_{\text{in,out}}(\Omega')]=\delta(\Omega-\Omega')$, $\hat{X}/\hat{P}_{\text{L,in/out}}^{\cos/\sin}$ are the cosine and sine components of the input/output light quadrature operators $\hat{X}/\hat{P}_{\text{L,in/out}}$, and $\hat{X}/\hat{P}_{\text{L,out}}(-\Omega)=\hat{X}/\hat{P}_{\text{L,out}}^{\dagger}(\Omega)$.
Using the Fourier transform convention
\begin{equation}
\hat{a}(t)=\frac{1}{\sqrt{2\pi}}\int_{-\infty}^{\infty}\hat{a}(\Omega)e^{-i\Omega t}d\Omega,\quad \hat{a}^{\dagger}(t)=\frac{1}{\sqrt{2\pi}}\int_{-\infty}^{\infty}\hat{a}^{\dagger}(\Omega)e^{i\Omega t}d\Omega,
\label{eq:FT}
\end{equation}
and defining $\hat{b}_{\pm,\text{out}} \equiv \hat{b}_{\pm \text{M,out}}e^{-i[2\theta_{\pm}-(\theta_{+}+\theta_{-})/2+\pi]}$, the phase-independent input/output relations in the frequency domain can be derived by plugging the Fourier transforms of Eqs.~(\ref{eq:S-IO},\ref{eq:M-IO}) into Eq.~\eqref{eq:two-mode frequency} and found to be  (in the lab frame)
\begin{eqnarray}
\hat{X}_{\text{L,out}}^{\cos}(\Omega)&=&\sqrt{1-\epsilon}\hat{X}_{\text{L,in}}^{\cos}(\Omega)+\sqrt{\epsilon}\hat{X}_{\text{in}}^{\cos'}(\Omega)-\frac{\sqrt{1-\epsilon}(\sqrt{\Gamma_{\text{SB}}}-\sqrt{\Gamma_{\text{SP}}})}{\sqrt{2}}\hat{P}_{\text{S}}(\Omega)-\frac{\sqrt{\Gamma_{\text{MB}}}-\sqrt{\Gamma_{\text{MP}}}}{\sqrt{2}}\hat{P}_{\text{M}}(\Omega),
\nonumber\\
\hat{X}_{\text{L,out}}^{\sin}(\Omega)&=&\sqrt{1-\epsilon}\hat{X}_{\text{L,in}}^{\sin}(\Omega)+\sqrt{\epsilon}\hat{X}_{\text{in}}^{\sin'}(\Omega)-\frac{\sqrt{1-\epsilon}(\sqrt{\Gamma_{\text{SB}}}-\sqrt{\Gamma_{\text{SP}}})}{\sqrt{2}}\hat{X}_{\text{S}}(\Omega)+\frac{\sqrt{\Gamma_{\text{MB}}}-\sqrt{\Gamma_{\text{MP}}}}{\sqrt{2}}\hat{X}_{\text{M}}(\Omega),
\nonumber\\
\hat{P}_{\text{L,out}}^{\cos}(\Omega)&=&\sqrt{1-\epsilon}\hat{P}_{\text{L,in}}^{\cos}(\Omega)+\sqrt{\epsilon}\hat{P}_{\text{in}}^{\cos'}(\Omega)+\frac{\sqrt{1-\epsilon}(\sqrt{\Gamma_{\text{SB}}}+\sqrt{\Gamma_{\text{SP}}})}{\sqrt{2}}\hat{X}_{\text{S}}(\Omega)+\frac{\sqrt{\Gamma_{\text{MB}}}+\sqrt{\Gamma_{\text{MP}}}}{\sqrt{2}}\hat{X}_{\text{M}}(\Omega),
\nonumber\\
\hat{P}_{\text{L,out}}^{\sin}(\Omega)&=&\sqrt{1-\epsilon}\hat{P}_{\text{L,in}}^{\sin}(\Omega)+\sqrt{\epsilon}\hat{P}_{\text{in}}^{\sin'}(\Omega)-\frac{\sqrt{1-\epsilon}(\sqrt{\Gamma_{\text{SB}}}+\sqrt{\Gamma_{\text{SP}}})}{\sqrt{2}}\hat{P}_{\text{S}}(\Omega)+\frac{\sqrt{\Gamma_{\text{MB}}}+\sqrt{\Gamma_{\text{MP}}}}{\sqrt{2}}\hat{P}_{\text{M}}(\Omega),
\label{eq:io relation t}
\end{eqnarray}
where $\Omega>0$, and 
$\hat{X}_{\text{S/M}}(\Omega)=(\hat{a}_{\text{S/M}}(\Omega)+\hat{a}^{\dagger}_{\text{S/M}}(\Omega))/\sqrt{2}$, $\hat{P}_{\text{S/M}}(\Omega)=(\hat{a}_{\text{S/M}}(\Omega)-\hat{a}^{\dagger}_{\text{S/M}}(\Omega))/(\sqrt{2}i)$ are the spin/mechanical quadrature operators in the lab frame. 
The joint input-output relations~\eqref{eq:io relation t} establish the EPR variables mapped into the output field, specifically that ($\hat{X}_{\text{S}}+g\hat{X}_{\text{M}}$, $-\hat{P}_{\text{S}}+g\hat{P}_{\text{M}}$) with $g=(\sqrt{\Gamma_{\text{MB}}}+\sqrt{\Gamma_{\text{MP}}})/[\sqrt{1-\epsilon}(\sqrt{\Gamma_{\text{SB}}}+\sqrt{\Gamma_{\text{SP}}})]$ are mapped to $\hat{P}_{\text{L,out}}^{\cos/\sin}$,  as stated in the main text.

\section{Stochastic Master Equation for conditional entanglement generation}

In this section, we introduce the standard mathematical theory necessary to describe a quantum system subject to continuous measurement, the Stochastic Master Equation (SME)~\cite{Wiseman2010}. In particular, this theory prescribes how the information gained by measurement should be used to update the (conditional) density matrix, which encodes the experimenter's knowledge of the state of the quantum system. We apply the SME to calculate the entanglement performance of dynamically stable conditional schemes. This serves as an important benchmark for our novel unconditional scheme; the results of this comparison are summarized in the main text, whereas additional details are given below in Subsection~\ref{sec:uncond-vs-cond}. Incidentally, as also discussed in Subsection~\ref{sec:uncond-vs-cond}, the conditional theory provides a shortcut to determining the scaling of our unconditional scheme.

The conditional state evolves according to the conditional stochastic master equation. A general form with $b$ decay channels and $c$ monitored channels ($c\leq b$) is given by~\cite{Vasilyev2013,Wiseman2010}
\begin{eqnarray}
\label{eq:general conditional equation}
d\hat{\rho} &=&-i\left[\hat{H},\hat{\rho}\right]dt+\sum\limits_{i=1}^{b}\mathcal{D}\left[\hat{J}_{i}\right]\hat{\rho} dt+\sum\limits_{i=1}^{c}\mathcal{H}\left[\hat{J}_{i}\right]\hat{\rho} dW_{i},
\\
I_i(t)dt &=& \langle \hat{J}_i + \hat{J}_i^\dag\rangle dt + dW_i,
\end{eqnarray}
where the Lindblad operator $\mathcal{D}[\hat{J}_{i}]\rho=\hat{J}_{i}\hat{\rho}\hat{J}_{i}^{\dagger}-\frac{1}{2}\hat{J}_{i}^{\dagger}\hat{J}_{i}\hat{\rho}-\frac{1}{2}\hat{\rho}\hat{J}_{i}^{\dagger}\hat{J}_{i}$ describes the system decoherence due to the coupling to the environment, and the operator $\mathcal{H}[\hat{J}_{i}]\hat{\rho}=(\hat{J}_{i}-\langle\hat{J}_{i}\rangle)\hat{\rho}+\hat{\rho}(\hat{J}_{i}^{\dagger}-\langle \hat{J}_{i}^{\dagger}\rangle)$ updates the density matrix conditioned on the observation of the homodyne photocurrent $I_i(t)$. The measurement terms proportional to $dW_{i}$, a Wiener increment of zero mean and $dW_{i}dW_{j}=\delta_{ij}dt$.

Considering our hybrid setup and subjecting the outgoing field to homodyne detection, the corresponding conditional master equation is given by~\cite{Stannigel2012}
\begin{eqnarray}
d\hat{\rho}&=&[\gamma_{\text{S,0}}(\bar{n}_{\text{S}}+1)]\mathcal{D}[\hat{a}_{\text{S}}]\hat{\rho} dt+[\gamma_{\text{S,0}}\bar{n}_{\text{S}}]\mathcal{D}[\hat{a}_{\text{S}}^{\dagger}]\hat{\rho} dt+[\gamma_{\text{M,0}}(\bar{n}_{\text{M}}+1)]\mathcal{D}[\hat{a}_{\text{M}}]\hat{\rho} dt+[\gamma_{\text{M,0}}\bar{n}_{\text{M}}]\mathcal{D}[\hat{a}_{\text{M}}^{\dagger}]\hat{\rho} dt
\nonumber\\
&&-\frac{\sqrt{1-\epsilon}R}{2}[\hat{a}_{\text{S}}\hat{a}_{\text{M}}-\hat{a}_{\text{S}}^{\dagger}\hat{a}_{\text{M}}^{\dagger},\hat{\rho}]dt+\mathcal{D}[\sqrt{\epsilon\Gamma_{\text{SB}}}\hat{a}_{\text{S}}]\hat{\rho}dt+\mathcal{D}[\sqrt{\epsilon\Gamma_{\text{SP}}}\hat{a}_{\text{S}}^{\dagger}]\hat{\rho} dt+\mathcal{D}[\hat{s}_{+}]\hat{\rho} dt+\mathcal{D}[\hat{s}_{-}]\hat{\rho} dt
\nonumber\\
&&+\mathcal{H}\left[e^{i\psi}\left(\hat{s}_{+}+\hat{s}_{-}\right)/\sqrt{2}\right]\hat{\rho} dW_{c}
+\mathcal{H}\left[-ie^{i\psi}\left(\hat{s}_{+}-\hat{s}_{-}\right)/\sqrt{2}\right]\hat{\rho} dW_{s},
\label{eq:conditional eq}
\end{eqnarray}
where $dW_{c}$ and $dW_{s}$ correspond to the cosine and sine components of the Wiener increments, the jump operators are defined as 
\begin{eqnarray}
\hat{s}_{+}=\sqrt{(1-\epsilon)\Gamma_{\text{SP}}}\hat{a}_{\text{S}}^{\dagger}+\sqrt{\Gamma_{\text{MB}}}\hat{a}_{\text{M}}
\nonumber\\
\hat{s}_{-}=\sqrt{(1-\epsilon)\Gamma_{\text{SB}}}\hat{a}_{\text{S}}+\sqrt{\Gamma_{\text{MP}}}\hat{a}_{\text{M}}^{\dagger},
\end{eqnarray}
and we choose the optimal detection phase $\psi=0$ for further calculation.

The conditional dynamics described by Eq.~\eqref{eq:conditional eq} in the long run reduces to Gaussian dynamics~\cite{Belavkin1992}, described by the (deterministic) Riccati equation for the second moments of the operators, and a stochastic equation for the first moments. These equations can be established by recognizing that Eq.~\eqref{eq:conditional eq} has the form Eq.~\eqref{eq:general conditional equation} for suitable $\hat{H}$,$\hat{J}_i$ and applying the procedure given in Ref.~\cite{Cernotik2015}, permitting us to efficiently solve for the steady state.

According to the global numerical minimization of the corresponding Riccati equation (in the dynamically stable regime), the optimal choice is $\theta_{\text{S}} \approx \pi/4$. The minimum of the EPR variance can be achieved for arbitrary $\theta_{\text{M}}$ by choosing the corresponding optimal $C_{\text{M}}$ for fixed $\theta_{\text{S}}$ and $C_{\text{S}}$. In the following, we consider the analytical solution for $\theta_{\text{S/M}}=\pi/4$ and determine the optimal $C_{\text{M}}$ and the resulting minimized EPR variance.

For $\theta_{\text{S}}=\theta_{\text{M}}=\pi/4$, the Riccati equation reads
\begin{eqnarray}
\frac{d}{dt}\Delta^{2}\hat{X}_{\text{S}}&=&-\frac{\gamma_{\text{S,0}}}{2}\Delta^{2}\hat{X}_{\text{S}}+\frac{\Gamma_{\text{S}}}{2}+\tilde{\gamma}_{\text{S,0}}-(2\sqrt{(1-\epsilon)\Gamma_{\text{S}}}\Delta^{2}\hat{X}_{\text{S}}+\sqrt{\Gamma_{\text{M}}}\langle \hat{X}_{\text{S}},\hat{X}_{\text{M}}\rangle)^{2},
\nonumber\\
\frac{d}{dt}\Delta^{2}\hat{X}_{\text{M}}&=&-\frac{\gamma_{\text{M,0}}}{2}\Delta^{2}\hat{X}_{\text{M}}+\frac{\Gamma_{\text{M}}}{2}+\tilde{\gamma}_{\text{M,0}}-(\sqrt{(1-\epsilon)\Gamma_{\text{S}}}\langle \hat{X}_{\text{S}},\hat{X}_{\text{M}}\rangle+2\sqrt{\Gamma_{\text{M}}}\Delta^{2}\hat{X}_{\text{M}})^{2},
\nonumber\\
\frac{d}{dt}\langle \hat{X}_{\text{S}},\hat{X}_{\text{M}}\rangle&=&-(2\sqrt{(1-\epsilon)\Gamma_{\text{S}}}\Delta^{2}\hat{X}_{\text{S}}+\sqrt{\Gamma_{\text{M}}}\langle \hat{X}_{\text{S}},\hat{X}_{\text{M}}\rangle)(\sqrt{(1-\epsilon)\Gamma_{\text{S}}}\langle \hat{X}_{\text{S}},\hat{X}_{\text{M}}\rangle+2\sqrt{\Gamma_{\text{M}}}\Delta^{2}\hat{X}_{\text{M}})
\nonumber\\
&&-\frac{\gamma_{\text{S,0}}+\gamma_{\text{M,0}}}{2}\langle \hat{X}_{\text{S}},\hat{X}_{\text{M}}\rangle-\sqrt{(1-\epsilon)\Gamma_{\text{S}}\Gamma_{\text{M}}}.
\label{eq: conditional eq}
\end{eqnarray}
In the hot motional bath limit ($\bar{n}_{\text{M}} \gg 1$), we have $-\frac{\gamma_{\text{M,0}}}{2}\Delta^{2}\hat{X}_{\text{M}}+\tilde{\gamma}_{\text{M,0}} \approx \tilde{\gamma}_{\text{M,0}}$, and the analytical solutions of this equation for the steady-state are
\begin{eqnarray}
\Delta^{2}\hat{X}_{\text{S}}&=&-\sqrt{C_{\text{M}}/(C_{\text{M}}+2)}/2-[(\sqrt{C_{\text{M}}r}+1/((2n_{\text{S}}+1)\sqrt{(C_{\text{M}}+2)r}))/(2\sqrt{C_{\text{S}}(1-\epsilon)})] \langle \hat{X}_{\text{S}},\hat{X}_{\text{M}}\rangle,
\nonumber\\
\Delta^{2}\hat{X}_{\text{M}}&=&(n_{\text{S}}+1/2)\sqrt{(C_{\text{M}}+2) \cdot (\sqrt{C_{\text{M}}(C_{\text{M}}+2)}r+1/(2n_{\text{S}}+1))^{2}/C_{\text{M}}+(1-\epsilon)C_{\text{S}}(C_{\text{S}}-2C_{\text{M}}r+2)}
\nonumber\\
&&-(n_{\text{S}}+1/2)[C_{\text{M}}r+2r-(1-\epsilon)C_{\text{S}}],
\nonumber\\
\langle \hat{X}_{\text{S}},\hat{X}_{\text{M}}\rangle&=&\sqrt{C_{\text{M}}r}(\sqrt{(C_{\text{M}}+2)/C_{\text{M}}}-2\Delta^{2}\hat{X}_{\text{M}})/\sqrt{{C_{\text{S}}(1-\epsilon)}},
\label{eq: conditional qnd sol}
\end{eqnarray}
where $r=\tilde{\gamma}_{\text{M,0}}/\tilde{\gamma}_{\text{S,0}}$ is the ratio of decoherence rates. Equation~\eqref{eq: conditional qnd sol} will be used to analyze the asymptotic behavior of the conditional case in the large spin cooperativity limit ($C_{\text{S}} \gg 1$) below.

\section{Steady-state entanglement optimization in spin-optomechanical system}\label{sec:ent-opt}

In this section, we compare the steady-state entanglement performances of unconditional and conditional schemes by global numerical optimization of the imbalanced EPR variance $\xi_{g}$. Furthermore, the asymptotic scaling functions of the optimized $\xi_g$ are derived in the limit of large spin cooperativity.

\begin{figure}[!h]
\centering{
\includegraphics[width=0.4\columnwidth]{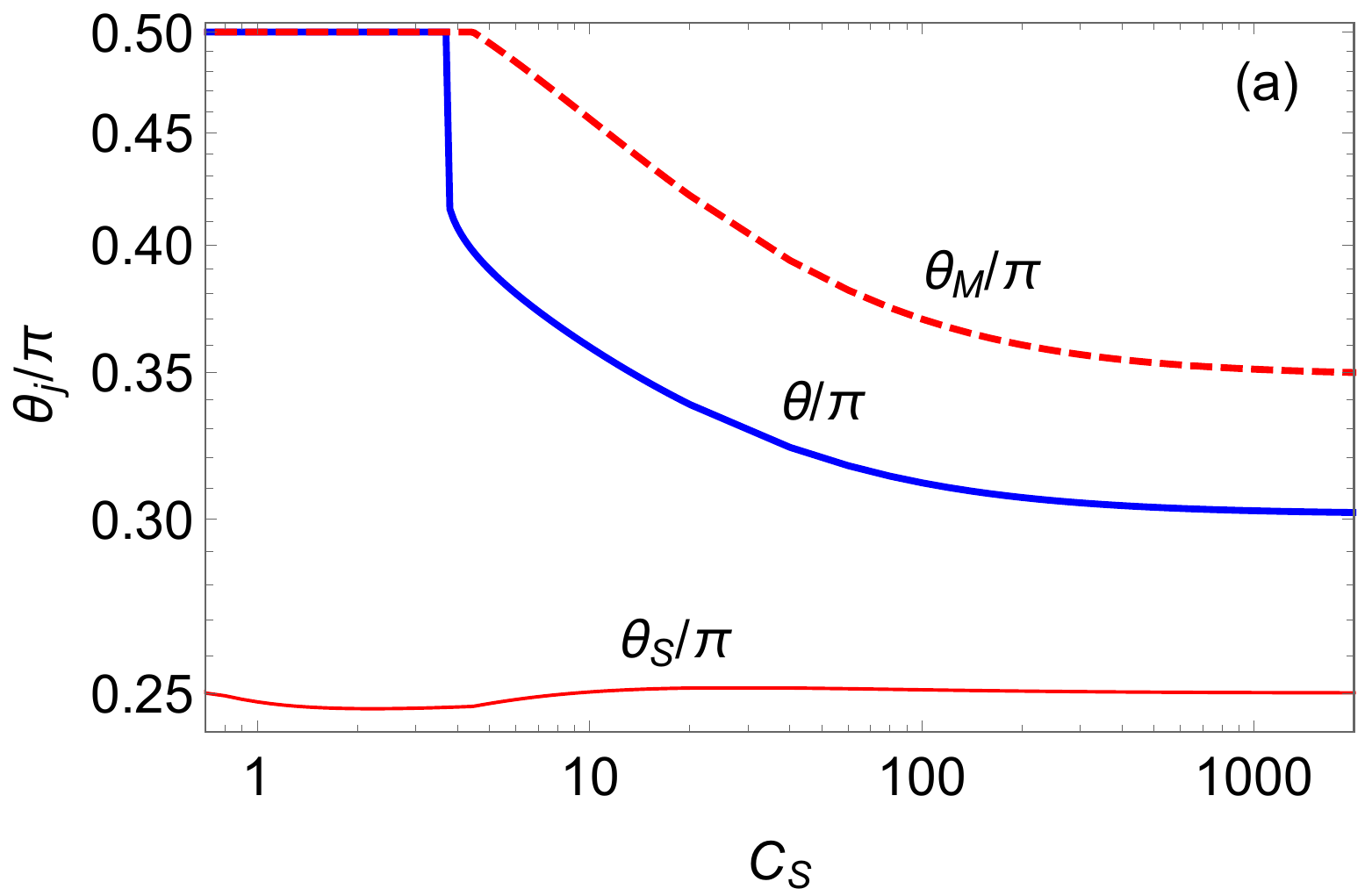}
\includegraphics[width=0.4\columnwidth]{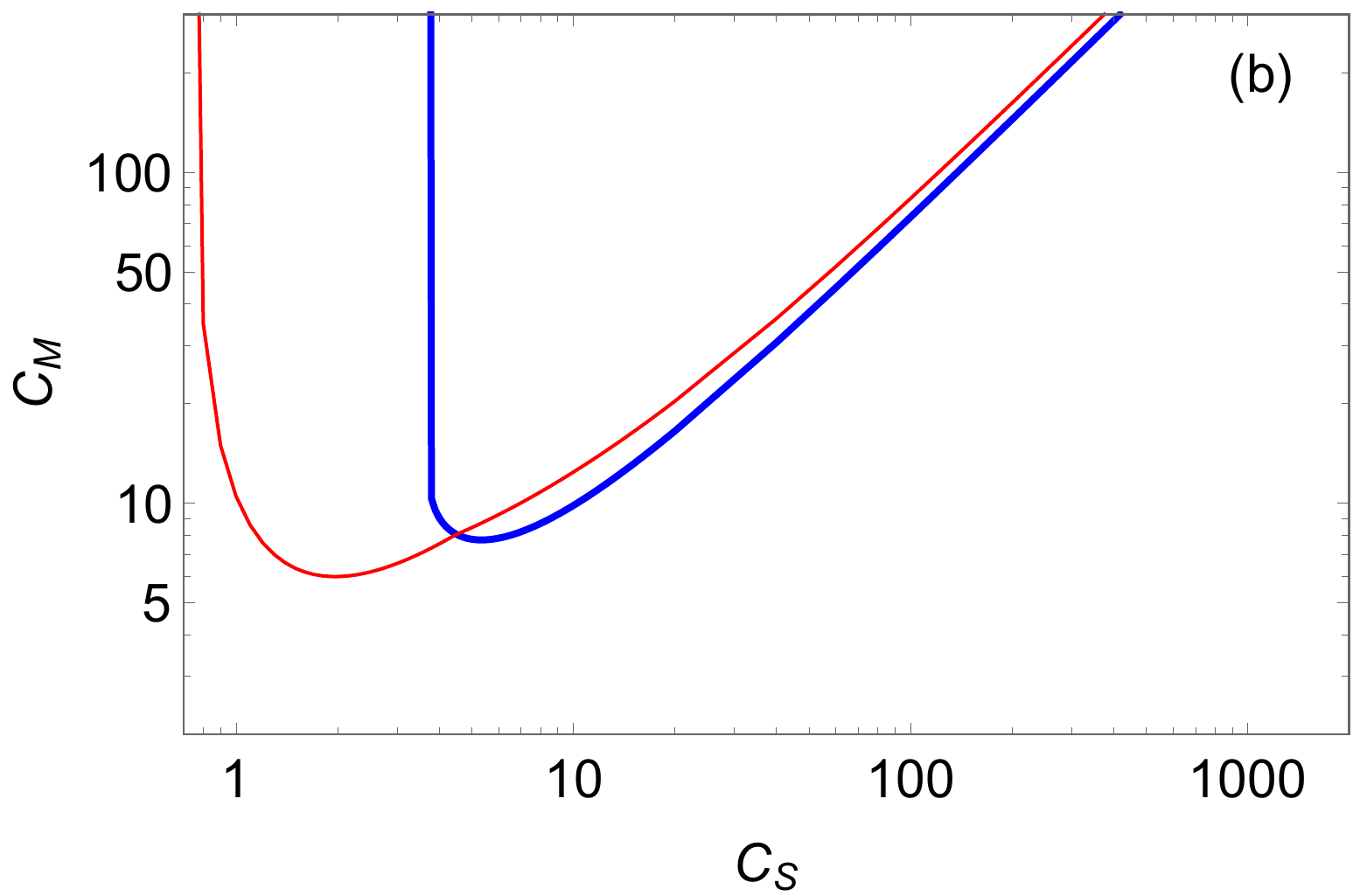}
}
\caption{Optimal (a) coupling angles $\theta_j$ and (b) mechanical cooperativity $C_{\text{M}}$ required to achieve the minimal EPR variance $\xi_{g}$ for the cases of symmetric ($\theta:=\theta_{\text{S}}=\theta_{\text{M}}$, blue) and asymmetric (red) coupling. The follow fixed parameters are assumed: transmission loss $\epsilon=0.1$, intrinsic linewidths $\gamma_{\text{S,0}}=2\pi \times 5\text{KHz}$, $\gamma_{\text{M,0}}=2\pi \times0.1\text{Hz}$, and thermal occupancies $\bar{n}_{\text{S}}=1$, $\bar{n}_{\text{M}}=10^{5}$.}
\label{fig:optimal theta and cm}
\end{figure} 

We now look at the details behind Fig.~\ref{fig:opt_ent} in the main text. Corresponding to the minimized $\xi_{g}$ for unconditional entanglement generation, as illustrated in Fig.~\ref{fig:optimal theta and cm}(a), the optimal interaction when allowing asymmetric coupling ($\theta_{\text{S}} \neq \theta_{\text{M}}$) shows $\theta_{\text{S}} \approx \pi/4$ in the regime of substantial entanglement (for smaller values of $C_\text{S}$, dynamical anti-broadening of the spin mode by tuning $\theta_{\text{S}}<\pi/4$ is seen to be optimal). When comparing with the symmetric coupling ($\theta_{\text{S}}=\theta_{\text{M}}=\theta$), i.e., the two oscillators are driven only by the common optical vacuum bath ($R=0$), $\theta_{\text{M}}> \theta$ shows that the optimal asymmetric coupling requires more beam-splitter interaction for the mechanical subsystem. In addition, optimal $C_\text{M}$ is larger for the asymmetric coupling when considering the values of $C_\text{S}$ permitting entanglement for both $R=0$ and $R\neq0$ as shown in Fig.~\ref{fig:optimal theta and cm}(b).

\subsection{Asymptotic scaling for symmetric unconditional entanglement}

For quantitative comparison between the schemes below, we now derive the asymptotic scaling function in the limit of large $C_{\text{S}}$. 
For the unconditional case of symmetric coupling ($R=0$), the expression for $\xi_{g}$ is given by plugging Eq.~(7) into Eq.~(8) in the main text; for $\theta \neq \pi/4$ and in the limit of large $C_{\text{S}}$ (i.e, $\gamma_{\text{S/M}} \gg \gamma_{\text{S/M,0}}$) we find,
\begin{equation}
\xi_{g} \approx -\frac{1}{\cos2\theta}\left[1+\frac{2(1-\epsilon+r)}{C_{\text{S}}(1-\epsilon)+C_{\text{M}}r}\right]+\tan2\theta \frac{4C_{\text{M}}C_{\text{S}}r(1-\epsilon)}{(C_{\text{M}}r+C_{\text{S}}(1-\epsilon))(C_{\text{M}}r+C_{\text{S}})},
\label{eq:ents_analytical}
\end{equation}
for given transmission loss $\epsilon$ and decoherence rates $\tilde{\gamma}_{j,0}$ ($j\in\{\text{S,M}\}$).
Minimizing Eq.~\eqref{eq:ents_analytical} and Taylor expanding in the limit of large $C_{\text{S}}$, we find the optimal mechanical cooperativity $C_{\text{M,opt}} \approx \sqrt{1-\epsilon}C_{\text{S}}/r$, and the optimal coupling angle
\begin{equation}
\sin 2\theta_{\text{opt}} \approx \frac{4(1-\epsilon)}{(\sqrt{1-\epsilon}+1)^{2}}\left(1-\frac{2(\sqrt{1-\epsilon}+r)}{\sqrt{1-\epsilon}+1} \frac{1}{C_{\text{S}}}\right).
\label{eq:optimal parameters}
\end{equation}
Plugging the optimized $C_{\text{M,opt}}$ and $\theta_{\text{opt}}$ back into Eq.~\eqref{eq:ents_analytical}, the leading contribution in the limit of large $C_{\text{S}}$ is
\begin{eqnarray}
\xi_{g}  \approx \sqrt{2(1+r)/C_{\text{S}}},
\label{eq:sym scaling}
\end{eqnarray}
when $\epsilon=0$. In the presence of transmission loss $\epsilon \neq 0$, the entanglement is lower-bounded by $\xi_{g}  \geq \sqrt{\epsilon}(1+\frac{\epsilon}{16}+...)$.

\subsection{Comparison between asymmetric unconditional and conditional schemes}\label{sec:uncond-vs-cond}

\begin{figure}[!h]
\centering{
\includegraphics[width=0.4\columnwidth]{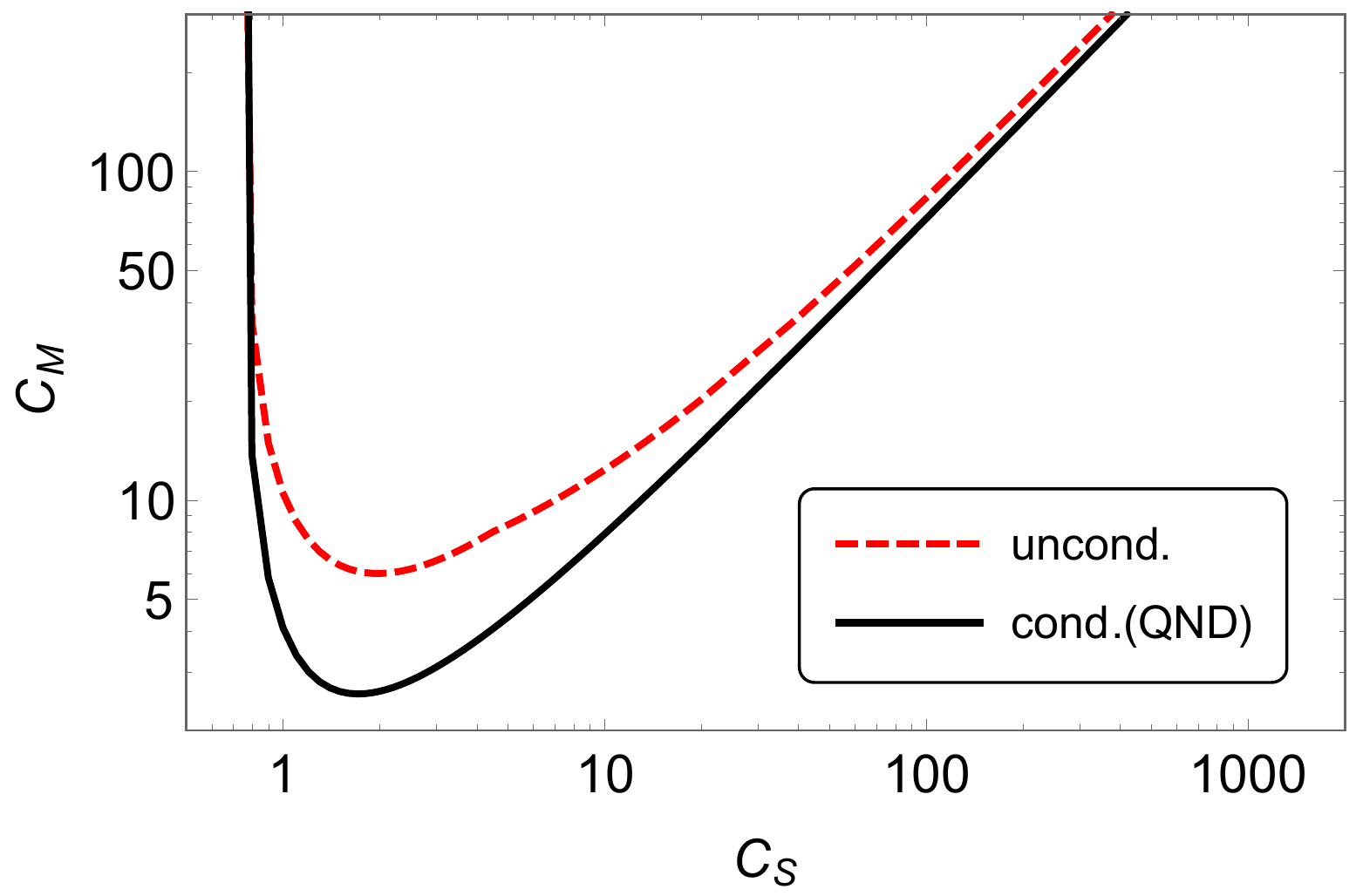}
}
\caption{
The optimal mechanical cooperativity $C_{\text{M}}$ for conditional QND and unconditional cases. The fixed parameters: transmission loss $\epsilon=0.1$, intrinsic linewidths $\gamma_{\text{S,0}}=2\pi \times 5\text{KHz}$, $\gamma_{\text{M,0}}=2\pi \times0.1\text{Hz}$, and thermal occupancies $\bar{n}_{\text{S}}=1$, $\bar{n}_{\text{M}}=10^{5}$.}
\label{fig:conditional ent and x}
\end{figure}

Next we compare the entanglement performance of the asymmetric unconditional scheme with that of the conditional scheme. 
The difference between the optimized unconditional entanglement and the conditional entanglement achievable by adding a continuous measurement of the output field while keeping all parameters the same is within a few percent~[inset of Fig.~\ref{fig:opt_ent} in the main text, red dashed curve]. 
We also consider the optimized conditional steady-state entanglement using QND interaction ($\theta_{\text{M}}=\theta_{\text{M}}=\pi/4$) and optimized 
mechanical cooperativity $C_{\text{M}}$ by means of the analytical solution shown in Eq.~\eqref{eq: conditional qnd sol}. 
Comparing again to the (separately) optimized unconditional scheme, we again find that they match within a few percent [inset of Fig.~\ref{fig:opt_ent} in the main text, red solid curve], while the required optimal $C_{\text{M}}$ in the unconditional case is larger than that of the conditional QND case~[Fig.~\ref{fig:conditional ent and x}]. 
In addition, the relative entanglement improvement of the conditional scheme over the unconditional scheme (referenced to the latter) and the difference between the required optimal $C_{\text{M}}$ for the two schemes approach zero as $C_{\text{S}}$ increases.

The asymptotic behavior of the optimized conditional entanglement (QND interaction) can be derived in the large spin cooperativity ($C_{\text{S}}$) limit. Substituting Eq.~\eqref{eq: conditional qnd sol} into Eq.~(8) in the main text, we minimize $\xi_g$ and perform a Taylor expansion for large $C_{\text{S}}$. The optimal mechanical cooperativity is then given by $C_{\text{M,opt}} \approx \sqrt{1-\epsilon}C_{\text{S}}/r$. By plugging $C_{\text{M,opt}}$ into Eq.~\eqref{eq: conditional qnd sol} and Eq.~(8) in the main text, the corresponding minimized conditional EPR variance can be expanded as 
\begin{eqnarray}
\xi_g \approx \sqrt{\frac{1+r+1/(2\bar{n}_{\text{S}}+1)}{2C_{\text{S}}}},
\end{eqnarray}
when $\epsilon=0$. In the presence of transmission loss, a lower bound is given by $\xi_{g}\geq\sqrt{\epsilon/(4-3\epsilon)}$ in the limit of small $\epsilon$. As the optimized performances of the conditional and asymmetric unconditional schemes in the large $C_{\text{S}}$ limit are essentially the same~[inset of Fig.~\ref{fig:opt_ent} in the main text, red curves],  the asymptotic behavior of the conditional case can also be applied to describe the unconditional case, and corresponds to the result stated in the main text.

\section{Link between unconditional entanglement and force sensitivity in hypothesis testing}

As stated in the final paragraph of the main text, 
we here present evidence that unconditional steady-state entanglement is linked to applying the hybrid system as a continuous  CQNC force sensor, e.g., for a mechanical force acting on the motional degree of freedom~\cite{Tsang2010,Wimmer2014,Bariani2015,Motazedifard2016} or a magnetic field acting on the spin system~\cite{Wasilewski2010,Sewell2012}. To illustrate the relation, we consider, for specificity, mechanical force sensing in the context of hypothesis testing, i.e., the task of determining whether a force of a prescribed waveform $F(t)=A_{0}f(t)=A_{0}\mathcal{F}^{-1}\{f(\Omega)\}$ is present or not on the mechanical oscillator (here $\mathcal{F}$ denotes the Fourier transform, and $A_{0}$ is the amplitude of the square-normalized waveform $f(t)$). The output field in the presence of the external force $F(t)$ can be decomposed as $\hat{P}_{\text{L,out}}(t)=\hat{f}_{\text{add}}(t)+A_{0}S(t)$, where $\hat{f}_{\text{add}}(t)$ contains the various contributions to the measurement noise of the hybrid sensor, and $S(t)=\sqrt{\gamma_{\text{M,0}}\Gamma}\chi_{\text{M}}(t)\ast f(t)$ is the signal of the normalized force $f(t)$ with $\Gamma$ and $\gamma_{\text{M,0}}$ being the corresponding readout and intrinsic mechanical damping rates, and $\chi_{\text{M}}(t)$ being the inverse Fourier transform of the mechanical susceptibility, whereas $\ast$ denotes convolution (detailed discussion will be given below). 

By projecting the output field $\hat{P}_{\text{L,out}}(t)$ onto a (real-valued) post-processing filter $G(t)$, the estimation of the force amplitude is given by
\begin{equation}
\hat{A}_{\text{est}}=\frac{\int_{-\infty}^{\infty}G(t)\hat{P}_{\text{L,out}}(t)dt}{\int_{-\infty}^{\infty}G(t)S(t)dt}=A_{0}+\frac{\int_{-\infty}^{\infty}G(t)\hat{f}_{\text{add}}(t)dt}{\int_{-\infty}^{\infty}G(t)S(t)dt}.\label{eq:A-est}
\end{equation}
The variance of the (zero-mean) noise term on the right-hand side of Eq.~\eqref{eq:A-est} determines the ability to resolve the presence (or absence) of the prescribed signal on top of the noise, i.e., the sensitivity. It can be calculated in the Fourier domain in terms of the symmetrized noise spectral density $N(\Omega)$, defined via $\langle\hat{f}_{\text{add}}^{\dagger}(\Omega)\hat{f}_{\text{add}}(\Omega')\rangle+\langle\hat{f}_{\text{add}}(\Omega)\hat{f}^{\dagger}_{\text{add}}(\Omega')\rangle:=2N(\Omega)\delta(\Omega-\Omega')$~\cite{Clerk2010}, as,
\begin{equation}
V=\langle(\hat{A}_{\text{est}}-A_{0})^{\dagger}(\hat{A}_{\text{est}}-A_{0})\rangle
=\frac{\int_{-\infty}^{\infty}d \Omega \int_{-\infty}^{\infty}d \Omega'G(\Omega)G^{\ast}(\Omega')\langle\hat{f}_{\text{add}}^{\dagger}(\Omega)\hat{f}_{\text{add}}(\Omega')\rangle}{(\int_{-\infty}^{\infty}G^{\ast}(\Omega)S(\Omega)d \Omega)^{2}}
=\frac{\int_{-\infty}^{\infty}d \Omega |G(\Omega)|^2 N(\Omega)}{(\int_{-\infty}^{\infty}G^{\ast}(\Omega)S(\Omega)d \Omega)^{2}};
\label{eq:sen_var}
\end{equation}
the symmetrized spectrum enters due to the evenness of $|G(\Omega)|^2$ (which follows from $G(-\Omega)=G^{\ast}(\Omega)$). 
Applying the matched filter function $G(\Omega)=S(\Omega)/N(\Omega)$ to extract the signal from the measurement record optimally~\cite{TurinJune1960}, the sensitivity~\eqref{eq:sen_var} equals
\begin{equation}
V=\left[\int_{-\infty}^{\infty}\frac{|S(\Omega)|^{2}}{N(\Omega)}d \Omega\right]^{-1}.
\label{eq:sen_var_optimal}
\end{equation}

Considering our hybrid system~\eqref{eq:Le-nophase} in the lab frame, the equations of motion in the frequency domain after adiabatic elimination of the optomechanical cavity are  (using the convention Eq.~\eqref{eq:FT})
\begin{eqnarray}
-i\Omega \hat{a}_{\text{M}}(\Omega)&=&-i\tilde{\Omega}_{\text{M}}\hat{a}_{\text{M}}(\Omega)-\frac{\gamma_{\text{M}}}{2}\hat{a}_{\text{M}}(\Omega)+\sqrt{\gamma_{\text{M,0}}}[\hat{a}_{\text{M,in}}(\Omega)+A_{0}f(\Omega)]+i[\sqrt{\Gamma_{\text{MB}}}\hat{b}_{\text{M,in}}(\Omega)+\sqrt{\Gamma_{\text{MP}}}\hat{b}_{\text{M,in}}^{\dagger}(-\Omega)],
\nonumber\\
-i\Omega \hat{a}_{\text{S}}(\Omega)&=&-i\Omega_{\text{S}}\hat{a}_{\text{S}}(\Omega)-\frac{\gamma_{\text{S}}}{2}\hat{a}_{\text{S}}(\Omega)+\sqrt{\gamma_{\text{S,0}}}\hat{a}_{\text{S,in}}(\Omega)+i[\sqrt{\Gamma_{\text{SB}}}\hat{b}_{\text{in}}(\Omega)+\sqrt{\Gamma_{\text{SP}}}\hat{b}_{\text{in}}^{\dagger}(-\Omega)],\label{eq:EOM-freq}
\end{eqnarray}
where we will again assume $\Omega_{\text{S}}=-\tilde{\Omega}_{\text{M}}<0$ 
(in terms of the effective resonance frequencies introduced in Section~\ref{sec:spin-sys})
and we have now included the external mechanical force $f(\Omega)$. Assuming that the system response is concentrated spectrally around its resonant frequency ($|\Omega_{\text{S}}| \gg \gamma_{\text{S}}$, $\tilde{\Omega}_{\text{M}} \gg \gamma_{\text{M}}$), the input-output relation for the hybrid system is given by (in absence of transmission loss, $\epsilon=0$) 
\begin{eqnarray}
\hat{P}_{\text{L,out}}(\Omega)&=&\hat{P}_{\text{L,in}}(\Omega)+\frac{\sqrt{\Gamma_{\text{SB}}}+\sqrt{\Gamma_{\text{SP}}}}{\sqrt{2}}\hat{a}_{\text{S}}^{\dagger}(-\Omega)+\frac{\sqrt{\Gamma_{\text{MB}}}+\sqrt{\Gamma_{\text{MP}}}}{\sqrt{2}}[\hat{a}_{\text{M}}(\Omega)+\sqrt{\gamma_{\text{M,0}}}\chi_{\text{M}}(\Omega)A_{0}f(\Omega)],
\label{eq: i/o relation f}
\end{eqnarray}
when $\Omega>0$, and $\hat{P}_{\text{L,out}}(-\Omega)=\hat{P}_{\text{L,out}}^{\dagger}(\Omega)$. Here $\chi_{\text{M}}(\Omega)=1/[\gamma_{\text{M}}/2+i(\tilde{\Omega}_{\text{M}}-\Omega)]$ is the mechanical susceptibility. The signals of the forces that can be read from Eq.~\eqref{eq: i/o relation f} are 
\begin{eqnarray}
S(\Omega)&=&(\sqrt{\Gamma_{\text{MB}}}+\sqrt{\Gamma_{\text{MP}}}) \sqrt{\gamma_{\text{M,0}}}\chi_{\text{M}}(\Omega)f(\Omega)/\sqrt{2},
\nonumber\\
\hat{f}_{\text{add}}(\Omega)&=&\hat{P}_{\text{L,out}}(\Omega)-S(\Omega).\label{eq:S-freq}
\end{eqnarray}

\begin{figure}[t]
\includegraphics[width=0.4\columnwidth]{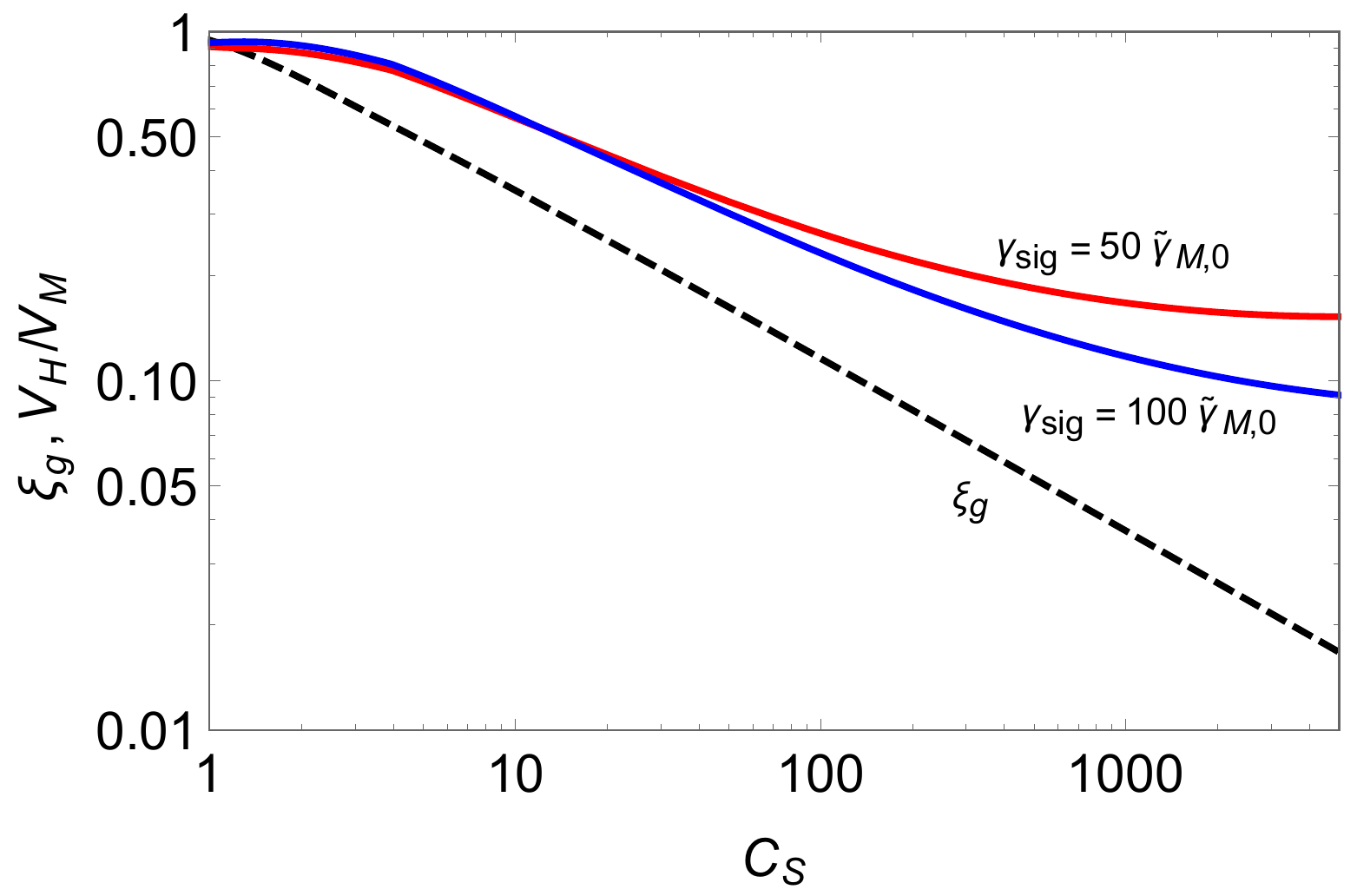}
\caption{Minimized unconditional EPR-variance $\xi_{g}$ (dashed) and the resulting sensitivity enhancement of the hybrid system compared with the SQL for mechanics-only sensing $V_{\text{H}}/V_{\text{M}}$ (solid) as a function of spin quantum cooperativity $C_{\text{S}}$ for fixed parameters $\gamma_{\text{S,}0}=2\pi \times 5 \text{kHz}, \bar{n}_{\text{S}}=1, \gamma_{\text{M,}0}\bar{n}_{\text{M}}= 2\pi \times 10 \text{kHz}$, and $\epsilon=0$.}
\label{fig:ent_sens}
\end{figure}

To illustrate the relation between entanglement and sensitivity performance, we consider the example of a force signal with a Lorentzian spectrum centered at the mechanical resonance $f(\Omega)=\sqrt{\gamma_{\text{sig}}}/[(\gamma_{\text{sig}}/2+i(\tilde{\Omega}_{\text{M}}-\Omega))]$, and $\int_{-\infty}^{\infty}|f(\Omega)|^{2}d\Omega=4\pi$ for $\tilde{\Omega}_{\text{M}}\gg\gamma_{\text{sig}}$.
We will benchmark the sensing enhancement provided by the hybrid system against the standard quantum limit (SQL) $V_{\text{M}}$, which is the minimal sensitivity~\eqref{eq:sen_var_optimal}
achievable in absence of the spin system ($\Gamma_{\text{SB/P}}=0$) when optimizing over the mechanical parameters $C_{\text{M}}$, $\theta_{\text{M}}$ for fixed decoherence rate $\tilde{\gamma}_{\text{M,0}}$ and signal bandwidth $\gamma_{\text{sig}}$. 
In order to link with entanglement, we use the parameters $C_{\text{M}}$ and $\theta_{\text{S/M}}$ given by optimizing the EPR variance $\xi_{g}$ for fixed decoherence rate $\tilde{\gamma}_{\text{S/M,0}}$ and signal bandwidth $\gamma_{\text{sig}}$ to evaluate the sensitivity of the hybrid system $V_{\text{H}}$ using Eqs.~(\ref{eq:sen_var_optimal}--\ref{eq:S-freq}). 
To the end of evaluating $V_{\text{M/H}}$, we note that for the mechanical mode alone, the added symmetrized noise spectrum is found from Eqs.~(\ref{eq:EOM-freq}--\ref{eq:S-freq}) to be 
\begin{eqnarray}
N_{\text{M}}(\Omega)=\frac{1}{2}+(|\chi_{\text{M}}(\Omega)|^{2}+|\chi_{\text{M}}(-\Omega)|^{2})\left[\frac{(\sqrt{\Gamma_{\text{MB}}}+\sqrt{\Gamma_{\text{MP}}})^{2}}{2}(\frac{\Gamma_{\text{MB}}+\Gamma_{\text{MP}}}{2}+\tilde{\gamma}_{\text{M,0}})-\frac{\gamma_{\text{M}}(\gamma_{\text{M}}-\gamma_{\text{M,0}})}{4}\right]
\end{eqnarray}
whereas for the the hybrid system, we find 
\begin{eqnarray}
N_{\text{H}}(\Omega)&=&\frac{1}{2}+h_{1}(|\chi_{\text{S}}(\Omega)|^{2}+|\chi_{\text{S}}(-\Omega)|^{2})+h_{2}(|\chi_{\text{M}}(\Omega)|^{2}+|\chi_{\text{M}}(-\Omega)|^{2})+h_{3}(|\chi_{\text{M}}(\Omega)|^{2}|\chi_{\text{S}}(-\Omega)|^{2}+|\chi_{\text{M}}(-\Omega)|^{2}|\chi_{\text{S}}(\Omega)|^{2})
\nonumber\\
&&+h_{4}(|\chi_{\text{M}}(\Omega)|^{2}|\chi_{\text{S}}(-\Omega)|^{2}(\Omega-\tilde{\Omega}_{\text{M}})^{2}+|\chi_{\text{M}}(-\Omega)|^{2}|\chi_{\text{S}}(\Omega)|^{2}(\Omega+\tilde{\Omega}_{\text{M}})^{2})
\end{eqnarray}
where the coefficients are given by 
\begin{eqnarray}
h_{1}&=&\frac{(\sqrt{\Gamma_{\text{SB}}}+\sqrt{\Gamma_{\text{SP}}})^{2}}{2}(\frac{\Gamma_{\text{SB}}+\Gamma_{\text{SP}}}{2}+\tilde{\gamma}_{\text{S,0}})-\frac{\gamma_{\text{S}}(\gamma_{\text{S}}-\gamma_{\text{S,0}})}{4},
\nonumber\\
h_{2}&=&\frac{(\sqrt{\Gamma_{\text{MB}}}+\sqrt{\Gamma_{\text{MP}}})^{2}}{2}(\frac{\Gamma_{\text{MB}}+\Gamma_{\text{MP}}}{2}+\tilde{\gamma}_{\text{M,0}})-\frac{\gamma_{\text{M}}(\gamma_{\text{M}}-\gamma_{\text{M,0}})}{4},
\nonumber\\
h_{3}&=&\frac{\gamma_{\text{M}}(\sqrt{\Gamma_{\text{SB}}}+\sqrt{\Gamma_{\text{SP}}})(\sqrt{\Gamma_{\text{MB}}}+\sqrt{\Gamma_{\text{MP}}})}{2}[R(\frac{\Gamma_{\text{SB}}+\Gamma_{\text{SP}}}{2}+\tilde{\gamma}_{\text{S,0}})-\frac{\gamma_{\text{S}}(\sqrt{\Gamma_{\text{SB}}\Gamma_{\text{MP}}}+\sqrt{\Gamma_{\text{SP}}\Gamma_{\text{MB}}})}{4}]
\nonumber\\
&&+\frac{R(\sqrt{\Gamma_{\text{MB}}}+\sqrt{\Gamma_{\text{MP}}})^{2}}{2}[R(\frac{\Gamma_{\text{SB}}+\Gamma_{\text{SP}}}{2}+\tilde{\gamma}_{\text{S,0}})-\frac{\gamma_{\text{S}}(\sqrt{\Gamma_{\text{SB}}\Gamma_{\text{MP}}}+\sqrt{\Gamma_{\text{SP}}\Gamma_{\text{MB}}})}{2}]
\nonumber\\
&&-\frac{R\gamma_{\text{S}}\gamma_{\text{M}}(\sqrt{\Gamma_{\text{SB}}}-\sqrt{\Gamma_{\text{SP}}})(\sqrt{\Gamma_{\text{MB}}}+\sqrt{\Gamma_{\text{MP}}})}{8}
\nonumber\\
h_{4}&=&-(\sqrt{\Gamma_{\text{SB}}}+\sqrt{\Gamma_{\text{SP}}})(\sqrt{\Gamma_{\text{MB}}}+\sqrt{\Gamma_{\text{MP}}})\frac{\sqrt{\Gamma_{\text{SB}}\Gamma_{\text{MP}}}+\sqrt{\Gamma_{\text{SP}}\Gamma_{\text{MB}}}}{2}+R\frac{(\sqrt{\Gamma_{\text{SB}}}-\sqrt{\Gamma_{\text{SP}}})(\sqrt{\Gamma_{\text{MB}}}+\sqrt{\Gamma_{\text{MP}}})}{2},
\end{eqnarray}
and $R=\sqrt{\Gamma_{\text{SB}}\Gamma_{\text{MP}}}-\sqrt{\Gamma_{\text{SP}}\Gamma_{\text{MB}}}$ as above.

We plot the sensing enhancement $V_{\text{H}}/V_{\text{M}}$ alongside the corresponding unconditional EPR variance $\xi_g$ in Fig.~\ref{fig:ent_sens}. Here it is seen that the improvement in unconditional entanglement with increasing spin cooperativity $C_{\text{S}}$ goes hand in hand with sub-SQL sensing enhancement for broadband force signals $\tilde{\Omega}_{\text{M}} \gg \gamma_{\text{sig}} \gg \tilde{\gamma}_{\text{M,0}}$. This is evidence that there is a link between unconditional entanglement generation and sub-standard quantum limit (SQL) sensitivity when applying the hybrid system as a continuous force sensor.

\bibliography{Entanglement}